\documentclass[prd,reprint,nofootinbib,eqsecnum,longbibliography]{revtex4-1}
\pdfoutput=1
\usepackage{amsmath,amssymb,bm}
\usepackage[colorlinks=true,citecolor=blue]{hyperref}
\usepackage{color, xcolor, enumerate}
\usepackage[normalem]{ulem}
\usepackage{longtable,array,multirow,graphicx,afterpage}

\usepackage{array,multirow}
\newcolumntype{L}[1]{>{\raggedright\let\newline\\\arraybackslash\hspace{0pt}}m{#1}}
\newcolumntype{C}[1]{>{\centering\let\newline\\\arraybackslash\hspace{0pt}}m{#1}}
\newcolumntype{R}[1]{>{\raggedleft\let\newline\\\arraybackslash\hspace{0pt}}m{#1}}

\newcommand{\Lie}{\mathcal{L}}
\newcommand{\bx}{\mathbf{x}}
\newcommand{\bz}{\mathbf{z}}

\newcommand{\subH}{ \textrm{\fontsize{6}{8}\selectfont{H}} }
\newcommand{\subB}{ \textrm{\fontsize{6}{8}\selectfont{B}} }
\newcommand{\subG}{ \textrm{\fontsize{6}{8}\selectfont{G}} }
\newcommand{\subSF}{ \textrm{\fontsize{6}{8}\selectfont{SF}} }
\newcommand{\hold}{\mathrm{hold}}
\newcommand{\subS}{\textrm{\fontsize{6}{8}\selectfont{S}}}
\newcommand{\subDW}{\textrm{\fontsize{6}{8}\selectfont{DW}}}

\newcommand{\bperp}{\mbox{\fontsize{6}{8}\selectfont{\pmb{$\perp$}}}}
\newcommand{\subTF}{\textrm{\fontsize{6}{8}\selectfont{TF}}}

\newcommand{\bfx}{{\bf{x}}}

\newcommand{\bfw}{{\bf{w}}}

\newcommand{\bfv}{{\bf{v}}}
\newcommand{\bfr}{{\bf{r}}}
\newcommand\redsout{\bgroup\markoverwith{\textcolor{red}{\rule[0.5ex]{2pt}{0.4pt}}}\ULon}

\def\be{\begin{equation}}
\def\ee{\end{equation}}

\begin{document}
\title{Self-forces on static bodies in arbitrary dimensions}

\author{Abraham I. Harte$^1$, \'Eanna \'E. Flanagan$^{2,3}$ and Peter Taylor$^{2,4}$}
\affiliation{$^1$Max Planck Institute for Gravitational Physics (Albert Einstein Institute), Am M\"{u}hlenberg 1, D-14476 Potsdam-Golm, Germany}
\affiliation{$^2$Cornell Center for Astrophysics and Planetary Science, Cornell
University, Ithaca, NY 14853}
\affiliation{$^3$Department of Physics, Cornell University, Ithaca, NY 14853}
\affiliation{$^4$School of Mathematical Sciences and Complex \& Adaptive Systems Laboratory,
University College Dublin, UCD, Belfield, Dublin 4, Ireland}

%
%
\newcount\hh
\newcount\mm
\mm=\time
\hh=\time
\divide\hh by 60
\divide\mm by 60
\multiply\mm by 60
\mm=-\mm
\advance\mm by \time
\def\hhmm{\number\hh:\ifnum\mm<10{}0\fi\number\mm}

\begin{abstract}

We derive exact expressions for the scalar and electromagnetic self-forces and self-torques acting on arbitrary static extended bodies in arbitrary static spacetimes with any number of dimensions. Non-perturbatively, our results are identical in all dimensions. Meaningful point particle limits are quite different in different dimensions, however. These limits are defined and evaluated, resulting in simple ``regularization algorithms'' which can be used in concrete calculations. In these limits, self-interaction is shown to be progressively less important in higher numbers of dimensions; it generically competes in magnitude with increasingly high-order extended-body effects. Conversely, we show that self-interaction effects can be relatively large in $1+1$ and $2+1$ dimensions. Our motivations for this work are twofold: First, no previous derivation of the self-force has been provided in arbitrary dimensions, and heuristic arguments presented by different authors have resulted in conflicting conclusions. Second, the static self-force problem in arbitrary dimensions provides a valuable testbed with which to continue the development of general, non-perturbative methods in the theory of motion. Several new insights are obtained in this direction, including a significantly improved understanding of the renormalization process. We also show that there is considerable freedom to use different ``effective fields'' in the laws of motion---a freedom which can be exploited to optimally simplify specific problems. Different choices give rise to different inertias, gravitational forces, and electromagnetic or scalar self-forces, but there is a sense in which none of these quantities are individually accessible to experiment. Certain combinations are observable, however, and these remain invariant under all possible field redefinitions.

\end{abstract}
\vskip 1pc

\maketitle

\section{Introduction and Summary}

Originally prompted by the discovery of the electron \cite{Sutherland, Abraham, Lorentz}, the past century has seen considerable effort devoted to understanding how the motions of charged particles might be affected by ``their own'' fields: What, for example, are the radiation-reaction forces? In what sense does self-interaction impart an effective inertia? While much has been learned over the years, ``self-force problems'' such as these have been notoriously subtle, and work on them continues to the present day.

Current interest has largely shifted to the gravitational variant of the self-force problem: How do the metric perturbations sourced by small masses affect their motion in general relativity? This is relevant for the anticipated observation of gravitational waves generated by extreme-mass-ratio inspirals---neutron stars or stellar-mass black holes orbiting and then falling into supermassive black holes \cite{PauReview, PauReview2}. The gravitational self-force is also relevant more broadly in gravitational wave astronomy in that it provides checks of, and inputs to, the post-Newtonian and effective-one-body approximation schemes \cite{PhysRevD.81.064004, PhysRevD.83.044044}.

Motivated by these developments, theoretical understanding of the self-force has improved enormously in the past two decades, and not only in gravitational contexts. In four spacetime dimensions, it is now understood how to rigorously formulate point particle limits, and what the equations of motion are in those limits \cite{GrallaWald, PoissonLR, BarackReview, PoundReview, HarteReview}. Non-perturbative results are available as well, describing motion and self-interaction for extended bodies in very general settings \cite{HarteReview}. All of this has been accomplished in generic spacetimes and for objects coupled to gravitational, electromagnetic, or scalar fields. Considerable effort has also been devoted to developing practical computational schemes with which to evaluate the physical consequences of the derived laws of motion \cite{BarackReview, WardellReview}, particularly for small (uncharged) masses in orbit around nearly-Kerr black holes.

For spacetime dimensions not equal to four, the self-force program is considerably less mature. Absent any rigorous derivations, a number of \textit{ad hoc} methods have been suggested to compute (mostly higher-dimensional) self-forces in various contexts \cite{TaylorFlanagan, Poisson5dSF, FrolovZelnikovSF, Shuryak, Kosyakov, Galtsov, Galakhov, Kazinski}. Although it is not possible to compare all of these methods directly, it is known that at least some of them are inequivalent. For example, the work of Beach, Poisson, and Nickel \cite{Poisson5dSF} suggested that the self-force on a charged particle in five spacetime dimensions might depend in an essential way on the details of that particle's internal structure, even if it were spherically symmetric. An analysis of the same system by Taylor and Flanagan \cite{TaylorFlanagan} utilized a different method and found conflicting results. Unexplained ambiguities arose in both cases, although these had very different characters. If the ambiguities in either approach were in some sense correct, they would represent surprising departures from the known behavior of the four-dimensional self-force. One motivation for this paper is to clarify these issues, and more generally to determine if the self-force depends in any essential way on dimensionality.

Besides matters of principle such as these, a more direct reason for considering non-standard numbers of dimensions is in connection with holographic dualities; it has been claimed in this context that the $4+1$ dimensional self-force problem can be used to understand jet quenching in $3+1$ dimensional quark-gluon plasmas \cite{Shuryak, Shuryak2}.

Separately, lower-dimensional self-force effects might be directly accessible to experiment: There are, for example, systems where liquid droplets bouncing on an oil bath generate surface waves, and those waves in turn affect the horizontal motion of the droplet \cite{PilotWave}. This is at least qualitatively a self-force problem in two spatial dimensions. There are also a variety of condensed matter systems which act as though they are confined to one or two spatial dimensions (see e.g., \cite{TopIns, QED2}), and if this type of confinement could be arranged for something analogous to an isolated charge---as has recently been suggested for deformed graphene \cite{Stegmann}---it might be relatively straightforward to measure self-interaction effects in a wide variety of geometries. Different spatial metrics and topologies could be explored by varying the confining surface, and external accelerations might be used to introduce nontrivial lapse functions.

We do not attempt to model any such systems here, but instead consider as a first step a ``standard'' self-force system in arbitrary dimensions: isolated extended bodies coupled to scalar or electromagnetic fields in fixed background spacetimes. Our treatment is exact except for the neglect of gravitational backreaction. We also assume that both the spacetime and the body of interest are static.

Although the staticity constraint might appear to be overly restrictive, it already allows for a number of interesting statements. The aforementioned disagreements in the existing literature \cite{Poisson5dSF, TaylorFlanagan} appear, for example, in the static regime. Focusing attention on the static problem can also highlight interesting features which are not otherwise apparent---even in cases where the dynamical equations are already known. Lastly, static systems provide a simple testing ground with which to develop new insights into more general self-interaction problems.

Before describing our results in these directions, we first remark on the status of the dynamical self-force problem in non-standard dimensions: Although it was alluded to briefly in \cite{HarteReview}, it does not appear to have been emphasized before that much of the existing non-perturbative work developed to describe the $3+1$ dimensional self-force \cite{HarteSyms, HarteScalar, HarteEM, HarteTrenorm, HarteGrav} generalizes immediately to other dimensions. One of its implications is that a result known as the Detweiler-Whiting prescription\footnote{The Detweiler-Whiting prescription originally arose as a regularization procedure which succinctly describes the motions of point particles in four spacetime dimensions. It was later shown to be the limit of an exact, non-singular identity which holds for generic extended bodies. Both the identity and its limit generalize to all even-dimensional spacetimes.} \cite{PoissonLR, DetweilerWhiting2003, HarteReview} generalizes and remains exact for fully-dynamical extended bodies in all even-dimensional spacetimes. A problem arises, however, if the number of spacetime dimensions is odd; a construction known as the Detweiler-Whiting Green function appears not to exist. While this does not appear to be a fundamental obstacle, it does imply that known results require some modification before being extended to the odd-dimensional dynamical setting. A possible solution to this problem is briefly discussed in \ref{Sect:noStatic}, although it is not our main theme.

We instead focus on the static self-force problem, in both odd and even-dimensional spacetimes. Our approach uses and builds upon the aforementioned non-perturbative techniques developed by Harte \cite{HarteSyms, HarteScalar, HarteEM, HarteTrenorm, HarteGrav}, which themselves were inspired by the work of Mathisson \cite{Mathisson} and especially Dixon \cite{Dix70a, Dix74, Dix79, DixonReview}. These techniques allow the bulk properties of extended bodies to be understood exactly in generic spacetimes, and automatically provide, e.g., precise definitions for all quantities which appear in the resulting laws of motion. One convenient feature of this approach is that a body's linear and angular momenta are treated as two aspects of a single mathematical structure, and consequently, the self-torque emerges ``for free'' with the self-force.

It is much more common in the self-force literature to employ perturbative methods (see e.g., \cite{PoundReview, GrallaHarteWald, GrallaWald}), which are perhaps more familiar. While these methods could also be applied in the present context, they typically require calculations which must be repeated almost from scratch in each new dimension, and the complexity of those calculations grows rapidly with the number of dimensions. No such problems arise for the non-perturbative approach adopted here. Our methods are almost completely agnostic to the number of dimensions, and are simpler than the perturbative approach even in $3+1$ dimensions.

The essential difficulty of the self-force problem is that the net force exerted on an object depends on the fields inside of it, but these fields can be almost arbitrarily complicated. In particular, the internal fields vary at least on lengthscales comparable to the body's size (and perhaps on much smaller scales as well). This makes it difficult to transform integral expressions for the net force---whose evaluation might appear to require detailed knowledge of an object's interior---into simple expressions which can be used without that knowledge.

The main points can be illustrated even in Newtonian gravity \cite{Dix79, HarteScalar, HarteReview}, although they are so simple in that case as to rarely be emphasized. Very briefly, consider a compact extended body in three-dimensional Euclidean space. If this body has mass density $\rho_m$, the Newtonian gravitational potential $\phi_g$ satisfies $\nabla^2 \phi_g = 4 \pi \rho_m$ and the net gravitational force is
\begin{equation}
  \bm{F} = -\int \rho_m (\bx) \bm{\nabla} \phi_g(\bx) d^3 x.
  \label{Fnewt}
\end{equation}
The integrand here can be arbitrarily complicated, and one might naively expect that the force depends in an essential way on these complications. That this is not the case follows from the observation that for any translation-invariant ``propagator'' $G(\bx, \bx')$, the net force $\bm{F}$ is invariant under all field replacements $\phi_g \rightarrow \hat{\phi}_g$ with the form
\begin{equation}
  \hat{\phi}_g (\bx) \equiv \phi_g (\bx) - \int \rho_m (\bx') G(\bx,\bx') d^3 \bx'.
  \label{phiHatNewt}
\end{equation}
In practice, this result is typically applied in the special case where $G(\bx,\bx') = - 1/|\bx-\bx'|$, which satisfies $\nabla^2 G(\bx,\bx') = 4 \pi \delta^3(\bx-\bx')$ and is therefore a Green function for the Newtonian field equation. Considering that case, $\hat{\phi}_g$ satisfies the vacuum field equation in a neighborhood of the body and may thus be interpreted as an effective ``external field.'' This is useful because external fields typically behave much more simply than physical ones: If all distances to other masses are sufficiently large, $\bm{\nabla} \hat{\phi}_g$ varies only slightly in the force integral \eqref{Fnewt}, and may therefore be pulled out of it to yield
\begin{equation}
        \bm{F} = - m \bm{\nabla} \hat{\phi}_g.
        \label{FnewtSimp}
\end{equation}
This is the foundation for most of Newtonian celestial mechanics. For our purposes, it is important to emphasize that it is the ``effective field'' $\hat{\phi}_g$ which appears in simple expressions for the force, not the physical field $\phi_g$. Except in special cases such as spherical symmetry, it is not correct to replace the right-hand side of \eqref{FnewtSimp} by $-m \bm{\nabla} \phi_g$. In a point particle limit, the map $\phi_g \mapsto \hat{\phi}_g$ becomes a type of regularization procedure; the ``force on a point particle'' can be described as the monopole force due to a point particle field which has been regularized in a particular way. This result should not be viewed as ``fundamental,'' but rather as a corollary to the $\phi_g \rightarrow \hat{\phi}_g$ invariance of $\bm{F}$.

Such comments suggest that it can be essential also in more complicated theories to express force laws in terms of fields which are distinct from the physical ones. Moreover, those fields should remain regular even in point particle limits (as long as such limits exist). The steps outlined above which provide the appropriate prescription in the Newtonian context also provide an outline for this paper: We i) generalize \eqref{Fnewt} for static, charged bodies in curved spacetimes with arbitrary dimension, ii) derive a result analogous to the $\phi_g \rightarrow \hat{\phi}_g$ invariance of Newtonian theory, and iii) show that for appropriate choices of effective field, the associated force integrals admit simple approximations similar to \eqref{FnewtSimp}. The result is a concrete prescription for computing self-interaction effects in arbitrary dimensions.

There are, of course, considerable differences between our problem and the Newtonian one. Perhaps the most significant of these is that forces do not necessarily remain fixed when replacing physical fields by effective fields. We nevertheless show that if the class of effective fields is chosen appropriately, the resulting changes have a special form which allows them to be absorbed into finite renormalizations of a body's stress-energy tensor. The Newtonian statement that forces remain invariant under replacements $\phi_g \rightarrow \hat{\phi}_g$ is therefore replaced by a statement that relativistic forces are preserved by simultaneous replacements involving both long-range fields \textit{and} stress-energy tensors (but not, e.g., charge distributions). This considerably generalizes the mass renormalization effect which has been discussed since the earliest work on electromagnetic self-interaction \cite{Sutherland, Abraham, Lorentz}.

Although the result that stress-energy tensors are renormalized by self-interaction has been recognized before \cite{HarteTrenorm,HarteReview}, we obtain several new features of this effect. In prior work on the dynamical self-force problem, two mechanisms were identified by which renormalizations could occur. One of these depended on a kind of ``temporal boundary term,'' and affected only a body's linear and angular momenta---essentially the monopole and dipole moments of its stress-energy tensor \cite{HarteScalar}. Although we find that monopole and dipole moments are also renormalized in the static problem, the mathematical mechanism by which this occurs is different and is identified here for the first time.

In dynamical settings, the quadrupole and higher multipole moments of a body's stress-energy tensor---but not its monopole and dipole---had previously been found to be renormalized via the dependence of a particular propagator on the background geometry  \cite{HarteTrenorm}. We show that this same mechanism also plays a role in static problems, but make it more precise by providing the first explicit, non-perturbative formulae for its effects.

Although the two renormalization mechanisms at work here appear to affect different quantities and to have different origins, we show that they are nevertheless ``compatible'' in the sense that a single non-perturbative formula can be obtained for a renormalized stress-energy tensor $\hat{T}_{\subB}^{ab}$. All stress-energy moments which appear in the laws of motion, including the momenta, then follow from Dixon's integral definitions \cite{Dix74} applied to $\hat{T}^{ab}_\subB$ (instead of their usual application to a body's ``bare'' stress-energy tensor $T^{ab}_\subB$). We also show that the difference between $\hat{T}^{ab}_\subB$ and $T^{ab}_\subB$ depends only on a body's charge density and functional derivatives of an appropriate propagator with respect to the geometric fields. Even though this difference at least roughly describes ``the stress-energy of the self-field,'' it is interesting to note that its support cannot extend significantly beyond that of $T^{ab}_\subB$. Characteristic magnitudes of the effective moments can therefore be estimated in the usual ways using only a body's size and effective mass.

These kinds of stress-energy renormalizations arise when replacing the true scalar or electromagnetic fields by effective equivalents which are related by equations similar to the Newtonian $\phi_g \mapsto \hat{\phi}_g$ map \eqref{phiHatNewt}. Given, e.g., a relativistic scalar field $\phi$, an appropriate propagator $G$ may be introduced and used to construct an effective field $\hat{\phi}$. While the class of allowed propagators is strongly constrained, it is far from unique. It is neither necessary nor sufficient, for example, that $G$ be a Green function. In general, each allowable propagator applied to the same physical system implies a different $\hat{\phi}$ and a different $\hat{T}_{\subB}^{ab}$. As a consequence, individual terms which one might want to identity as \textit{the} self-force or \textit{the} gravitational force involve some degree of choice---a fact which seems to have been missed in the existing literature (even in four spacetime dimensions\footnote{The dominant effect in the point particle limit in four spacetime dimensions is a degeneracy between the inertia term (mass times acceleration), and the piece of the self-force which is proportional to acceleration.}). A key insight of this paper is that in general, scalar or electromagnetic self-forces cannot be divorced from inertial forces or gravitational extended-body effects; it is only particular combinations of these quantities which are physically unambiguous. While these remarks imply that additional care can be required when interpreting self-force results, the freedom to choose different propagators also opens up new possibilities for practical computations: One can choose whichever propagator is simplest for the problem at hand. We illustrate the usefulness of this explicitly in Rindler spacetimes, where different propagator choices result in very different levels of computation.

Our conclusions on the general nature of the static self-force may be stated as follows:  Except for the methods used in certain existence results presented in appendix \ref{Sect:Hadamard}, all of our non-perturbative arguments are independent of dimension. The dimension of spacetime is therefore irrelevant to any foundational aspects of the problem. In particular,  there is no more dependence on a body's internal structure in higher dimensions than there is in four spacetime dimensions.  Nevertheless, there is a sense in which dependence on internal structure does arise, in both four spacetime dimensions and in higher and lower dimensions, via renormalization of body parameters.  This issue is discussed in detail in section \ref{Sect:ST}.

Additionally, we find no obstacle to constructing well-behaved point particle limits. Dimension does, however, affect the details of the point particle limits which can be meaningfully considered. This can be understood by noting that the self-energy of a charge distribution depends on its size in a dimension-dependent way. Noting that a body's self-energy cannot significantly exceed its mass without violating positive-energy conditions, dimension-dependent bounds may be obtained which relate the relative magnitudes of different types of forces. We show more specifically that the leading-order electric or scalar self-force in an $n+1$ dimensional spacetime can at most be comparable in size to extended-body effects which involve a body's $2^{(n-2)}$-pole moments (for $n \geq 2$). In the usual $n=3$ case, it follows that the self-force is at most comparable to ordinary dipole effects. For larger $n$, quadrupole or higher moments must be taken into account as well. In lower dimensions, the self-force can instead compete even with leading-order test-body effects.

This paper is organized as follows: Section \ref{Sect:PhysProb} describes the overall setup for the problems we consider, including the ``holding field'' which we take to be a primary observable. Our core non-perturbative results are derived in section \ref{Sect:mf}, which defines generalized momenta for extended bodies and obtains the associated forces. A class of identities is derived there which allows self-interaction to be taken into account in relatively simple ways. Renormalization effects are derived as well. Next, section \ref{Sect:Multipole} describes how to convert integral expressions for the generalized force into series involving a body's multipole moments. A center-of-mass is defined, as well as a split of the generalized momentum into linear and angular components. Forces and torques necessary to hold an object fixed are obtained to all multipole orders. Approximations are first considered in section \ref{Sect:pp}, which discusses what could be meant by a point particle limit. These limits are subsequently defined and an associated algorithm is derived which can be used to compute the limiting force and torque.  Renormalization of a body's mass and stress-energy quadrupole in the point particle limit are explicitly computed in section \ref{Sect:bodyR}. Section \ref{Sect:Examples} compares the approach used here to others in the literature, and applies our ideas explicitly by giving examples of calculations in Rindler and Schwarzschild-Tangherlini spacetimes. Lastly, section \ref{Sect:noStatic} speculates on how to generalize this work to dynamical settings.

Several additional results have been placed in appendices. Notations and conventions used throughout this paper are explained in appendix \ref{app:notation}. Appendix \ref{Sect:Hadamard} discusses Hadamard Green functions and parametrices, and shows that the latter are explicit examples of the type of propagator whose existence we require. Appendix \ref{Sect:DW} shows that in even spacetime dimensions where the Detweiler-Whiting prescription is valid for dynamical charges, specializing it to static systems results in a prescription which is consistent with our \textit{a priori} static results derived in section \ref{Sect:mf}. Appendix \ref{Sect:RindlerAppendix} supplements section \ref{Sect:Rindler} by providing an alternative derivation of the self-force in Rindler spacetime.
Finally, appendix \ref{Sect:VarDerivs} computes the variational derivatives of the Hadamard parametrix, for use in the renormalization computations of section \ref{Sect:bodyR}.

\section{The setting: static extended bodies in static spacetimes}
\label{Sect:PhysProb}

The systems we consider consist of a spatially-compact body $B$ embedded in a static, $n+1$ dimensional spacetime (with $n \geq 1$). Rather than releasing this object and letting it fall freely, we instead imagine that it is held in place and is internally stationary: There must exist a timelike vector field $\tau^a$ such that the spacetime metric $g_{ab}$ and the body's stress-energy tensor $T^{ab}_\subB$ satisfy
\begin{equation}
  \mathcal{L}_\tau g_{ab} = \mathcal{L}_\tau T^{ab}_{\subB} = 0,
    \label{Static}
\end{equation}
where $\mathcal{L}_\tau$ denotes the Lie derivative with respect to $\tau^a$. This generically requires the imposition of external forces and torques, and it is these quantities which represent the main physical unknowns.

\subsection{General description of the goal}
\label{Sect:genDescription}

Forces exerted via direct mechanical contact with other objects are difficult to describe generically, so we instead suppose that $B$ is endowed with some kind of charge, and that forces can be imposed by applying external ``holding fields'' which interact with that charge. The required holding \textit{force} can then be translated into a required holding \textit{field}. A central aim of this paper is to determine those holding fields which are consistent with\footnote{No externally-imposed field can \textit{imply} stationarity without a precise specification for a body's internal composition. Even in elementary Newtonian mechanics, it is only the behavior of certain bulk degrees of freedom which can be described generically. We nevertheless specialize to those cases where the internal degrees of freedom are stationary whenever the bulk is stationary.}  the staticity assumption \eqref{Static}.

This type of computation is very simple if $B$ is small and its self-fields are weak: A precise limit may then be found in which the worldtube of such a body can be replaced by a single worldline. Staticity implies that the unit velocity of this worldline is $u^a = \tau^a/N$, where $\tau^a$ is the static Killing field and
\begin{equation}
  N \equiv \sqrt{ - \tau_a \tau^a}.
  \label{NDef}
\end{equation}
Differentiating $u^a$ using Killing's equation, the body's acceleration is seen to be
\begin{equation}
  u^b \nabla_b u_a = \nabla_a \ln N,
  \label{accel}
\end{equation}
which suggests that $\ln N$ is in some ways analogous to an ordinary Newtonian potential. Applying the Lorentz force law for a body with mass $m$ and electric charge $Q$ finally shows that such an acceleration can be maintained by imposing an electromagnetic holding field $F_{ab}^\hold$ which satisfies
\begin{align}
   Q F_{ab}^\hold u^b = m \nabla_a \ln N.
  \label{HoldingF0}
\end{align}

Our aim is to generalize this equation. In particular, we would like to understand what happens when a body's self-field can no longer be neglected. One complication which then arises is that the Lorentz force law cannot be applied as it was in \eqref{HoldingF0}. That would make sense only if the field were approximately constant throughout $B$, which would be an unreasonably severe restriction.

Another potential obstacle to understanding self-interaction is that it can strongly affect internal stresses while producing very little net force; interesting effects can thus depend on delicate cancellations. Moreover, if the net self-force is small---as it is in many applications---it can be understood only in combination with other similarly-small effects. Indeed, we shall see in section \ref{Sect:pp} that generalizing \eqref{HoldingF0} to allow for nontrivial self-fields generically requires that we also generalize it to allow for finite-size effects.

Our approach exactly describes the forces and torques acting on arbitrarily-structured extended bodies, so all such effects are automatically taken into account. It is only at the end of our discussion where specific approximations are adopted and the relative magnitudes of different terms can be examined.

\subsection{Spacetime geometry}

Before proceeding, it is useful to more precisely describe the geometry of our setup and to briefly collect some of its properties: The background spacetime is assumed to have the form $(\Sigma \times I, g_{ab})$, where $\Sigma$ is an $n$-dimensional manifold and $I \subseteq \mathbb{R}$ an open interval. In all regions of interest, the timelike Killing field $\tau^a$ is assumed to be static in the sense that it satisfies the Frobenius condition $\tau_{[a} \nabla_b \tau_{c]} = 0$. Contracting this with $\tau^a$ while using \eqref{NDef} provides the useful identity
\begin{equation}
  \nabla_a \tau_b = -2 \tau_{[a} \nabla_{b]} \ln N.
  \label{gradTau}
\end{equation}

We define a time coordinate $t$ via $\tau^a = \partial/\partial t$, so the constant-$t$
hypersurfaces $\Sigma_t$ are orthogonal to $\tau^a$ and diffeomorphic to $\Sigma$. If $\tau^a$ is used to evolve between these hypersurfaces, the associated shift
vector vanishes and $N$ is the lapse. The intrinsic geometry on each $\Sigma_t$ is described by the spatial metric
\begin{equation}
  h_{ab} \equiv g_{ab} + \tau_a \tau_b/N^2,
  \label{hDef}
\end{equation}
and the spatial Ricci tensor $R^{\bperp}_{ab}$ can be related to the spacetime Ricci tensor $R_{ab}$ via
\begin{subequations}
\label{RicciProjections}
        \begin{gather}
                R_{ab} \tau^a \tau^b = N D^2 N , \qquad R_{bc} h^{b}{}_{a} \tau^c = 0,
        \\
                h^{c}_a h^d_b R_{cd} = R^{\bperp}_{ab} - N^{-1} D_a D_b N,
        \end{gather}
\end{subequations}
where $D_a$ denotes the covariant derivative associated with $h_{ab}$ and $D^2 \equiv h^{ab} D_a D_b$ is the associated Laplacian. We shall also have occasion to use a directed surface element on $\Sigma_t$, which can be written as
\begin{equation}
        dS_{a}=-N^{-1}\tau_{a}\,d V_{\bperp}
        \label{dS}
\end{equation}
in terms of the $n$-dimensional volume element $d V_{\bperp}$ associated with $h_{ab}$. If $n$ spatial coordinates $\bx$ are introduced in addition to the time coordinate $t$, the coordinate components of the metric take the form
\begin{equation}
  g_{\mu\nu} dx^\mu dx^\nu = - N^2(\bx) dt^2 + h_{ij}(\bx) dx^i dx^j.
  \label{metric}
\end{equation}
in terms of $x^\mu = (t,\bx)$.

Lastly, note the overall scale of $\tau^a$, and therefore $t$, is at least locally irrelevant. Physical quantities must therefore be invariant under all rescalings
\begin{equation}
  t \rightarrow \alpha t, \qquad N \rightarrow \alpha^{-1} N
  \label{tScale}
\end{equation}
by positive constants $\alpha$.

\subsection{Stress energy conservation and the field equations}

Embedded in the static spacetime $(\Sigma \times I, g_{ab})$ is a body $B$ whose stress-energy tensor $T^{ab}_\subB$ is static in the sense of \eqref{Static}, and is also contained in a worldtube whose spatial sections have compact support. The gravitational influence of $B$ is ignored in the sense that no relation is imposed between $g_{ab}$ and $T^{ab}_\subB$. It is also assumed that $B$ is locally isolated, meaning that there are neighborhoods of it in which the total stress-energy tensor $T^{ab}_\mathrm{tot}$ can be split into three parts:
\begin{equation}
  T^{ab}_\mathrm{tot} = T^{ab}_\subB + T^{ab}_\mathrm{fld} + T^{ab}_\mathrm{bkg}.
\end{equation}
$T^{ab}_\mathrm{fld}$ denotes the stress-energy tensor
associated with any non-gravitational fields---either scalar or electromagnetic---which couple to $B$, while the ``background'' stress-energy $T^{ab}_\mathrm{bkg}$ is assumed to be non-interacting in the sense that $\nabla_b T^{ab}_\mathrm{bkg} = 0$. The background stress-energy is included here for reasons of generality, but plays no further role in our discussion (except perhaps to act implicitly as a source for $g_{ab}$). Forces and torques on $B$ are instead derived using local stress-energy conservation in the form
\begin{equation}
  \nabla_b T^{ab}_\mathrm{tot} = \nabla_b ( T^{ab}_\subB + T^{ab}_\mathrm{fld}) = 0.
  \label{StressCons0}
\end{equation}

We specialize to cases where $B$ generates an electromagnetic
field sourced by a current density $J^a$, or a massless linear scalar field sourced by a charge density $\rho$. These densities are assumed to be smooth and stationary, and also to have supports bounded by that of $T^{ab}_\subB$. The scalar fields we consider explicitly satisfy the wave equation\footnote{Our derivation easily generalizes for nonzero field masses and curvature couplings. We omit these possibilities for brevity and also to minimize differences between the scalar and electromagnetic problems.}
\begin{equation}
  \nabla^a \nabla_a \phi = - \omega_n \rho
  \label{BoxPhi}
\end{equation}
in a neighborhood of $B$, where $\omega_n$ is the convenient constant
\begin{equation}
        \omega_n \equiv \frac{2 \pi^{ \frac{n}{2} } }{ \Gamma (\frac{n}{2}  ) },
\end{equation}
equal to the area of a unit sphere in $n$-dimensional Euclidean space. If $\phi$ is stationary in the sense that $\mathcal{L}_\tau \phi = 0$, the hyperbolic equation \eqref{BoxPhi} reduces to the elliptic field equation
\begin{equation}
  D^a(N D_a \phi) = - \omega_n \rho N.
  \label{phiEqn}
\end{equation}
The left-hand side here is equal to $N \nabla^a \nabla_a \phi$ acting on a static field; the overall factor of $N$ is used to obtain a differential operator which is spatially self-adjoint---a property which is crucial for our later development.

An equation very similar to \eqref{phiEqn} can also be derived for static electromagnetic fields $F_{ab}$. Consider a vector potential $A_a$ which satisfies $F_{ab} = 2 \nabla_{[a} A_{b]}$, and suppose that there are some static fields $J$ and $\Phi$ such that
\begin{equation}
  J^a = J \tau^a, \qquad A_a = N^{-2} \Phi \tau_a.
  \label{EMsimps}
\end{equation}
Although they can be weakened, these assumptions automatically exclude, e.g., current loops and external magnetic fields. They nevertheless encompass most physical systems which are commonly considered, and also provide a simple link between the electromagnetic and scalar problems. Assuming them, local charge conservation $\nabla_a J^a = 0$ follows automatically from the stationarity of $J$. The Maxwell equation $\nabla_b F^{ab} = \omega_n J^a$ also reduces in this case to
\begin{align}
  D^a ( N^{-1} D_a \Phi )  = -\omega_n J N,
  \label{MaxwellPhi}
\end{align}
and it is easily verified that the resulting $A_a$ satisfies the Lorenz gauge condition $\nabla^a A_a = 0$. Comparing \eqref{phiEqn} and \eqref{MaxwellPhi} shows that in this static context, the electric potential $\Phi$ and the scalar potential $\phi$ satisfy field equations whose differential operators differ only in the substitution $N \rightarrow N^{-1}$.

Allowing for the presence of both scalar and electromagnetic charge, the stress-energy conservation equation \eqref{StressCons0} reduces to\footnote{Our normalization convention for the scalar and electromagnetic fields is such that the Lagrangian density is $\rho \phi + J_a A^a - (\nabla \phi)^2/(2 \omega_n) - F_{ab} F^{ab}/(4 \omega_n) + (\mbox{matter terms})$.}
\begin{align}
  \nabla_b T_\subB^{ab} = \rho \nabla^a \phi - J \nabla^a \Phi .
  \label{StressCons}
\end{align}
The scalar field and charge density remain invariant under the time rescalings \eqref{tScale}, while the electromagnetic quantities instead rescale via
\begin{equation}
  \Phi \rightarrow \alpha^{-1} \Phi , \qquad J \rightarrow \alpha J.
  \label{EMScale}
\end{equation}

\subsection{Self-fields and holding fields}
\label{Sect:ExtFields}

As stated above, one of the main goals of this paper is to generalize \eqref{HoldingF0}, thus obtaining those external fields which hold $B$ fixed. This is ambiguous, however, in the absence of certain additional specifications. In the scalar case, we require a functional which maps charge densities $\rho$ onto ``self-fields'' $\phi_\mathrm{self}[\rho]$. These can reasonably be called self-fields only if
\begin{equation}
  D^a (N D_a \phi_\mathrm{self}[\rho] ) = - \omega_n \rho N
  \label{phiSelf}
\end{equation}
in a neighborhood of the body, and also if $\phi_\mathrm{self}[0] = 0$. Physically, $\phi_\mathrm{self}[\rho]$ represents the field which arises when $B$ is added to the system. In many applications, it is most naturally described by supplementing \eqref{phiSelf} with physically-appropriate boundary conditions---for example decay at infinity. More explicitly, there will usually be some time-independent Green function $G_\mathrm{self}(x,x')$ for which
\begin{equation}
  \phi_\mathrm{self} [\rho] (x) = \int_{\Sigma_t} \rho(x') N(x') G_\mathrm{self}(x,x') dV_{\bperp}'.
  \label{phiSelfConv}
\end{equation}
Regardless, all that is needed at this point is that \textit{some} choice has been made for $\phi_\mathrm{self}[\rho]$.


The ``holding field'' $\phi_\hold$ is now defined to be everything which is not contained in the self-field,
\begin{equation}
   \phi_\hold \equiv \phi - \phi_\mathrm{self}[\rho] ,
   \label{phiExtDef}
\end{equation}
and it is this quantity that our derivation eventually constrains. It
follows from \eqref{phiEqn} and \eqref{phiSelf} that $D^a (N D_a
\phi_\hold) = 0$ in a neighborhood of $B$. An analogous splitting and choice of self-field is also assumed to have been made for the electromagnetic field: $\Phi = \Phi_\mathrm{self}[J] + \Phi_\hold$.

\section{Momentum and force}
\label{Sect:mf}

Following standard practice in, e.g., Newtonian celestial mechanics, we consider only the ``bulk'' degrees of freedom associated with $B$---namely its ``linear and angular momenta.'' The body's remaining aspects are to be ignored as much as possible. The particular notion of momentum employed here is originally due to Dixon\footnote{More precisely, Dixon considered extended objects potentially coupled to electromagnetic fields in curved, four-dimensional spacetimes. The linear and angular momenta used here correspond to his in a purely gravitational setting (and generalized for arbitrary $n$). Including the missing electromagnetic terms is straightforward, but omitted for simplicity.} \cite{Dix70a, Dix74, Dix79, DixonReview}, who obtained it as a part of a comprehensive theory of multipole moments for extended bodies in general relativity. It was found in \cite{HarteSyms} and subsequent work \cite{HarteScalar, HarteEM, HarteTrenorm, HarteReview, Bobbing} to be useful to re-express Dixon's linear and angular momenta in terms of a single ``generalized momentum'' which lives in a particular abstract vector space (and not a tangent space anywhere in spacetime).

For the systems considered in this paper, it is convenient to define the generalized momentum at time $t$ by
\begin{equation}
          P_t(\xi) \equiv \int_{\Sigma_t} T_\subB^{ab} \xi_a d S_b,
  \label{PDef}
\end{equation}
where the $\xi^a$ are vector fields drawn from a particular vector space $K_\subG$ of ``generalized Killing fields'' with dimension
\begin{equation}
  \dim K_\subG = \frac{1}{2} (n+1) (n+2).
  \label{dimK}
\end{equation}
For each $t$, the generalized momentum $P_t(\cdot)$ is a linear operator on $K_\subG$, and can therefore be interpreted as a vector in the dual space $K_\subG^*$. It follows from \eqref{dimK} that this vector has $\frac{1}{2} (n+1) (n+2)$ components, physically corresponding to $n+1$ components of linear momentum and $\frac{1}{2} n (n+1)$ components of angular momentum. Explicit decompositions into linear and angular momenta are described in section \ref{Sect:Momenta}, although significant conceptual and calculational simplifications result by  delaying this for as long as possible.

The particular space of generalized Killing fields considered here is not immediately important. Indeed, it plays no role in our discussion until section \ref{Sect:Multipole}, and even there, only a few of its properties are needed: First, $K_\subG$ includes all Killing vectors which may exist, and is equal to the space of Killing vector fields in maximally-symmetric spacetimes. More generally, $K_\subG$ also includes vector fields which are not Killing. In those cases, it requires as part of its specification a ``frame.'' This consists of a timelike worldline $\mathcal{Z}$ and a foliation of the spacetime---really only a foliation of a sufficiently large neighborhood of $\mathcal{Z}$---into a family of hypersurfaces. All $\xi^a \in K_\subG$ are then Killing on $\mathcal{Z}$,
\begin{equation}
        \Lie_\xi g_{ab} |_\mathcal{Z} = \nabla_a \Lie_\xi g_{bc} |_\mathcal{Z} = 0,
        \label{LieXiZ}
\end{equation}
and this implies that the Killing transport equation
\begin{equation}
  \left[ \tau^c \nabla_c (\nabla_a \xi_b) = - R_{abc}{}^{d} \tau^c \xi_d \right]_\mathcal{Z},
  \label{KillingTransport}
\end{equation}
is satisfied on $\mathcal{Z}$. It is natural in the static systems considered here to let the foliation coincide with the $\Sigma_t$, and also to let $\mathcal{Z}$ be an orbit of $\tau^a$. Precisely which orbit is not immediately important, although a particularly useful choice is discussed in section \ref{Sect:Momenta} wherein $\mathcal{Z}$ is identified with the body's ``center-of-mass worldline.''

The final property of the generalized Killing fields which we require is that they ``preserve separations from $\mathcal{Z}$.'' Making this precise requires the concept of a separation vector $X^a (x,x')$ between two events $x$ and $x'$, which is naturally defined via the exponential map\footnote{Note that $X_a (x,x') = - \nabla_a X(x,x')$, where $X(x,x')$ is the world function on $(\Sigma \times I ,g_{ab})$, a biscalar equal to one half of the squared geodesic distance between $x$ and $x'$ as computed by $g_{ab}$ \cite{PoissonLR, Synge, FriedlanderWave}. We reserve the more conventional symbol $\sigma(x,x')$ for the spatial world function associated with $(\Sigma,h_{ab})$.}:
\begin{equation}
  \exp_x X^a(x,x') = x'.
  \label{XDef}
\end{equation}
Letting $z_t \equiv \mathcal{Z} \cap \Sigma_t$ and choosing any $x' \in \Sigma_t$ which is not too far from $z_t$, it can now be shown \cite{HarteSyms} that
\begin{align}
  \Lie_\xi X^a (z_t,x') = 0
  \label{LieX}
\end{align}
for all $\xi^a \in K_\subG$, where the Lie derivative is understood to act separately on both arguments: $\Lie_\xi X^a = \xi^b \nabla_b X^a - X^b \nabla_b \xi^a + \xi^{b'} \nabla_{b'} X^a$. This provides a sense in which the generalized Killing fields ``preserve separations from $\mathcal{Z}$.'' Further details may be found in \cite{HarteSyms, HarteReview, Bobbing}.

Now that the generalized momentum $P_t$ has been defined, our next task is to compute its time derivative. This is most easily obtained by first considering the difference in momentum between two discrete times $t$ and $t'$. If $t' > t$, it follows from \eqref{PDef}, Gauss' theorem, and the compact spatial support of $T^{ab}_\subB$ that
\begin{align}
        P_{t'} (\xi)-P_t (\xi) &= \oint_{ \partial \Omega(t,t') } \!\! T_\subB^{ab} \xi_a d S_b
        \nonumber
        \\
        &= \int_{\Omega(t,t')} \!\! \nabla_b (T_\subB^{ab} \xi_a) dV
        \nonumber
        \\
        &= \int_t^{t'} \!\! dT \left[ \int_{\Sigma_T} \!\! \nabla_b (T_\subB^{ab} \xi_a) N d V_{\bperp} \right],
\end{align}
where $dV$ is the spacetime volume element associated with $g_{ab}$ and $\Omega(t,t')$ denotes a worldtube which encloses the body between $\Sigma_t$ and $\Sigma_{t'}$. Applying stress-energy conservation \eqref{StressCons} while taking the limit $t' \rightarrow t$ finally shows that
\begin{equation}
  \frac{d P_t(\xi) }{d t} = \int_{\Sigma_t} \left( \frac{1}{2} T^{ab}_\subB \Lie_\xi g_{ab} + \rho \Lie_\xi \phi - J \Lie_\xi \Phi \right) N dV_{\bperp}.
  \label{GenForce}
\end{equation}
This describes the rate of change of generalized momentum, and may therefore be interpreted as a ``generalized force.'' The term involving $\Lie_\xi g_{ab}$ encodes gravitational forces and torques, while those involving $\Lie_\xi \phi$ and $\mathcal{L}_\xi \Phi$ respectively encode scalar and electromagnetic forces and torques. Although it is common to ignore the gravitational component of this equation (which first appears at quadrupole order\footnote{Gravitational dipole effects are kinematical, arising via the translation from generalized momenta to ordinary linear and angular momenta expressed as tensors on spacetime. See section \ref{Sect:Momenta}.}) when $n=3$, its relative importance can change significantly in different numbers of dimensions.

One result which may be deduced immediately from \eqref{GenForce} is that changes in $P_t(\xi)$ measure the degree to which $\xi^a$ generates symmetries. In the static cases considered here, $\tau^a$ generates an exact symmetry, and like any Killing field, it is also an element of $K_\subG$. Hence,
\begin{equation}
  E \equiv - P_t(\tau)
  \label{energy}
\end{equation}
must be independent of $t$. It is naturally interpreted as the body's total energy as seen by static observers. Similar conservation laws hold for every other Killing field $\Xi^a$ which may exist where $\Lie_\Xi \phi = \Lie_\Xi \Phi = 0$. More generally though, $P_t(\xi)$ is not necessarily constant. Although the physical system is assumed to be static, time dependence can arise in the momentum via time dependence in $\xi^a$; even in flat spacetime, boost-type Killing fields depend on $t$.

Regardless of symmetry, the generalized force \eqref{GenForce} simplifies significantly if $B$ is a small test body in the sense that all fields $g_{ab}$, $\phi$, and $\Phi$ vary slowly throughout each of its spatial cross-sections. This assumption results in multipole expansions for the force and torque in the sense obtained by Dixon \cite{Dix70a, Dix74, Dix79} (see also \cite{HarteReview} and section \ref{Sect:Multipole} below). If self-interaction is significant, however, fields vary rapidly inside $B$ and additional techniques must be applied. We suppose in particular that the gravitational self-interaction is negligible while the scalar and electromagnetic self-interaction is not. The latter two cases are nearly identical, so the relevant steps are described in section \ref{Sect:ForceScalar} by temporarily assuming that $J = 0$. Those changes which are required to understand the electromagnetic problem are then explained in section \ref{Sect:ForceEM}.

\subsection{Scalar forces}
\label{Sect:ForceScalar}

Understanding self-interaction associated with $\phi$ is equivalent to approximating the scalar portion
\begin{equation}
  \int_{\Sigma_t} \rho N \Lie_\xi \phi \,d V_{\bperp}
  \label{Fsc}
\end{equation}
of the total force \eqref{GenForce}. The immediate difficulty with simplifying this integral is that $\rho$ could be arbitrarily complicated, and the field equation \eqref{phiEqn} implies that $\phi$ necessarily inherits any such complications. The approach we take is to identify a specific field $\phi_\subS$ which i) includes most of the difficult, small-scale structure which might be present in $\phi$, and ii) exerts a force which can be computed directly and then subtracted out. We refer to the result as an ``$S$-field\footnote{This terminology is inspired by Detweiler and Whiting \cite{DetweilerWhiting2003}, who introduced what they called a ``singular field'' for point particles. Our $\phi_\subS$ plays a similar role both physically and mathematically, although it is not singular for the extended objects considered here. We therefore compromise by referring to it as an $S$-field, where the ``$S$'' no longer stands for ``singular.''}.''  The class of possible $S$-fields adopted below suggest that they are a kind of self-field, and in some cases, $\phi_\subS$ can indeed be equal to the $\phi_\mathrm{self}$ introduced in section \ref{Sect:ExtFields}. In other cases, however, the two fields may be very different. The $S$-field should be viewed more generally as a computational tool, while the self-field is instead a physical object.

The first step to defining $\phi_\subS$ is to demand that it be a sum of ``elementary self-fields'' associated with each infinitesimal charge element in $B$. Mathematically, this idea is expressed by introducing a two-point propagator $G(x,x')$ on $\Sigma \times \Sigma$ which generates the $S$-field
\begin{equation}
  \phi_\subS (x) \equiv \int_{\Sigma} \rho(x') N(x') G(x,x') d V_{\bperp}'.
  \label{phiSDef}
\end{equation}
While this prescribes $\phi_\subS$ only on the spatial manifold $\Sigma$, we shall often use its natural (time-independent) extension to the spacetime manifold $\Sigma \times I$. When convenient, similar extensions are also used for the propagator. Physically, $G(x,x')$ might describe what could be meant by ``the field at $x$ as generated by charge at $x'$.'' Most potential choices for this propagator are not beneficial, however; they do not generate $S$-fields which simplify the force integral \eqref{Fsc}.

In order to find propagators which do simplify this integral, we first demand that $\phi_\subS$ satisfy a reciprocity relation in the sense that
\begin{equation}
        G(x,x') = G(x',x).
        \label{symassumption}
\end{equation}
This implies that the total force ``exerted by'' $\phi_\subS$ can be written as
\begin{equation}
          \int_{\Sigma_t} \! \rho N \mathcal{L}_\xi \phi_\subS \,d V_{\bperp} = \frac{1}{2} \int_{\Sigma_t} \! d V_{\bperp} \int_{\Sigma_t}  \! d V_{\bperp}' (\rho N) (\rho' N') \Lie_\xi G.
          \label{SForce}
\end{equation}
The benefit of this expression is that it relates the generalized force exerted by $\phi_\subS$ to the symmetries of $G$, and these can be controlled independently of any specific properties of $B$. Note in particular that if $\mathcal{L}_\Xi G$ vanishes for some $\Xi^a \in K_\subG$, the $\Xi^a$-component of the generalized force due to $\phi_\subS$ must also vanish. This observation can be used to immediately see, e.g., that the spatial forces and torques exerted by ordinary Newtonian self-fields must vanish in Euclidean space \cite{HarteScalar, HarteReview}: $G$ in that context is conventionally chosen to be a Green function which is invariant under all translations and rotations.

Much less obviously, the generalized force due to $\phi_\subS$ can be simplified even in generic cases where $\mathcal{L}_\xi G \neq 0$. This occurs, for example, if $G = G[N,h_{ab}]$ is restricted to be a bidistribution which depends only on the spacetime geometry on $\Sigma$, and if this dependence is quasilocal in the sense that for fixed $x$ and $x'$, the functional derivatives\footnote{Functional derivatives are sometimes defined with respect to particular coordinates $x^i$ so that, e.g., the variation of some functional $F[N]$ is $\delta F = \int (\delta F/\delta N') \delta N' d^n x'$. The definition adopted here is slightly different, avoiding coordinates by demanding a similar integral but with $d^n x'$ replaced by $dV_{\bperp}'$.} $\delta G(x,x')/\delta N(x'')$ and $\delta G(x,x')/\delta h_{a''b''} (x'')$ have compact support in $x''$.  The invariance of $\phi_\subS$ under time rescalings with the form \eqref{tScale} then implies that
\begin{equation}
  G \rightarrow \alpha G
  \label{GScaleScalar}
\end{equation}
when $N \rightarrow \alpha^{-1} N$.

More substantially, the definitions of the Lie and functional derivatives together with diffeomorphism-invariance imply that
\begin{align}
  \Lie_\psi G(x,x') = \int_{\Sigma} d V_{\bperp}'' \Bigg[ \frac{ \delta G(x,x') }{ \delta h_{a''b''}(x'') } \Lie_\psi h_{a''b''} (x'')
  \nonumber
  \\
  ~ + \frac{ \delta G(x,x') }{ \delta N(x'') } \Lie_\psi N (x'') \Bigg]
  \label{LieGspace}
\end{align}
for any vector field $\psi^a$ on $\Sigma$.  In this equation all quantities are tensor fields on $\Sigma$, and in particular the indices $a''$, $b''$ are spatial.
However, the result (\ref{LieGspace}) may be used to compute Lie derivatives of $G$ on spacetime
with respect to arbitrary spacetime vector fields $\xi^a$, as follows. If $x$ and $x'$ are points on $\Sigma \times I$ which lie on a single hypersurface $\Sigma_t$, the time-independence of the spacetime extension of the propagator implies that $\Lie_\xi G(x,x')  = \Lie_\psi G(x,x')$, where $\psi^a = h^{a}{}_{b} \xi^b|_{\Sigma_t}$ can be translated into a vector field on $\Sigma$.
Hence $\Lie_\xi G(x,x')$ is given by the right hand side of (\ref{LieGspace}) with this $\psi$.
We now reinterpret this right hand side in terms of tensor fields on spacetime.
First, $N$ and $h_{ab}$ can be extended to tensor fields on spacetime in the natural way
by demanding that $\Lie_\tau N = \Lie_\tau h_{ab} =0$ and $h_{ab} \tau^a =0$.
We similarly extend the functional derivatives, so that
\begin{equation}
  \frac{\delta G(x,x')}{\delta h_{a''b''}(x'')} \tau_{b''}(x'') = 0.
  \label{dGdhOrth}
\end{equation}
With these conventions, the $n$-dimensional Lie derivatives with respect to $\psi^a$ coincide
with $n+1$-dimensional Lie derivatives with respect to $\xi^a$, evaluated on $\Sigma_t$.
The final result is
\begin{align}
  \mathcal{L}_\xi G(x,x') = \int_{\Sigma_t} d V_{\bperp}'' \Bigg[ \frac{ \delta G(x,x') }{ \delta h_{a''b''}(x'') } \mathcal{L}_\xi h_{a''b''} (x'')
  \nonumber
  \\
  ~ + \frac{ \delta G(x,x') }{ \delta N(x'') } \mathcal{L}_\xi N (x'') \Bigg]
  \label{LieG0}
\end{align}
for arbitrary vector fields $\xi^a$ and for all $x,x' \in \Sigma_t$.
Here all quantities are tensors on spacetime, and $a''$, $b''$ are spacetime indices.
This is the result we need to interpret the generalized force exerted by $\phi_\subS$.

Although it is clear from \eqref{SForce} and \eqref{LieG0} that this force cannot vanish in general, there is a sense in which it is nevertheless ``ignorable.'' It can be removed by appropriately redefining---or renormalizing---our description of $B$. To see this, first note that Lie derivatives of $N$ and $h_{ab}$ can be translated in part into Lie derivatives of $g_{ab}$. Using
\begin{equation}
N\,\mathcal{L}_{\xi}N=-\tfrac{1}{2}\tau^{a}\tau^{b}\mathcal{L}_{\xi}g_{ab}+\frac{d}{dt}(\xi^a \tau_a)
\end{equation}
and \eqref{dGdhOrth}, we obtain
\begin{align}
  \mathcal{L}_\xi G = \int_{\Sigma_t} \!\! d V_{\bperp}'' \left( \frac{ \delta G }{ \delta h_{a''b''} } - \frac{\tau^{a''} \tau^{b''} }{2 N''} \frac{ \delta G }{ \delta N'' } \right) \mathcal{L}_\xi g_{a''b''}
  \nonumber
  \\
  ~ - \frac{d}{dt} \int_{\Sigma_t} d S_{a''} \xi^{a''} \frac{ \delta G }{ \delta N'' }.
  \label{LieG}
\end{align}
The force due to $\phi_\subS$ therefore splits into two distinct components. One of these is linear in $\mathcal{L}_\xi g_{ab}$, and recalling that the gravitational force in \eqref{GenForce} is also linear in $\mathcal{L}_\xi g_{ab}$, that portion of the scalar force can be interpreted as just another component of the gravitational force; \textit{it acts to renormalize} $T^{ab}_\subB$. Physically, this might be interpreted as a consequence of the ``gravitational mass distribution'' of the $S$-field.

The remaining portion of the force due to $\phi_\subS$ is a total time derivative. Noting that the generalized force is itself a total time derivative of the generalized momentum, time derivatives which are linear in $\xi^a$ but otherwise independent of $t$ can always be ``removed'' by renormalizing $P_t$. This physically accounts for the inertia of the body's self-field, but via a different mathematical mechanism from the one \cite{HarteScalar, HarteReview} which arises in dynamical contexts.

Together, these observations imply that the generalized force due to $\phi_\subS$ can be entirely eliminated by changing the definitions of $P_t$ and $T^{ab}_\subB$. Doing so results in a generalized force in which the $S$-field does not explicitly appear:
\begin{equation}
  \frac{d \hat{P}_t }{dt} = \int_{\Sigma_t} \left( \frac{1}{2} \hat{T}^{ab}_\subB \mathcal{L}_\xi g_{ab} + \rho \mathcal{L}_\xi \hat{\phi} \right) N d V_{\bperp}.
  \label{GenForceFin}
\end{equation}
This is identical in form to our original expression \eqref{GenForce}, although the momentum, stress-energy tensor, and scalar field have all been shifted from their original definitions. The physical scalar field $\phi$ has been replaced by
\begin{align}
  \hat{\phi} \equiv \phi - \phi_\subS,
  \label{phiHatDef}
\end{align}
and to compensate, a self-field contribution has been added to the momentum
\begin{align}
\label{PHat}
  \hat{P}_t \equiv  P_t & + \frac{1}{2} \int_{\Sigma_t} \! \! dS_a \xi^a
  \nonumber
  \\
  &~ \times \left[ \int_{\Sigma_t} \! \! d V_{\bperp}' \int_{\Sigma_t} \! \! d V_{\bperp}'' (\rho' N') (\rho'' N'') \frac{\delta G}{ \delta N} \right],
\end{align}
and also to the stress-energy tensor
\begin{align}
  \hat{T}^{ab}_\subB \equiv T^{ab}_\subB + \frac{1}{N} \int_{\Sigma_t} \!\! d V_{\bperp}' \int_{\Sigma_t} \!\! d V_{\bperp}'' (\rho' N') (\rho'' N'')
  \nonumber
  \\
   ~ \times \Big( \frac{\delta G }{ \delta h_{ab} } - \frac{\tau^a \tau^b}{2 N} \frac{\delta G}{\delta N} \Big).
   \label{THat}
\end{align}
An approximate version of this renormalized stress-energy tensor is computed explicitly in section \ref{Sect:bodyR}. Regardless, using \eqref{PDef} and \eqref{dGdhOrth}, the renormalized momentum can be written exactly as
\begin{equation}
  \hat{P}_t = \int_{\Sigma_t} \hat{T}^{ab}_\subB \xi_a dS_b,
  \label{PHatSimp}
\end{equation}
which is identical to the definition \eqref{PDef} for $P_t$ except for the replacement $T^{ab}_\subB \rightarrow \hat{T}^{ab}_\subB$. The same renormalized stress-energy therefore controls both the effective inertia and the effective gravitational force. Also note that the quasilocality of the functional derivatives implies that $\hat{T}^{ab}_\subB$ has compact support even though the stress-energy tensor associated with $\phi_\subS$ does not.

These results are exact. They represent a \textit{class} of identities which hold for any $S$-fields $\phi_\subS$ which have been generated via \eqref{phiSDef} using geometrically-constructed, symmetric propagators $G$ which satisfy \eqref{GScaleScalar}. These constraints on the propagator are very weak---and may be weakened even further---so many possibilities exist. The problem is now to find useful examples. In particular, the mapping $\phi \mapsto \hat{\phi}$ should remove those field variations which had initially made \eqref{Fsc} so difficult to approximate.  Once an example with this property is identified, simple point particle limits follow directly\footnote{Some effort is still required to convert expressions involving generalized momenta into ordinary multipole approximations for the force and torque. These same steps arise, however, even in test body limits. The benefit of \eqref{GenForceFin} is that it allows the well-understood manipulations associated with the test body regime to be immediately generalized.}.

Our selection criterion for $G$ is that the associated $\hat{\phi}$ should be easy to compute and largely independent of the body's internal structure. One way to enforce this is to demand that $\hat{\phi}$ satisfy the vacuum field equation in a neighborhood of $B$, which follows if $G$ is a Green function. Letting
\begin{equation}
  D^a (N D_a G ) = - \omega_n \delta_\Sigma (x,x'),
  \label{GDef}
\end{equation}
it is implied by \eqref{phiEqn}, \eqref{phiSDef}, and \eqref{phiHatDef} that
\begin{equation}
  D^a (N D_a \hat{\phi}) = 0,
  \label{phiHatFieldEqn}
\end{equation}
where the Dirac distribution here is understood to be the natural one on $(\Sigma, h_{ab})$. The utility of this choice can be motivated by considering point particles. Although point particles properly arise only as limits of extended objects, and are discussed more fully in section \ref{Sect:pp}, it suffices here to naively consider fields $\phi$ which are sourced by pointlike, distributional charge densities. No matter how singular such fields might be, it follows from the general theory of elliptic partial differential equations---see the remark following theorem 6.6 in \cite{Agmon}---that any $\hat{\phi}$ satisfying \eqref{phiHatFieldEqn} is everywhere smooth as long as $N(\bx)$ and $h_{ij}(\bx)$ are themselves smooth with respect to some spatial coordinates $\bx$. Variations in $\hat{\phi}$ therefore occur over much larger scales than those associated with $B$, and all force integrals simplify as they do for small test bodies.
%

The elliptic regularity result used to motivate \eqref{GDef} may be generalized to obtain a somewhat larger class of useful propagators: It is known that if a smooth elliptic differential operator acting on a field results in a smooth (but not necessarily vanishing) source, that field must still be smooth \cite{Agmon}. Furthermore, convolving a singular distribution with something smooth results in something else which is also smooth. This suggests that point particle limits remain simple if $S$-fields are defined more generally in terms of \textit{parametrices}---bidistributions $G$ which satisfy
\begin{equation}
    D^a (N D_a G ) = - \omega_n \delta_\Sigma (x,x') + \mathcal{S} (x,x')
    \label{GDefparam}
\end{equation}
for some smooth $\mathcal{S}$. The renormalized field $\hat{\phi}$ which appears in \eqref{GenForceFin} then satisfies
\begin{equation}
  D^a (N D_a \hat{\phi} ) = - \int_{\Sigma_t} \rho(x') N(x') \mathcal{S}(x,x') d V_{\bperp}',
\end{equation}
the right-hand side of which is smooth even in a point particle limit. The Green functions considered above now correspond to special parametrices for which $\mathcal{S} = 0$. It is useful, however, to allow $\mathcal{S} \neq 0$ in general; there are important cases for which such propagators are significantly simpler to construct.

To summarize, we have shown that the generalized force acting on a static, extended scalar charge satisfies \eqref{GenForceFin}, where $\hat{T}_\subB^{ab}$ is determined by \eqref{THat}, $\hat{P}_t$ by \eqref{PHatSimp}, and $\hat{\phi}$ by \eqref{phiSDef} and \eqref{phiHatDef}. Each of these definitions depends on a choice of propagator. This is not fixed uniquely, but is instead constrained to have the following properties:
\begin{enumerate}
        \item \label{functionalAssume} The propagator $G[N,h_{ab}] (x,x')$ is a bidistribution on $\Sigma \times \Sigma$ which depends functionally only on the lapse and the spatial metric.

        \item It depends on $N$ and $h_{ab}$ only quasilocally in the sense that for fixed $x,x' \in \Sigma$, the functional derivatives
                \begin{equation*}
                        \frac{\delta G(x,x')}{\delta N(x'')}, \qquad  \frac{\delta G(x,x')}{\delta h_{a''b''}(x'')}
                \end{equation*}
                have compact support in $x''$.

        \item It transforms appropriately under time rescalings generated by any constant $\alpha > 0$ and any spatial diffeomorphism $\varphi: \Sigma \to \Sigma$,
                \begin{subequations}
                        \begin{eqnarray*}
                                G[\alpha^{-1} N,h_{ab}]  &=& \alpha G[N,h_{ab}], \\
                                G[\varphi_*N, \varphi_* h_{ab}] &=& \varphi_* G[N,h_{ab}].
                        \end{eqnarray*}
                \end{subequations}

        \item The propagator is symmetric, $G(x,x') = G(x',x)$.

        \item \label{paramAssume} It is a parametrix for the field equation, meaning that
                \begin{equation*}
                        D^a (N D_a G)+ \omega_n \delta_\Sigma(x,x'),
                \end{equation*}
                is a smooth function on $\Sigma \times \Sigma$.
\end{enumerate}
While it is not obvious that propagators satisfying these assumptions exist at all, we show in appendix \ref{Sect:Hadamard} that they do, and that one example is the well-known Hadamard parametrix. Other examples exist as well, and the physical interpretation of the resulting ambiguities is discussed in section \ref{Sect:Ginvariance}.

Before proceeding further, it is important to note that these assumptions can be weakened considerably. The simplest such modification is to remove the functional dependence on $N$ and $h_{ab}$, which is useful if, for example, a particular parametrix is known in a given geometry, but not in any nearby geometries. It is then sufficient to demand that $G(x,x') = G(x',x)$, and that
\begin{align}
  \Lie_\psi G(x,x') = \int_{\Sigma} d V_{\bperp}'' \Big[ G_{(h)}^{a''b''} (x,x';x'') \Lie_\psi h_{a''b''} (x'')
  \nonumber
  \\
  ~ + G_{(N)} (x,x';x'') \Lie_\psi N (x'') \Big]
  \label{LieGen}
\end{align}
for all $\psi^a$ on $\Sigma$, where the three-point coefficients\footnote{If these coefficients exist, they are not unique. One possible freedom is that divergence-free terms with compact support in $x''$ can always be added to $G_{(h)}^{a''b''} (x,x';x'')$.} $G_{(h)}^{a''b''} (x,x';x'')$ and $G_{(N)}(x,x';x'')$ have compact support in $x''$ for fixed $x$ and $x'$.

Although we do not exploit them in this paper, even more general maps $\phi \mapsto \hat{\phi}$ can be considered. The two-point propagators used to construct $\phi_\subS$ can easily be replaced, for example, by symmetric, geometrically-constructed $p$-point propagators for any $p \geq 2$. The simple subtraction \eqref{phiHatDef} can also be replaced by certain nonlinear maps which continuously ``flow'' from $\phi$ through some family $\hat{\phi}_\lambda$ of effective potentials---thus inducing an associated flow of multipole moments. Another possibility is to consider propagators which depend on non-geometrical fields. Allowing a two-point propagator to depend on $N$, $h_{ab}$, and $\phi$ would, for example, result in renormalizations of $\rho$ as well as $T^{ab}_\subB$. These types of generalizations aren't particularly interesting for the problem considered here, but can be essential when discussing nonlinear theories or the coupled gravitoscalar or gravitoelectromagnetic self-force problems.

\subsection{Electromagnetic forces}
\label{Sect:ForceEM}

It is evident from the force \eqref{GenForce} and the field equations \eqref{phiEqn} and \eqref{MaxwellPhi} that the electromagnetic and scalar interactions considered here are almost identical. All arguments used in the scalar case may therefore be repeated essentially verbatim. This results in an effective electromagnetic potential
\begin{subequations}
\label{phiSDefEM}
\begin{gather}
\label{MaxwellPhiHat}
  \hat{\Phi} \equiv \Phi - \Phi_\subS,
  \\
  \Phi_\subS(x) \equiv \int_{\Sigma} J(x') N(x') \mathcal{G}(x,x') d V_{\bperp}',
\end{gather}
\end{subequations}
where $\mathcal{G}[N,h_{ab}] (x,x')$ is a symmetric propagator on $\Sigma \times \Sigma$ which scales as
\begin{equation}
  \mathcal{G} \rightarrow \alpha^{-1} \mathcal{G}
  \label{GscaleEM}
\end{equation}
under the time reparameterizations \eqref{tScale}. If this $\mathcal{G}$ depends only quasilocally on $N$ and $h_{ab}$, the appropriate modification of \eqref{GenForceFin} is
\begin{equation}
  \frac{d \hat{P}_t }{dt}  = \int_{\Sigma_t} \left( \frac{1}{2} \hat{T}^{ab}_\subB \mathcal{L}_\xi g_{ab} - J \mathcal{L}_\xi \hat{\Phi} \right) N d V_{\bperp},
  \label{GenForceFinEM}
\end{equation}
where the effective momentum and stress-energy are
\begin{align}
  \hat{P}_t \equiv \int_{\Sigma_t} \hat{T}^{ab}_\subB \xi_a dS_b,
\end{align}
and
\begin{align}
  \hat{T}^{ab}_\subB \equiv T^{ab}_\subB - \frac{1}{N} \int_{\Sigma_t} \!\! d V_{\bperp}' \int_{\Sigma_t} \!\! d V_{\bperp}'' (J' N') (J'' N'')
  \nonumber
  \\
  ~ \times \Big( \frac{\delta \mathcal{G} }{ \delta h_{ab} } - \frac{\tau^a \tau^b}{2 N} \frac{\delta \mathcal{G} }{\delta N} \Big).
  \label{THatem}
\end{align}

As in the scalar case, it can be convenient to narrow down the class of propagators even further by demanding that $\mathcal{G}$ be a parametrix for Maxwell's equations in the sense that [cf. \eqref{MaxwellPhi}]
\begin{equation}
  D^a ( N^{-1} D_a \mathcal{G} ) = - \omega_n \delta_\Sigma (x,x') + \mathcal{S}(x,x')
  \label{GDefEM}
\end{equation}
for some smooth $\mathcal{S}(x,x')$. The effective electromagnetic field
\begin{equation}
  \hat{F}_{ab} \equiv 2 \nabla_{[a} \hat{A}_{b]} = 2 \nabla_{[a} ( N^{-2} \hat{\Phi} \tau_{b]})
\end{equation}
then satisfies the vacuum Maxwell equation up to a smooth source term:
\begin{equation}
  D^a ( N^{-1} D_a \hat{\Phi} ) = - \int_{\Sigma_t} J(x') N(x') \mathcal{S}(x,x') d V_{\bperp}'.
\end{equation}

We have thus far considered the scalar and electromagnetic cases separately. This has been only for notational simplicity, and there is no obstacle to allowing both $\rho$ and $J$ to be simultaneously nonzero. Their effects merely add.

\subsection{Measurable quantities are independent of choice of propagator}
\label{Sect:Ginvariance}

The identities \eqref{GenForceFin} and \eqref{GenForceFinEM} allow scalar and electromagnetic forces to be computed using effective fields $\hat{\phi}$ and $\hat{\Phi}$ which can be considerably simpler than their physical counterparts. These fields are obtained from $\phi$ and $\Phi$ using two-point propagators $G$ and $\mathcal{G}$ which satisfy the five assumptions listed at the end of section \ref{Sect:ForceScalar} [with minor modifications in the electromagnetic case to be consistent with \eqref{GscaleEM} and \eqref{GDefEM}]. Many propagators can be written down which satisfy these assumptions. They might be Green functions or more general parametrices. In some cases, assumption \ref{paramAssume} can even be relaxed to allow something else entirely. This lack of uniqueness provides an interesting flexibility which does not appear to have been noted in other self-force contexts: It can allow one's computational methods to be tailored to the details of whichever particular problem might be at hand. We provide an example of this in section \ref{Sect:Rindler} and appendix \ref{Sect:RindlerAppendix}, where the force on a uniformly-accelerated charge in flat spacetime is obtained using two different propagators---one of which results in much less computation than the other.

Although it can be useful to consider different definitions for the effective fields, the physical interpretations of these fields must be considered with care. In general, different choices for the propagators $G$ and $\mathcal{G}$ give rise
to different momenta ${\hat P}_t$ and different stress-energy tensors ${\hat T}_\subB^{ab}$. Scalar, electromagnetic, inertial, and gravitational forces also individually depend on the choices for $G$ and $\mathcal{G}$. This splitting of the force into components is associated with unphysical aspects of our description, and may be interpreted as a kind of gauge freedom\footnote{There are two different components of this gauge freedom: First, the choice of propagator in a given spacetime affects the renormalized fields via \eqref{phiSDef} and \eqref{phiHatDef}, and thus the scalar and electromagnetic self-forces. Second, the choice of propagator in nearby spacetimes, as encoded in the variational derivatives in \eqref{LieG0}, can affect the renormalizations \eqref{PHat} and \eqref{THat} of the stress-energy tensor and the generalized momentum.}.  While the details of a particular problem sometimes provide selection principles which can reduce this freedom ``by convention,'' it cannot be avoided in general.

A natural question is then to ask for observable quantities which remain invariant under all possible propagator transformations. As noted in section \ref{Sect:ExtFields}, a natural quantity to consider here is the total force which is required to hold $B$ in place. We assume for simplicity that the space $K_\subG$ of generalized Killing fields does not depend on propagator transformations\footnote{We relax this assumption slightly in section \ref{Sect:Momenta}, where the reference worldline $\mathcal{Z}$ used to construct $K_\subG$ is chosen to coincide with the center-of-mass worldline determined by $\hat{T}^{ab}_\subB$.}.  Then, from (\ref{phiExtDef}), (\ref{phiHatDef}) and (\ref{GenForceFin}) and their electrostatic analogs,
 the generalized holding force is explicitly
\begin{align}
  \mathcal{F}_\hold &\equiv \int_{\Sigma_t} \left( \rho \Lie_\xi \phi_\hold - J \Lie_\xi \Phi_\hold \right) N d V_{\bperp}
  \nonumber
  \\
  &= \frac{ d \hat{P}_t }{ dt } - \int_{\Sigma_t} \bigg( \frac{1}{2} \hat{T}^{ab}_\subB \Lie_\xi g_{ab} + \rho \Lie_\xi \hat{\phi}_\mathrm{self}
  \nonumber
  \\
  & \qquad \qquad \qquad \qquad ~ - J \Lie_\xi \hat{\Phi}_\mathrm{self} \bigg) N dV_{\bperp}.
  \label{FholdDef}
\end{align}
Here
\begin{equation}
  \hat{\phi}_\mathrm{self} \equiv \phi_\mathrm{self} - \phi_\subS, \qquad \hat{\Phi}_\mathrm{self} \equiv \Phi_\mathrm{self} - \Phi_\subS
  \label{phiHatSelfDef}
\end{equation}
involve the self-fields introduced in section \ref{Sect:ExtFields}. The first equality in \eqref{FholdDef} shows that $\mathcal{F}_\hold$ cannot depend on $G$ or $\mathcal{G}$. The second equality shows that the holding force can nevertheless be written in terms of quantities which do (individually) depend on these propagators. It is sometimes convenient to discuss these latter quantities on their own, in which case we call, e.g.,
\begin{equation}
  \int_{\Sigma_t} \rho N \Lie_\xi \hat{\phi}_\mathrm{self} \,dV_{\bperp}
  \label{SFscalar}
\end{equation}
``the'' scalar self-force, and
\begin{equation}
  \frac{1}{2} \int_{\Sigma_t} \hat{T}^{ab}_\subB N \Lie_\xi g_{ab} \,dV_{\bperp}
  \label{GravForce}
\end{equation}
``the'' gravitational force. We emphasize that such expressions are unique only in connection with specific propagators, a lack of recognition of which has led to some confusion in the literature---see section \ref{Sect:Examples}.

\section{Multipole expansions}
\label{Sect:Multipole}

In the typical contexts where problems of motion are considered, charge distributions and stress-energy tensors are not known in detail. Nevertheless, our final expression \eqref{FholdDef} for the generalized holding force $\mathcal{F}_\hold$ is an integral involving precisely these quantities. Following standard practice in Newtonian gravity or elementary electrostatics, progress is made by introducing multipole moments. If, for example, the renormalized field $\hat{\phi}$ varies sufficiently slowly throughout the body $B$, the force \eqref{SFscalar} can be accurately approximated using only a finite number of multipole moments $q^{a_1 \cdots a_p}$. Retaining these moments is significantly simpler than retaining the infinite number of degrees of freedom associated with the complete charge density $\rho$.

Except in very special cases, this type of multipole approximation cannot be applied directly to the bare force \eqref{GenForce}---$\phi$ and $\Phi$ inherit all lengthscales present in $\rho$ and $J$, and therefore do not vary slowly. If $G$ and $\mathcal{G}$ are well-chosen, however, the same comments do not apply to the effective fields $\hat{\phi}$ and $\hat{\Phi}$ appearing in \eqref{GenForceFin} and \eqref{GenForceFinEM}. We assume from now on that these hatted fields can be approximated throughout $B$ using an appropriately-defined low-order Taylor series.

\subsection{Covariant Taylor series}
\label{Sect:Taylor}

The type of Taylor series adopted here is easily explained: If some quantity is to be expanded about an origin $z_t$, use $g_{ab}$ to construct Riemann normal coordinates about $z_t$, and then compute an elementary Taylor expansion in these coordinates. With some additional work, equivalent constructions can also be described without any explicit reference to coordinates \cite{HarteTrenorm, HarteReview}. Adopting the second viewpoint, scalar fields have expansions with the form\footnote{The equality sign and infinite upper limit here are formal. The series doesn't necessarily converge in practice, and we shall only ever use a finite number of terms.}
\begin{equation}
  \hat{\phi}(x') = \sum_{p=0}^\infty \frac{1}{p!} X^{a_1} \cdots X^{a_p} \hat{\phi}_{, a_1 \cdots a_p} (z_t),
  \label{TaylorPhi}
\end{equation}
where $X^a = X^a (z_t,x')$ is the separation vector between $x'$ and $z_t$ defined by \eqref{XDef}. The coefficients $\hat{\phi}_{,a_1 \cdots a_k}(z_t)$ are tensor fields which reduce to partial derivatives in a Riemann normal coordinate system with origin $z_t$, and are known as tensor extensions \cite{HarteTrenorm, Dix74, Dix79} of $\hat{\phi}$. These extensions can be defined explicitly via
\begin{equation}
  \hat{\phi}_{,a_1 \cdots a_p} (y) \equiv \left[ \frac{ \partial^p \hat{\phi} ( \exp_y Y^b ) }{ \partial Y^{a_1} \cdots \partial Y^{a_p} }   \right]_{Y^c = 0}.
  \label{phiExtendDef}
\end{equation}
The first few examples are explicitly $\hat{\phi}_{,} = \hat{\phi}$ and
\begin{subequations}
        \label{phiExtensions}
        \begin{gather}
                \hat{\phi}_{,a} = \nabla_a \hat{\phi} , \quad \hat{\phi}_{,a b} = \nabla_{b} \nabla_{a} \hat{\phi},
                \\
                \hat{\phi}_{,abc} = \nabla_{(a} \nabla_b \nabla_{c)} \hat{\phi} .
         \end{gather}
\end{subequations}
Higher-order extensions can be more complicated, involving $R_{abc}{}^{d}$ and its derivatives contracted into derivatives of $\hat{\phi}$. Regardless, it is clear from \eqref{phiExtendDef} that
\begin{equation}
  \hat{\phi}_{,a_1 \cdots a_p} = \hat{\phi}_{,(a_1 \cdots a_p)}
  \label{phiSyms}
\end{equation}
for all $p$.

We also need an expansion for $g_{ab}$. Again demanding that this reduce to an  elementary Taylor expansion in Riemann normal coordinates, it may be shown that \cite{HarteTrenorm}
\begin{equation}
  g_{a'b'} = \nabla_{a'} X^a \nabla_{b'} X^b \sum_{p=0}^\infty \frac{1}{p!} X^{c_1} \cdots X^{c_p} g_{ab,c_1 \cdots c_p}.
  \label{Taylorg}
\end{equation}
Noting that frame components of $X^a$ can be interpreted as Riemann normal coordinate functions, $\nabla_{b'} X^a$ reduces (non-perturbatively) to the identity matrix $\delta^\mu_\nu$ in those coordinates. The appearance of this gradient in \eqref{Taylorg} therefore ``corrects'' the naive coordinate expansion by appropriately transporting lowered indices at $z_t$ to lowered indices at $x'$. The metric extensions here require a similar type of transport, but in the ``opposite direction.'' This is accomplished by
\begin{equation}
  H^{b'}{}_{a}(x',z_t) \equiv (\nabla_{b'} X^a)^{-1},
  \label{HDef}
\end{equation}
which also reduces to the identity in Riemann normal coordinates. Using it, extensions of the metric can be computed from
\begin{align}
  g_{ab,c_1 \cdots c_p} \equiv \left[ \frac{ \partial^p (H^{a'}{}_{a} H^{b'}{}_{b} g_{a'b'})}{ \partial Y^{c_1} \cdots \partial Y^{c_p} } \right]_{Y^d = 0}.
  \label{MetricExtensions}
\end{align}
The resulting tensors have the symmetries $g_{ab,c_1 \cdots c_p} = g_{(ab),c_1 \cdots c_p} = g_{ab,(c_1 \cdots c_p)}$ and, for all $p \geq 1$ \cite{HarteTrenorm},
\begin{equation}
  g_{a(b,c_1 \cdots c_p)} = g_{(ab,c_1 \cdots c_{p-1}) c_p} = 0.
  \label{gSyms}
\end{equation}
The zeroth extension is the metric itself, $g_{ab,} = g_{ab}$, and the first extension vanishes: $g_{ab,c} = 0$. All higher-order metric extensions involve the curvature, which is evident from the first nontrivial examples \cite{HarteReview}
\begin{subequations}
        \label{gExtensions}
        \begin{gather}
                g_{ab,c_1 c_2} = \frac{2}{3} R_{a(c_1 c_2)b}, \quad g_{ab,c_1 c_2 c_3} = \nabla_{(c_1} R_{|a|c_2 c_3)b},
                \\
                g_{ab,c_1 c_2 c_3 c_4} = \frac{6}{5} \nabla_{(c_1 c_2} R_{|a|c_3 c_4) b} + \frac{16}{15} R_{a(c_1 c_2}{}^{d} R_{|b|c_3 c_4)d} .
        \end{gather}
\end{subequations}

Recalling that the holding force \eqref{FholdDef} does not involve $\hat{\phi}$ and $g_{ab}$ on their own, but rather their Lie derivatives with respect to elements of $K_\subG$, we now develop Taylor series for these Lie derivatives.

If the origin $z_t$ about which a Taylor expansion is performed lies on the worldline $\mathcal{Z}$ used to construct the space $K_\subG$ of generalized Killing vectors,  \eqref{LieX} and \eqref{TaylorPhi} immediately imply that
\begin{equation}
  \mathcal{L}_\xi \hat{\phi} (x') = \sum_{p=0}^\infty \frac{1}{p!} X^{a_1} \cdots X^{a_p} \mathcal{L}_\xi \hat{\phi}_{, a_1 \cdots a_p}(z_t)
  \label{LiePhi}
\end{equation}
whenever $x' \in \Sigma_t$. This is the first point in our discussion where any properties of the generalized Killing fields have been used.

An analogous expansion for the metric is significantly more complicated to derive, but may be shown to be \cite{HarteTrenorm}
\begin{align}
  \mathcal{L}_\xi g_{a'b'} = \sum_{p=2}^\infty \frac{1}{p!} \stackrel{ (p) }{ A } \!\! {}^{ab}{}_{a'b'} X^{c_1} \cdots X^{c_p} \mathcal{L}_\xi g_{ab,c_1 \cdots c_p}
  \label{Lieg}
\end{align}
when $\xi^a \in K_\subG$ and $z_t \in \mathcal{Z}$. The $p$-dependent transport operator which appears here is explicitly
\begin{align}
  \stackrel{ (p) }{ A } \!\! {}^{ab}{}_{a'b'} \equiv \nabla_{(a'} X^{a} \nabla_{b')} X^{b} + \frac{2}{p-1} \stackrel{(p)}{\Theta} \! {}^{ab}{}_{df} \tau^d H_{(a'}{}^{f} \nabla_{b')} t,
  \label{Adef}
\end{align}
where
\begin{align}
  \stackrel{(p)}{\Theta} \! {}^{abcd} \equiv (p-1) \int_0^1 s^{p-2} \nabla^{f''} X^a_s \nabla^{h''} X^b_s
  \nonumber
  \\
  ~ \times \nabla_{f''} X_s^{(c} \nabla_{h''} X^{d)}_s ds .
  \label{ThetaDef}
\end{align}
The integrand in this last expression is to be evaluated along an affinely-parameterized geodesic $\gamma''(s)$ satisfying\footnote{Primes and double primes here are not derivatives with respect to $s$, but instead are attached to indices associated with different points in spacetime.} $\gamma''(0) = z_t$ and $\gamma''(1) = x'$, and the $X^a_s$ are separation vectors between $z_t$ and $\gamma''(s)$. The integral is normalized to match the notation in \cite{Dix74}, and also so that its flat-spacetime limit,
\begin{equation}
        \stackrel{(p)}{\Theta} \! {}^{abcd} \rightarrow g^{a(c} g^{d)b},
        \label{ThetaLimit}
\end{equation}
is independent of $p$.

Although complicated, the details of these expressions are rarely needed in practice. The important point to note in \eqref{Lieg} is that the Taylor expansion for the metric starts only at $p=2$. This corresponds to quadrupole order when evaluating a force, and follows from the fact \eqref{LieXiZ} that the generalized Killing fields can fail to be Killing only for second and higher order deviations from $z_t$. There is no analogous symmetry which is guaranteed to hold for $\hat{\phi}$, so the scalar field expansion \eqref{LiePhi} can be nontrivial even at monopole order.

\subsection{The multipole force}
\label{Sect:multipole}

Multipole expansions for the generalized force now follow immediately
from the Taylor expansions just derived and from the integral forces
obtained in section \ref{Sect:mf}. Using \eqref{GenForceFin} and
\eqref{GenForceFinEM} together with \eqref{LiePhi} and \eqref{Lieg},
we obtain
\begin{align}
  \frac{d \hat{P}_t}{dt}  = \sum_{p=0}^\infty \frac{1}{p!} (N q^{a_1 \cdots a_p} \mathcal{L}_\xi \hat{\phi}_{,a_1 \cdots a_p} - Q^{a_1 \cdots a_p} \mathcal{L}_\xi \hat{\Phi}_{,a_1 \cdots a_p})
  \nonumber
  \\
  ~ + \frac{1}{2} \sum_{p=2}^\infty \frac{1}{p!} N \hat{I}^{a_1 \cdots a_p bc} \mathcal{L}_\xi g_{bc,a_1 \cdots a_p},
  \label{GenForceMultipole}
\end{align}
where the $2^p$-pole scalar and electromagnetic moments which appear here are explicitly
\begin{subequations}
  \label{chargeMoments}
        \begin{align}
        q^{a_1 \cdots a_p} \equiv \frac{1}{N} \int_{\Sigma_t} X^{a_1} \cdots X^{a_p} \rho' N' d V_{\bperp}',
        \\
        Q^{a_1 \cdots a_p} \equiv \int_{\Sigma_t} X^{a_1} \cdots X^{a_p} J' N' d V_{\bperp}'.
        \end{align}
\end{subequations}
It is clear that these moments are purely spatial and also symmetric in all indices. Their differing normalizations guarantee that they remain invariant under time rescalings with the form \eqref{tScale} [cf. \eqref{EMScale}].

While a relatively simple expression for the $2^p$-pole moments $\hat{I}^{a_1 \cdots a_p bc}$ of $\hat{T}_{\subB}^{ab}$ is easily suggested by comparing  \eqref{GenForceFin}, \eqref{Lieg}, and \eqref{GenForceMultipole}, additional consideration of the index symmetries \eqref{gSyms} associated with the metric extensions shows that some components of that expression do not couple to the force \cite{Dix74, HarteTrenorm}. A less obvious definition which takes this into account is obtained by first defining the auxiliary $2^p$-pole moment
\begin{align}
  \hat{J}^{a_1 \cdots a_p bc} \equiv \frac{1}{N} \int_{\Sigma_t} X^{a_1} \cdots X^{a_{p-2}} X^{ [a_{p-1} } X^{[b}
  \nonumber
  \\
  ~ \times \stackrel{(p)}{A} \!\! {}^{a_p]c]}{}_{a'b'}  \hat{T}^{a'b'}_\subB N' d V_{\bperp}',
  \label{Jmoments}
\end{align}
where the notation indicates independent antisymmetrizations on the index pairs $(a_{p-1},a_p)$ and $(b,c)$. The moments appearing in \eqref{GenForceMultipole} can then be defined by
\begin{align}
  \hat{I}^{a_1 \cdots a_p bc} \equiv 4 \left( \frac{p-1}{p+1} \right) \hat{J}^{(a_1 \cdots a_{p-1} | b | a_p ) c}.
  \label{Imoments}
\end{align}
These are separately symmetric in their first $p$ and final two indices,  satisfy
\begin{equation}
  \hat{I}^{(a_1 \cdots a_p b) c} = 0
\end{equation}
for all $p \geq 2$, and are partially spatial in the sense that
\begin{equation}
  \tau_{a_1} \hat{I}^{a_1 \cdots a_{p-2} [a_{p-1} [a_p b ]c ]} = 0
\end{equation}
for all $p \geq 3$. Except for the substitution $T^{ab}_\subB \rightarrow \hat{T}_{\subB}^{ab}$ and the overall factor of $1/N$ inserted for convenience in \eqref{Jmoments}, these stress-energy moments are identical to those originally derived by Dixon \cite{Dix74}.

For both the charge and stress-energy multipole moments, there are important cases in which the given definitions result in tensors which are still ``more complicated'' than necessary; some of their components decouple from $d \hat{P}_t/dt$ for particular classes of fields. For example, it follows from \eqref{LieXiZ} and \eqref{gExtensions} that if the vacuum Einstein equation $R_{ab} = 0$ holds, traces of $\hat{I}^{a_1 \cdots a_p bc}$ decouple at least for $p=2, 3$. The stress-energy quadrupole and octupole moments can therefore be replaced, in vacuum backgrounds, by their trace-free counterparts. It follows from \eqref{phiExtensions} that similar comments also apply to $q^{a b}$ and $q^{abc}$ whenever $\nabla^a \nabla_a \hat{\phi} = 0$. It is not clear, however, if these types of simplifications can be continued to higher multipole orders.

\subsection{Center of mass}
\label{Sect:Momenta}

The multipole expansion \eqref{GenForceMultipole} for the generalized force is useful only if it can be adequately approximated by low-order truncations of the infinite sums which appear there. Whether or not this is possible depends not only on the nature of the physical system and the choice of propagators, but also on the worldline $\mathcal{Z}$ about which our expansions have been performed. If a useful truncation is obtained for one particular $\mathcal{Z}$, the same cannot necessarily be said for worldlines which differ by distances comparable to any lengthscales associated with $g_{ab}$ or $\hat{\phi}$. It is therefore essential that $\mathcal{Z}$ be appropriately ``centered'' on $B$ so that the higher multipole moments remain as small as possible. The interpretation we adopt is more specifically that $\mathcal{Z}$ should be a ``center-of-mass worldline'' for $B$.

Even for freely-falling, uncharged test bodies in special relativity, the center-of-mass is a nontrivial concept. The typical approach has the following flavor: First, an antisymmetric angular momentum tensor $S^{ab}(z_t)$ is defined with respect to an arbitrary origin $z_t$. Space-space components of this tensor contain information physically associated with an object's ``spin'' and ``orbital'' angular momentum, while time-space components instead describe the ``mass dipole vector.'' A center of mass can then be defined roughly as that worldline for which the dipole moment vanishes. Part of the subtlety with this definition in generic (not necessarily static) contexts is that it is unclear which frame should be used to split the angular momentum tensor into spacelike and timelike components. Different possibilities---sometimes interpreted as different ``observers'' for whom the dipole moment appears to vanish---result in different worldlines. These choices are often referred to as ``spin supplementary conditions.'' While some have nicer properties than others, it is mainly a matter of convenience which particular worldline is used to represent an extended worldtube. It is important, however, that all possibilities be confined to a sufficiently small region. This has indeed been established in very simple cases \cite{DixonSR, CostaHelical}, and recent progress has been made on extending it \cite{CostaReview}, although a general result using the definitions given here is not known. We nevertheless adopt a center-of-mass definition which broadly follows this tradition and assume that the resulting $\mathcal{Z}$ remains near $B$ in a suitable sense.

The first step in carrying out this procedure is to define an angular momentum tensor. From the perspective of the formalism discussed in section \ref{Sect:mf}, this is naturally associated with certain components of the generalized momentum $\hat{P}_t$. More precisely, an angular momentum\footnote{Even after a spin supplementary condition has been applied so $S^{ab}$ is purely spatial, it is only when $n=3$ that angular momentum can be equivalently described as a vector quantity.} $S^{ab} = S^{[ab]}$ and a linear momentum $p^a$ may be introduced implicitly at $z_t$ by demanding that
\begin{equation}
  \hat{P}_t(\xi) = p^a(t) \xi_a(z_t) + \frac{1}{2} S^{ab}(t) \nabla_a \xi_b (z_t)
  \label{PtopS}
\end{equation}
for all $\xi^a \in K_\subG$. The linear momenta are therefore those components which are associated with generalized Killing fields which appear to be purely translational at $z_t$, meaning that $\nabla_a \xi_b(z_t) = 0$. The remaining angular components are instead associated with those vector fields which appear to generate pure Lorentz transformations at $z_t$, in the sense that $\xi_a (z_t) = 0$.

The left-hand side of \eqref{PtopS} is well-defined because any $\xi^a \in K_\subG$ is uniquely determined by $\xi_a(z_t)$ and $\nabla_a \xi_b = \nabla_{[a} \xi_{b]} (z_t)$ for any $z_t \in \mathcal{Z}$ \cite{HarteSyms}. More explicitly, the $X^a$ and $H^{a'}{}_{a}$ defined by \eqref{XDef} and \eqref{HDef} may be used to show that
\begin{equation}
  \xi^{a'} = H^{a'}{}_{b} ( -\nabla_a X^b \xi^a + X^a \nabla_a \xi^b)
\end{equation}
for any $x' \in \Sigma_t$. This determines a generalized Killing field for each 1-form $\xi_a(z_t)$ and each 2-form $\nabla_a \xi_b (z_t)$. Varying over all such possibilities while using \eqref{PHatSimp} and \eqref{PtopS} shows that the linear and angular momenta are explicitly
\begin{subequations}
        \label{momenta}
        \begin{align}
          \label{padef}
                p_a = - \int_{\Sigma_t} (H_{a'b} \nabla_a X^b) \hat{T}^{a'b'}_\subB dS_{b'},
                \\
                S^{ab} = 2 \int_{\Sigma_t} ( X^{[a} H_{a'}{}^{b]} ) \hat{T}^{a'b'}_\subB dS_{b'}.
        \end{align}
\end{subequations}
These coincide with Dixon's momenta \cite{Dix70a, Dix74} up to the replacement $T^{ab}_\subB \rightarrow \hat{T}_{\subB}^{ab}$. As in all other parts of our discussion, explicit integrals such as \eqref{momenta} are included for completeness but play no role in what follows.

 The next step is to understand how $p_a$ and $S^{ab}$ evolve in time. One way to accomplish this is to deduce from the assumed stationarity of the system that $\mathcal{L}_\tau p_a = \mathcal{L}_\tau S^{ab} = 0$, or equivalently
 \begin{equation}
  \frac{ Dp^a }{dt} = p^b \nabla_b \tau^a, \qquad \frac{ DS^{ab} }{ dt } = - 2 S^{c[a} \nabla_c \tau^{b]}.
  \label{ForceFromTau}
\end{equation}
The momenta can therefore be computed for all time given only their initial values and $\nabla_a \tau^b$. These are not, however, the interesting observables. As explained in sections \ref{Sect:ExtFields} and \ref{Sect:Ginvariance}, we instead seek those holding fields which must be applied in order to maintain \eqref{ForceFromTau}.

Appropriate holding fields can be obtained by deriving alternative evolution equations for the momenta in terms of the previously-derived evolution equation \eqref{GenForceMultipole} for $\hat{P}_t$. Recalling that generalized Killing fields satisfy the Killing transport equation \eqref{KillingTransport} on $\mathcal{Z}$, direct differentiation of \eqref{PtopS} yields
\begin{align}
  \frac{D p^a}{dt} = \frac{1}{2} R_{bcd}{}^{a} S^{bc} \tau^d + F^a, \quad
  \frac{D S^{ab} }{ dt }  = 2 p^{[a} \tau^{b]} + N^{ab},
  \label{Papapetrou}
\end{align}
where the force $F^a$ and torque $N^{ab} = N^{[ab]}$ are defined implicitly via
\begin{equation}
  \frac{d}{dt} \hat{P}_t(\xi) = F_a \xi^a + \frac{1}{2} N^{ab} \nabla_a \xi_b.
  \label{PDotToFN}
\end{equation}
Note that if $d \hat{P}_t / dt$ is negligible, \eqref{Papapetrou} reduces to the well-known Mathisson-Papapetrou equations traditionally used to describe spinning particles in curved spacetimes. One feature of the formalism described here is that the Mathisson-Papapetrou terms  $\frac{1}{2} R_{bcd}{}^{a} S^{bc} \tau^d$ and $2 p^{[a} \tau^{b]}$ have a clear geometrical origin: They arise because the decomposition of $K_\subG$ into ``generators of Lorentz transformations'' $\oplus$ ``generators of translations'' is meaningful only with respect to  a preferred point. Applying a time derivative varies the relevant point (in time), and vector fields which appear purely translational at, e.g., $z_t$ do not necessarily have the same character at $z_{t+dt}$. This change results in a mixing of the linear and angular momenta over time\footnote{This interpretation of the Mathisson-Papapetrou terms is only minimally related to the definition of $K_\subG$, and applies also in maximally-symmetric spacetimes where all relevant vector fields are genuinely Killing. It extends even to some non-relativistic settings \cite{HarteReview}.}. Following \cite{Dix79}, the force $F_a$ and torque $N^{ab}$ defined here exclude kinematic effects such as these. They depend only on the dynamics of the generalized momentum.

Combining \eqref{GenForceMultipole} with \eqref{PDotToFN} while varying over all generalized Killing fields now results in explicit multipole series for the force and torque. We find it useful below to split these series into ``gravitational,'' ``holding,'' and ``self'' components in the sense that
\begin{subequations}
\label{FNmultipole}
\begin{align}
  F_a &= F_a^\mathrm{grav} + F_a^\mathrm{self} + F_a^\hold,
  \label{Fmultipole}
  \\
  N^{ab} &= N^{ab}_\mathrm{grav} + N^{ab}_\mathrm{self} + N^{ab}_\hold.
\end{align}
\end{subequations}
The gravitational force and torque which appear here are explicitly
\begin{subequations}
\label{FNgrav}
\begin{align}
  F_a^\mathrm{grav} \equiv & ~ \frac{1}{2} \sum_{p=2}^\infty \frac{N}{p!} \hat{I}^{b_1 \cdots b_p cd} \nabla_a g_{cd,b_1 \cdots b_p}  ,
\\
  N^{ab}_\mathrm{grav} \equiv & ~ \sum_{p=2}^\infty \frac{2N}{p!} ( \hat{I}^{c_1 \cdots c_p d [a} g^{b]}{}_{d,c_1 \cdots c_p}
  \nonumber
  \\
  & ~ + \frac{p}{2} \hat{I}^{c_1 \cdots c_{p-1} [a |dh|} g_{dh,c_1 \cdots c_{p-1}}{}^{b]} ) ,
\end{align}
\end{subequations}
while the holding forces and torques are
\begin{subequations}
\label{FNhold}
\begin{gather}
  F_a^\hold \equiv \sum_{p=0}^\infty \frac{1}{p!} ( N q^{b_1 \cdots b_p} \nabla_a \phi_{,b_1 \cdots b_p}^\hold
   - Q^{b_1 \cdots b_p} \nabla_a \Phi_{,b_1 \cdots b_p}^\hold ),
  \label{Fhold}
  \\
  N^{ab}_\hold \equiv \sum_{p=1}^\infty \frac{2}{(p-1)!} ( N q^{c_1 \cdots c_{p-1} [a} \phi^\hold_{,c_1 \cdots c_{p-1} }{}^{b]}
  \nonumber
  \\
   ~ - Q^{c_1 \cdots c_{p-1} [a}  \Phi^\hold_{,c_1 \cdots c_{p-1} }{}^{b]}   ).
   \label{Nmultipole}
\end{gather}
\end{subequations}
The self-force $F_a^\mathrm{self}$ and self-torque $N^{ab}_\mathrm{self}$ are identical in form to the holding force and holding torque except for the substitutions $\phi^\hold_{, b_1 \cdots b_p} \rightarrow \hat{\phi}^\mathrm{self}_{, b_1 \cdots b_p}$ and $\Phi^\hold_{, b_1 \cdots b_p} \rightarrow \hat{\Phi}^\mathrm{self}_{, b_1 \cdots b_p}$ [where $\hat{\phi}_\mathrm{self}$ and $\hat{\Phi}_\mathrm{self}$ are defined by \eqref{phiHatSelfDef}].
Each of these expressions scales like
\begin{equation}
  F_a^{\ldots} \rightarrow \alpha^{-1} F_a^{\ldots}, \qquad N^{ab}_{\ldots} \rightarrow \alpha^{-1} N^{ab}_{\ldots}
\end{equation}
under time reparameterizations with the form \eqref{tScale}.

Equation \eqref{PDotToFN} and the conservation of the renormalized energy $\hat{E} \equiv - \hat{P}_t(\tau)$ [cf. \eqref{energy}] imply that at least one component of these equations always vanishes:
\begin{equation}
  F_a \tau^a + \frac{1}{2} N^{ab} \nabla_a \tau_b = 0.
  \label{eCons}
\end{equation}
If any additional symmetries are present, similar constraints may be associated with them as well.

We now have two independent sets of evolution equations for $p_a$ and $S_{ab}$, namely \eqref{ForceFromTau} and \eqref{Papapetrou}. Equating them results in the consistency conditions
\begin{subequations}
\label{consistency}
\begin{align}
  F^a = p^b \nabla_b \tau^a - \frac{1}{2} R_{bcd}{}^{a} S^{bc} \tau^d,
  \label{Fconsistency}
  \\
  N^{ab} = -  2 ( p^{[a} \tau^{b]} + S^{c[a} \nabla_c \tau^{b]}).
  \label{Nconsistency}
\end{align}
\end{subequations}
Contracting the second of these equations with $\tau_b$ further shows that the momentum must be related to the unit velocity $u^a \equiv \tau^a/N$ via
\begin{align}
   p^a = (-p_b u^b) u^a + \frac{1}{N} ( u^b S_{b}{}^{c} \nabla_c \tau^a + N^{a}{}_{b} u^b )
  \nonumber
  \\
  ~ + S^{ab} \nabla_b \ln N,
  \label{MomVelRelation}
\end{align}
thus implying that $p^a$ need not be parallel to $u^a$. If these two vectors are indeed non-parallel, $B$ is said to possess a ``hidden momentum'' \cite{Bobbing, CostaReview}.

The momentum-velocity relation \eqref{MomVelRelation} holds for any $\mathcal{Z}$ which is an orbit of $\tau^a$; we have not yet imposed a center-of-mass condition which would single out one particular orbit. A common spin supplementary condition may nevertheless be imposed which does single out a particular worldline while also removing one component of the hidden momentum: Let the ``mass dipole moment'' vanish for static observers in the sense that
\begin{equation}
  S_{ab}(z_t) \tau^b(z_t) = 0.
  \label{Centroid}
\end{equation}
We call those points $\mathcal{Z} = \cup_t z_t$ which are consistent with this constraint the center-of-mass worldline. Applying it from now on, the momentum and velocity are seen to be related via
\begin{align}
   p^a = m u^a + S^{ab} D_b \ln N + \frac{1}{N} N^{a}{}_{b} u^b ,
  \label{MomVelRelation2}
\end{align}
where we have introduced the mass
\begin{equation}
  m \equiv - p_a u^a.
  \label{mDef}
\end{equation}

The consistency equations \eqref{consistency} can now be simplified significantly. First note from \eqref{gradTau}  that
\begin{align}
  R_{bcd}{}^{a} \tau^d = - 2 \nabla_{[b} \nabla_{c]} \tau^a = - 2 u_{[b} \nabla_{c]} \nabla^a N,
\end{align}
so the Mathisson-Papapetrou ``force'' $\frac{1}{2} R_{bcd}{}^{a} S^{bc} \tau^d$  vanishes on account of \eqref{Centroid}. Use of \eqref{MomVelRelation2} also shows that
\begin{equation}
  p^b \nabla_b \tau_a = m D_a N - u_a u^b N_{b}{}^{c} D_c \ln N,
\end{equation}
so \eqref{Fconsistency} reduces to
\begin{equation}
  F_a = m D_a N - u_a u^b N_{b}{}^{c} D_c \ln N.
\end{equation}
The mixing here between forces and torques can be eliminated by recalling \eqref{eCons}, which finally results in
\begin{equation}
  h^{b}{}_{a} F_b = m D_a N .
  \label{Fconsistency2}
\end{equation}
Our consistency condition for the body's translational degrees of freedom therefore reduces to the simple statement that the total spatial force is what one would expect for an uncharged monopole test particle with worldline $\mathcal{Z}$. The torque balance equation \eqref{Nconsistency} simplifies similarly; using \eqref{gradTau}, \eqref{Centroid}, and \eqref{MomVelRelation2}, it reduces to
\begin{equation}
  h^{c}{}_a h^{d}{}_b N_{cd} = 0.
  \label{Nconsistency2}
\end{equation}
That the spatial components of the net torque must vanish is again what might have been expected on elementary grounds. Such simplicity in exact equations might be viewed as additional evidence that the definitions for $F_a$ and $N_{ab}$ are ``physically appropriate.'' Note in particular that even though the spin can cause $p^a$ to differ from $m u^a$, it plays no explicit role in \eqref{Fconsistency2} or \eqref{Nconsistency2}.

\subsection{Holding forces}

We have emphasized in sections \ref{Sect:genDescription} and \ref{Sect:Ginvariance} that the primary observable here is the generalized holding force $\mathcal{F}_\hold$, as defined by \eqref{FholdDef}. Moreover, $\mathcal{F}_\hold$ can be decomposed into an (ordinary) holding force $F_a^\hold$ and a holding torque $N^{ab}_\hold$ using an equation analogous to \eqref{PDotToFN}. Doing so results in the multipole expansions \eqref{FNhold}, which relate these quantities to $\phi_\hold$ and $\Phi_\hold$. Equations \eqref{Fconsistency2} and \eqref{Nconsistency2} further show that consistency with the staticity assumption is maintained only if
\begin{subequations}
\label{FNconsistency}
\begin{gather}
  h^{b}{}_{a} F_b^\hold = m D_a N - h^{b}{}_{a} (F_b^\mathrm{grav} + F_b^\mathrm{self} ),
  \label{Fconsistency3}
  \\
  h^{c}{}_{a} h^{d}{}_{b} N_{cd}^\hold = - h^{c}{}_{a} h^{d}{}_{b} ( N_{cd}^\mathrm{grav} + N_{cd}^\mathrm{self} ).
  \label{Nconsistency3}
\end{gather}
\end{subequations}
Multipole expansions for the gravitational force and torque which appear here are given by \eqref{FNgrav}, while the self-force and self-torque are obtained by the replacements $\phi^\hold_{, b_1 \cdots b_p} \rightarrow \hat{\phi}^\mathrm{self}_{, b_1 \cdots b_p}$ and $\Phi^\hold_{, b_1 \cdots b_p} \rightarrow \hat{\Phi}^\mathrm{self}_{, b_1 \cdots b_p}$ in \eqref{FNhold}. These are some of our main results. They apply for essentially all static, extended charge distributions in which the center-of-mass condition \eqref{Centroid} has been applied.

\subsection{Monopole approximation}

As a simple example of these equations, consider a purely-electric charge (so $\rho = 0$) in an approximation where all forces and torques are ignored except for those which couple to the monopole moments $m$ and $Q$. It then follows from \eqref{FNgrav} and \eqref{FNhold} that all of $N_{ab}$ is negligible and \eqref{Nconsistency3} is trivially satisfied. The balance of forces associated with \eqref{Fconsistency3} is more interesting, implying that the electromagnetic holding field $F_{ab}^\hold = 2 \nabla_{[a} ( \tau_{b]} N^{-2} \Phi_\hold)$ must be related to the body's location and its effective self-field $\hat{F}_{ab}^\mathrm{self} = F_{ab}^\mathrm{self} - F_{ab}^\subS$ via
\begin{equation}
  \frac{1}{N} F_a^\hold = Q F_{ab}^\hold u^b = m D_a \ln N - Q \hat{F}_{ab}^\mathrm{self} u^b.
  \label{HoldingF1}
\end{equation}
This generalizes the elementary result \eqref{HoldingF0} to allow for nontrivial self-interaction. Recall that even if $F_{ab}^\mathrm{self}$ is the self-field of a point charge\footnote{As with any consistent discussion of Maxwell theory, the formalism here does not make sense for charges which are ``truly'' pointlike. Distributional charge distributions nevertheless arise as limits of smooth charge distributions. See section \ref{Sect:pp}.}, its hatted counterpart $\hat{F}_{ab}^\mathrm{self}$ remains smooth at the charge's location---at least when the $\mathcal{G}$ used to compute it is a parametrix (and also satisfies the electromagnetic analogs of the remaining properties summarized at the end of section \ref{Sect:ForceScalar}). Incidentally, \eqref{MomVelRelation2} implies that the momentum and velocity satisfy the elementary relation $p^a = m u^a$ when the dipole and higher-order moments are neglected.

The simple truncation used to obtain \eqref{HoldingF1} is instructive, but is not necessarily physically appropriate. We next discuss more carefully what a ``point particle limit'' might mean and what its implications are. It is only at this stage in our discussion where the number of dimensions starts to play any explicit role\footnote{The number of dimensions has appeared implicitly via our demonstration in appendix \ref{Sect:Hadamard} that there exist propagators which satisfy the constraints of section \ref{Sect:mf}. The final conclusion of that argument---that appropriate propagators do indeed exist---is nevertheless independent of $n$.}. The neglect of gravitational multipole couplings in \eqref{HoldingF1} will be seen, e.g., to be generically consistent only for $n<4$. Similarly, neglecting the electromagnetic dipole moment is generically consistent only for $n < 3$.

\section{Point particle limits}
\label{Sect:pp}

The equations derived above are very general. In many practical applications, however, one would like to specialize them to cases where the body $B$ is sufficiently small that its internal structure---or equivalently its higher multipole moments---can be effectively ignored. Making this statement precise requires an approximation which strongly depends on the number of spatial dimensions $n$. In fact we shall see that in higher dimensions, forces associated with higher multipole moments can scale in the same way as the leading-order self-force and therefore cannot be ignored (even for ``spherically-symmetric'' bodies).

In this section, we describe a set of assumptions on a one-parameter family of bodies $B_\lambda$ which has the interpretation that the limit $\lambda \rightarrow 0^+$ is a ``point particle'' limit. These assumptions are detailed in section \ref{Sect:Scalings} and some of their consequences are derived in section \ref{Sect:limitoutside}.

\subsection{Motivating and defining an appropriate one-parameter family}
\label{Sect:Scalings}

Fixing a particular static spacetime and a particular center-of-mass worldline $\mathcal{Z}$, a family of bodies $B_\lambda$ can describe a point particle limit if the spatial support of each member ``shrinks down'' to $\mathcal{Z}$ as $\lambda \rightarrow 0^+$. Physically, the overall linear dimensions of $B_\lambda$ are assumed to be proportional to $\lambda$ at least when this parameter is sufficiently small.

More precisely, fix coordinates $(t,\bx)$ so that the metric components have the explicitly-static form \eqref{metric} with $N(\bx)$ and $h_{ij}(\bx)$ both smooth. We then assume that
\begin{subequations}
\label{Scalings}
        \begin{gather}
            \rho(\bx;\lambda) = \lambda^\beta \tilde{\rho} ( [\bx - \bm{z}]/      \lambda; \lambda ),
            \\
                J (\bx;\lambda) = \lambda^\beta \tilde{J} ( [\bx - \bm{z} ]/\lambda; \lambda ),
                \\
            T_\subB^{\mu\nu} (\bx;\lambda) = \lambda^\gamma \tilde{T}_\subB^{\mu\nu} ( [\bx - \bm{z} ]/\lambda; \lambda ),
        \end{gather}
\end{subequations}
where $\bm{z}$ is the coordinate location of $\mathcal{Z}$ (for all $t$). The constants $\beta$ and $\gamma$ are to be fixed below, while the functions $\tilde{\rho}$, $\tilde{J}$, and $\tilde{T}_\subB^{\mu\nu}$ are assumed to have compact support in their first argument and to be smooth in both arguments. Also suppose that $\tilde{\rho} ( \tilde{\bx}, 0 )$, $\tilde{J}( \tilde{\bx}, 0)$, and $\tilde{T}_\subB^{\mu\nu} ( \tilde{\bx}, 0 )$ exist and are nonzero at least for some $\tilde{\bx} \neq 0$. More specifically, denote the largest $| \tilde{\bx} |$ for which they are nonvanishing by $\tilde{\mathcal{R}} > 0$. It then follows that $B_\lambda$ is contained in the ball $| \bx - \bm{z} | < \mathcal{R} \equiv \lambda \tilde{\mathcal{R}} $ as $\lambda \rightarrow 0^+$, thus providing a sense in which---as claimed---the family shrinks linearly towards $\bm{z}$. It can be shown that if assumptions like these are valid in one static coordinate system with the form \eqref{metric}, they are also valid in all other smoothly-related static coordinate systems \cite{GrallaHarteWald}.

The nontrivial issue is now to choose the scaling parameters $\beta$ and $\gamma$ associated via \eqref{Scalings} with the charge densities and the stress-energy tensor. We can constrain these parameters by imposing three physical requirements as $\lambda \rightarrow 0^+$:
\begin{enumerate}
        \item The self-energy does not exceed the total mass. \label{scalingCons1}
        \item The mass density remains finite. \label{scalingCons2}
        \item The electromagnetic or scalar self-interactions are more significant than the gravitational self-interaction. \label{scalingCons3}
\end{enumerate}
The first two of these assumptions are very mild, while the last specializes our discussion to a particular physical regime.

More specifically, requirement \ref{scalingCons1} is inspired by the need to exclude negative energy densities and similar pathologies. If a body's dominant electric multipole moment is its total charge $Q$, its self-energy is expected to be of order $Q^2/\mathcal{R}^{n-2}$ [see, e.g., \eqref{mShift} at least for $n>2$]. The scalings \eqref{Scalings} then imply that\footnote{This scaling law and others discussed in this section are valid up to possible factors of $\ln \lambda$.}
\begin{equation}
  \frac{ Q^2 }{ m \mathcal{R}^{n-2} } \sim \lambda^{2 (\beta + 1) - \gamma },
  \label{selfEnergy}
\end{equation}
which remains finite as $\lambda \rightarrow 0^+$ only if
\begin{equation}
  \gamma \leq 2 (\beta + 1).
  \label{rhoScaling}
\end{equation}
Requirement \ref{scalingCons2} additionally implies that $m/\mathcal{R}^n \sim \lambda^\gamma$ cannot diverge, so
\begin{equation}
  \gamma \geq 0, \qquad \beta \geq -1,
  \label{TScaling}
\end{equation}
where the second of these inequalities results from combining the first with \eqref{rhoScaling}. Lastly, the ratio of the gravitational and electric self-energies is typically of order $(m/Q)^2 \sim \lambda^{2 ( \gamma - \beta) }$, which tends to zero as $\lambda \to 0^+$ when
\begin{equation}
  \gamma > \beta.
  \label{smallGravSF}
\end{equation}

Although relations (\ref{rhoScaling}), (\ref{TScaling}) and (\ref{smallGravSF}) do not determine $\beta$ and $\gamma$ uniquely, they do imply that the importance of the self-force relative to multipolar forces {\it diminishes} with increasing $n$. To see this, first note that the inertial force $m D_a N$ appearing in \eqref{Fconsistency3} scales like $\lambda^{\gamma + n}$. In many cases, it is this which the leading-order holding force must balance. If not, the leading-order holding force instead counteracts the self-force, which scales like $Q^2 \sim \lambda^{2 (\beta + n)}$. Hence,
\begin{equation}
        \nabla \Phi_\hold \sim \lambda^{ \min (\gamma - \beta, \beta + n)}.
        \label{phiHoldScale}
\end{equation}
If $\beta + n < \gamma - \beta$ for some $n$, the inequality must reverse in higher dimensions---implying that the self-force is subdominant for sufficiently large $n$. In the latter cases, the $2^p$-pole gravitational force as well as the $2^p$-pole scalar and electromagnetic holding forces are expected to generically scale like
\begin{equation}
        (\mbox{$2^p$-pole force or torque}) \sim \lambda^{\gamma + n + p}.
\end{equation}
These effects are large compared to the leading-order self-force for all multipole orders
\be
        p < p_\subSF \equiv n + 2 \beta -\gamma.
\label{pMax}
\ee
From \eqref{rhoScaling} it follows that the critical multipole order
satisfies $p_\subSF \ge n-2$. If $n = 2$, the monopole holding force
and the self-force can therefore appear at the same order; the latter
is not necessary small. If $n=3$, the self-force can no longer be
quite this large, but is at most comparable to the dipole holding
force. Moving to $n=4$, the leading-order self-force can be as large
as ordinary quadrupole effects, but no larger\footnote{These
  conclusions can also be motivated using the language of effective
  field theory: Consider for example the quadrupole coupling of an object moving
in a generic spacetime.  The action can then contain a term $\int d\tau c_{abcd} R^{abcd}$, where $R^{abcd}$ is the Riemann tensor and $c_{abcd}$ are body parameters. Self-field effects will renormalize $c_{abcd}$ by a term proportional to $ {\cal R}^{4-n} q^2$,
where $q$ is the charge and ${\cal R}$ the size of the body, by dimensional analysis.  This term remains important as ${\cal R} \to 0$ for $n \ge 4$.}.

Similar comments also apply to the self-torque. This first arises from a monopole-dipole coupling, and therefore scales like $Q^2 \mathcal{R} \sim \lambda^{2(\beta+n)+1}$. Spatial components of self-torques can consequently be comparable to ordinary $2^{(p_\subSF + 1)}$-pole torques.

If a particular observable---say the holding force---is to be understood up to some given accuracy, it follows from \eqref{pMax} that the self-force can be ignored for sufficiently large $n$. If, however, the self-force is considered ``interesting'' on its own, its effects can be meaningfully interpreted only in combination with all extended-body terms up to order $p_\subSF$. These statements are actually independent of conditions \ref{scalingCons2} and \ref{scalingCons3} above.

We now specialize the discussion further by considering those families whose mass densities do not vary appreciably as $\lambda \rightarrow 0^+$. This can be motivated by noting the density of solid matter does not change very much except under severe conditions, and we do not want our limit to implicitly impose those conditions or to strongly vary the underlying material. Thus setting $\gamma = 0$, it follows from \eqref{rhoScaling} and \eqref{smallGravSF} that $\beta \in [-1 , 0)$. It is convenient for the development of simple Taylor expansions that the scaling exponents be integers, so consider
\begin{equation}
  \beta = -1, \qquad \gamma = 0.
  \label{scaleExps}
\end{equation}
Fractional self-energies then have finite limits as $\lambda \rightarrow 0^+$. If $n \geq 2$, holding fields scale like $\lambda^1$, $2^p$-pole forces scale like $\lambda^{n+p}$, scalar and electromagnetic self-forces are of order $\lambda^{2 (n-1)}$, and gravitational self-forces are of order $\lambda^{2n}$. The self-interaction is also ``as large as possible'' in the sense that
\begin{equation}
  p_\subSF = n-2.
\end{equation}
The minimum exponent in \eqref{phiHoldScale} reverses if $n=1$, in which case the leading-order holding force and self-force both scale like $\lambda^0$.

We note that our scaling exponents \eqref{scaleExps} differ (in the comparable $n=3$ case) from those considered in \cite{GrallaHarteWald}; their choices violate our conditions \ref{scalingCons2} and \ref{scalingCons3}.

\subsection{Evaluating the point particle limit}
\label{Sect:limitoutside}

Assuming a one-parameter family of bodies $B_\lambda$ which satisfy \eqref{Scalings} and \eqref{scaleExps}, holding fields can be determined by appropriate truncations of the multipole series \eqref{FNgrav}, \eqref{FNhold}, and \eqref{FNconsistency}. Self-force effects first arise at the scaling order $\lambda^{2(n-1)}$, which corresponds when $n \neq 1$ to the multipole order $p_\subSF = n-2$. It follows, e.g., that the monopole approximation is all that's needed when $n=2$, in which case \eqref{HoldingF1} holds for purely-electric charges up to error terms of order $\lambda^3$. Similarly, the first influence of the electric self-torque in $2+1$ dimensions occurs via
\begin{equation}
  Q_{[a} (F_{b]c}^\hold + \hat{F}_{b]c}^\mathrm{self}) u^c + O(\lambda^4) =  0,
\end{equation}
which provides one algebraically-independent constraint on $F_{ab}^\hold$. Analogous expressions for larger $n$ are easily obtained from the general multipole series, and involve additional multipole moments of the body's charge distribution. If $n \geq 4$, the stress-energy multipole moments $\hat{I}^{c_1 \cdots c_p ab}$ must be taken into account as well as the charge moments when including leading-order self-force effects. Understanding the leading-order self-torque generically requires the consideration of quadrupole or higher stress-energy moments for all $n \geq 3$.

One complication which remains when evaluating a holding force or holding torque is the computation of the effective self-fields $\hat{\phi}_\mathrm{self}$ and $\hat{\Phi}_\mathrm{self}$. These are related via \eqref{phiHatSelfDef} to the physical self-fields discussed in section \ref{Sect:ExtFields}, so computing them requires knowledge of the $S$-fields defined by \eqref{phiSDef} and \eqref{phiSDefEM}. These $S$-fields in turn depend on spatial bidistributions $G$ and $\mathcal{G}$ which must satisfy properties \ref{functionalAssume}-\ref{paramAssume} listed at the end of section \ref{Sect:ForceScalar} (or their electromagnetic analogs). Whatever these propagators are---the Hadamard parametrices described in appendix \ref{Sect:Hadamard} provide one possibility---suppose that they are fixed. If $\lambda$ is sufficiently small that $\bx \neq \bz$ lies outside of $B_\lambda$, it then follows from \eqref{phiSDef} and \eqref{chargeMoments} that the scalar $S$-field is
\begin{equation}
  \phi_\subS (\bx; \lambda) =  q N(\bz) G (\bx,\bz) + O(\lambda^n),
\end{equation}
where the net charge satisfies
\begin{equation}
        q = \lambda^{n-1} \tilde{q} + O(\lambda^n)
        \label{qScale}
\end{equation}
for some $\lambda$-independent $\tilde{q}$. If the physical self-field $\phi_\mathrm{self}$ is associated with a Green function $G_\mathrm{self}$ as in \eqref{phiSelfConv}, it also follows that
\begin{align}
  \hat{\phi}_\mathrm{self} (\bx; \lambda) =  \lambda^{n-1} \big\{ \tilde{q} N(\bz) \big[ G_\mathrm{self} (\bx,\bz) - G (\bx,\bz) \big]  \big\}
  \nonumber
  \\
  ~ + O(\lambda^n)  .
  \label{phiSelfpp}
\end{align}
Similarly, the effective electromagnetic self-field is
\begin{align}
  \hat{\Phi}_\mathrm{self} (\bx; \lambda) =  \lambda^{n-1} \big\{ \tilde{Q} \big[ \mathcal{G}_\mathrm{self} (\bx,\bz) - \mathcal{G} (\bx,\bz) \big]  \big\}
   + O(\lambda^n)  .
  \label{phiSelfppEM}
\end{align}
Recalling that $G$ and $\mathcal{G}$ are parametrices, the elliptic regularity results discussed in section \ref{Sect:ForceScalar} imply that $\hat{\phi}_\mathrm{self}(\bx;\lambda)$ and $\hat{\Phi}_\mathrm{self}(\bx;\lambda)$ are smooth even in the point particle limit, and even at the body's limiting location. More precisely, the quantities in braces in \eqref{phiSelfpp} and \eqref{phiSelfppEM} smoothly extend to $\bx = \bz$, and their derivatives do so as well. It is these fields which determine the ``point particle self-force.''

Our discussion thus far has considered generic extended bodies, certain families of extended bodies, and finally the forces and torques which apply to members of those families whose sizes tend to zero. The results of this final step may be summarized as an algorithm which can be applied to understand self-interaction for ``effective point particles'': Compute point particle self-fields in the usual way and then regularize them by subtracting off $S$-fields generated by appropriate propagators. The resulting regularized fields evaluated at the particle's location then determine forces and torques via ordinary test particle expressions. An infinite regularization therefore emerges as the limit of exact and finite results obeyed by nonsingular extended bodies.

Similar difference-type regularizations have been considered in the $n=3$ self-force literature at least since the work of Dirac \cite{Dirac}, eventually culminating in the Detweiler-Whiting scheme heuristically proposed in \cite{DetweilerWhiting2003} and later derived and generalized in \cite{HarteScalar, HarteEM, HarteGrav}. We have provided a derivation---and not merely an assertion---for the static analog of this type of scheme, shown that it is valid in arbitrary dimensions, extended it to allow for more general propagators, and provided precise definitions for all relevant parameters in terms of a body's internal properties. Our results are also valid to all multipole orders.

It is worth noting, however, that a superficially-distinct type of regularization has often been  considered in the prior literature on the point particle self-force: Instead of subtracting off an appropriate field from the physical one, a type of averaging procedure is instead applied directly to the gradient of the physical field. In practice, most such schemes have actually been hybrids involving both subtractions and surface averages. The clearest example of this type is due to Quinn and Wald \cite{QuinnWald, Quinn}, although see also, e.g., \cite{PoissonLR, Poisson5dSF}. It does not appear to be widely known that there are in fact correct regularizations which involve only averages, and that these are completely equivalent to the difference-type regularizations discussed above. We now derive such a scheme for static charges in arbitrary dimensions.

The notion of average considered here is that of an appropriate integral over a closed $n-1$ dimensional surface which surrounds the body of interest. Consider again a finite extended body $B$ with nonsingular charge density $\rho$. Our basic starting point is a generalization of the Kirchhoff representation \cite{PoissonLR} for $\phi_\mathrm{self}$: Using \eqref{symassumption} and \eqref{GDefparam} to integrate $[D^a (N D_a \phi_\mathrm{self}) + \omega_n \rho N] G = 0$ by parts over an $n$-volume $\mathcal{B} \subset \Sigma$ which encompasses a spatial section of $B$,
\begin{align}
  \phi_\mathrm{self} = \frac{1}{ \omega_n } \oint_{\partial \mathcal{B}} ( G D_{a'} \phi_\mathrm{self} - \phi_\mathrm{self}' D_{a'} G) N' dS_{\bperp}^{a'}
  \nonumber
  \\
  ~ + \frac{1}{ \omega_n } \int_\mathcal{B} (\omega_n \rho' N' G + \phi_\mathrm{self}' \mathcal{S} ) dV'_{\bperp},
\end{align}
where $dS_{\bperp}^a$ denotes the natural $n-1$ dimensional surface element on $\partial \mathcal{B}$. Now applying \eqref{phiSDef} and \eqref{phiHatSelfDef}, the effective self-field is seen to satisfy
\begin{align}
  \hat{\phi}_\mathrm{self} = \frac{1}{ \omega_n } \oint_{\partial \mathcal{B}} [ G D_{a'} \phi_\mathrm{self} - \phi_\mathrm{self}' D_{a'} G] N' dS_{\bperp}^{a'}
  \nonumber
  \\
  ~ + \frac{1}{ \omega_n } \int_\mathcal{B} \phi_\mathrm{self}' \mathcal{S} dV'_{\bperp}.
  \label{phiHatSelfsurf}
\end{align}
This is exact for all $\mathcal{B}$, all propagators satisfying the conditions summarized at the end of section \ref{Sect:ForceScalar}, and for all extended objects. If $G$ is a Green function and not a more general parametrix, $\mathcal{S} = 0$ and the effective field reduces purely to an integral involving $\phi_\mathrm{self}$ and $D_a \phi_\mathrm{self}$ on the surrounding surface $\partial \mathcal{B}$. It is therefore a kind of averaging map $\phi_\mathrm{self} \mapsto \hat{\phi}_\mathrm{self}$.

Now specializing to leading-order effects in the point particle limit, a similar average holds even if $\mathcal{S} \neq 0$. This follows because \eqref{qScale} and the discussion in appendix \ref{Sect:Hadamard} imply that for $\tilde{q} \neq 0$, the interior self-field is of order $\lambda$ if $n \neq 2$ or of order $\lambda \ln \lambda$ otherwise. The smoothness of $\mathcal{S}$ therefore implies that the contribution of the body's interior to the volume integral in \eqref{phiHatSelfsurf} scales like $\lambda^{n+1}$ or $\lambda^{n+1} \ln \lambda$. Contributions to the volume integral arising from the exterior of $B_\lambda$ can be kept to a similar magnitude if the radius of $\partial \mathcal{B}$ scales like $\lambda$. In practice, it is useful to impose a somewhat slower tendency to zero so that $\partial \mathcal{B}$ is both very large compared to $B_\lambda$ and very small as seen by exterior observers whose scales don't change with $\lambda$. Regardless, these arguments imply that the volume integral can contribute only at orders higher than the dominant terms in $\hat{\phi}_\mathrm{self}$. Self-forces and self-torques can therefore be computed using
\begin{align}
  \hat{\phi}_\mathrm{self} = \frac{1}{\omega_n} \lim_{\partial \mathcal{B} \rightarrow 0} \oint_{\partial \mathcal{B} } ( G D_{a'} \phi'_\mathrm{self} - \phi'_\mathrm{self} D_{a'} G ) N' dS^{a'}_{\bperp}
  \label{phiHatGradSurface}
\end{align}
through leading order, where $\phi_\mathrm{self}$ is a point particle self-field and the limit implies that $\partial \mathcal{B}$ is an $n-1$ sphere whose radius is sent to zero. This is entirely equivalent to \eqref{phiHatSelfDef}, and therefore returns a nonsingular result. Analogous expressions for the electromagnetic potential are obtained by the obvious replacements $\hat{\phi}_\mathrm{self} \rightarrow \hat{\Phi}_\mathrm{self}$ and $G \rightarrow \mathcal{G}$.

It is possible to obtain more explicit averaging integrals for specific propagator choices---perhaps written in terms of Riemann normal coordinates---although this can be computationally challenging. Simplifications can sometimes be found by choosing $\partial \mathcal{B}$ to have special properties, although the complexity of \eqref{phiHatGradSurface} still grows rapidly with increasing $n$. We therefore omit any such calculations here.

\section{Explicit renormalizations of body parameters}
\label{Sect:bodyR}

It is shown in section \ref{Sect:mf} that the stress-energy tensor which appears in the expressions for the holding force and holding torque is not the usual stress-energy tensor $T^{ab}_\subB$, but is instead the renormalized $\hat{T}^{ab}_\subB$. This renormalization can affect a body's effective mass $m$, its effective stress-energy quadrupole $\hat{J}^{abcd}$, and so on. In this section, we compute the leading-order renormalizations of the mass and quadrupole moment in the point particle limit, extending previous renormalization computations for $n=3$ which were given in \cite{HarteTrenorm}.  We assume that $n > 2$ and specialize to the scalar case for concreteness. We also choose the propagator $G$ to be the Hadamard parametrix $G_\subH$ defined in appendix \ref{Sect:Hadamard} and summarized by \eqref{GHadamardSC}.

The explicit computations in this section have two purposes.  First, the detailed results serve to
illustrate and make concrete the rather formal theoretical framework developed in this paper.
Second, as discussed in section \ref{Sect:Ginvariance}, different choices for the propagator $G$ give rise to different scalar forces, different gravitational forces, and so on---it is only appropriate sums which remain invariant. The quadrupole renormalization computed here gives an explicit illustration of this degeneracy. As noted above, quadrupole forces can first be competitive with the ``ordinary'' self-force when $n=4$, and in this number of dimensions, one simple type of propagator freedom can be parametrized by the arbitrary lengthscale $\ell$ which is used to construct the Hadamard parametrix \eqref{GHadamardSC}. This $\ell$ is associated with a non-uniqueness of the gravitational force which exactly cancels similar $\ell$-dependencies in the scalar self-force and in the inertial force.

We assume a one-parameter family $B_\lambda$ of matter configurations satisfying the point particle scaling assumptions \eqref{Scalings} and \eqref{scaleExps}. For any well-behaved spatial coordinates $\bx$ such that the bodies' mass centers are located at $\bx = 0$, it follows that
\begin{subequations}
\label{ffamily0}
        \begin{gather}
                \label{ffamily}
                \rho(\bx;\lambda) = \lambda^{-1} {\tilde \rho}(\bx/\lambda;\lambda)
                \\
                T_\subB^{\mu\nu}(\bx;\lambda) = {\tilde T}_\subB^{\mu\nu} (\bx/\lambda;\lambda)
                \label{ffamily1}
        \end{gather}
\end{subequations}
for some smooth $\tilde{\rho}$ and $\tilde{T}^{\mu\nu}_\subB$. We suppose in particular that $(t,\bx)$ are appropriately-centered Fermi normal coordinates, so $N(0) = 1$ and $h_{ij}(0) = \delta_{ij}$.
It is convenient to introduce rescaled coordinates
\begin{equation}
\tilde{\bx} \equiv \bx/\lambda,
\label{rescaled}
\end{equation}
so that, for example, the function $\lambda \rho$ is a smooth function of $\tilde{\bx}$ and $\lambda$ (but not necessarily of $\bx$ and $\lambda$).

The renormalizations we compute depend linearly on the renormalized stress-energy tensor, which may be expanded in two parts:
\be
        {\hat T}_\subB^{\mu\nu}(\bfx;\lambda) = {\tilde T}_\subB^{\mu\nu}(\bfx/\lambda;\lambda) +
        \delta {\tilde T}_\subB^{\mu\nu}(\bfx/\lambda;\lambda).
\label{deltaTdef}
\ee
The first term here is the bare (unrenormalized) body stress-energy tensor which appears in the scaling assumption (\ref{ffamily1}), while $\delta {\tilde T}_\subB^{\mu\nu}$ instead quantifies the stress-energy renormalization due to a body's $S$-field. We call the mass and quadrupole components arising from $\delta {\tilde T}_\subB^{\mu\nu}$ the mass and quadrupole renormalizations. Note that although ${\tilde T}_\subB^{\alpha\beta}$ is smooth in its arguments, by assumption, we will find below that $\delta {\tilde T}_\subB^{\alpha\beta}$ can depend logarithmically on $\lambda$ for fixed $\bx/\lambda$.

We can derive an expression for the stress-energy renormalization $\delta {\tilde T}_\subB^{\alpha\beta}$ by writing the renormalization prescription (\ref{THat})
in the Fermi normal coordinates $(t,\bx)$, changing to the rescaled radial coordinates (\ref{rescaled}),
and using the scaling assumption (\ref{ffamily0}) and $d V_{\bperp} = \sqrt{h} d^n x$.  The result is
\begin{align}
        \delta {\tilde T}_{\subB}^{\mu\nu}(\tilde{\bx},\lambda) = \frac{\lambda^{2n-2}}{N(\lambda \tilde{\bx})} \int d^n \tilde{\bfv}
\sqrt{h(\lambda \tilde{\bfv} )}
\int d^n \tilde{\bfw} \sqrt{h(\lambda \tilde{\bfw} )}
\nonumber \\
\times
N(\lambda \tilde{\bfv} ) \, N(\lambda \tilde{\bfw} )\,
{\tilde \rho}(\tilde{\bfv},\lambda) \,{\tilde \rho}(\tilde{\bfw},\lambda) \nonumber \\
        ~ \times \left[
        \frac{ \delta G_\subH( \lambda \tilde{\bfv},\lambda \tilde{\bfw}) }{\delta h_{\mu\nu}(\lambda \tilde{\bfx})}
        - \frac{u^\mu u^\nu}{2 N(\lambda \tilde{\bx})}
        \frac{\delta G_\subH(\lambda \tilde{\bfv},\lambda \tilde{\bfw}) }{\delta N(\lambda \tilde{\bfx})}
        \right] .
        \label{big}
\end{align}
Note that in this expression, the indices $\mu,\nu$ refer to components with respect to the original Fermi coordinates $(t,\bx)$, not to the rescaled coordinates $(t,\tilde{\bx})$.
To leading order\footnote{Note that the leading-order contributions to $\hat{T}^{ab}_\subB - T^{ab}_\subB$ are not necessarily sufficient to determine all corrections which might be comparable to or larger than self-force effects.} in $\lambda$, the expression (\ref{big}) can be simplified:
we can replace ${\tilde \rho}(\tilde{\bfv},\lambda)$ and ${\tilde \rho}(\tilde {\bfw},\lambda)$
by their $\lambda \to 0$ limits, which we write simply as ${\tilde \rho}(\tilde{\bfv})$ and ${\tilde \rho}(\tilde{\bfw})$.
Similarly we can replace the instances of $h$ and $N$ with $h(0)=1$ and $N(0)=1$.

The evaluation of the variational derivatives which appear on the third line of (\ref{big}) is discussed in appendix \ref{Sect:VarDerivs}.  We show there that the variational derivatives formally consist of infinite series of line integrals of derivatives of Dirac delta distributions, of successively higher derivative orders. Evaluating $\delta {\tilde T}_{\subB}^{\mu\nu}$ pointwise is therefore impractical\footnote{No such problems appear if $G$ is identified with a propagator obtained by truncating the Hadamard series at a sufficiently high finite order.  Indeed, this is often more practical than attempting to use the full Hadamard parametrix.}. Nevertheless, we are primarily interested in the lowest moments of the stress-energy tensor, rather than its pointwise values. For this purpose, truncated versions of the series suffice. We argue in appendix \ref{Sect:VarDerivs} that it suffices to compute $\delta G_\subH/\delta h_{\mu\nu}$ and $\delta G_\subH/\delta N$ only through terms involving second and fewer derivatives of Dirac distributions. We also ignore all terms which enter at subleading orders in $\lambda$. Those terms which remain are given in \eqref{vd1} and \eqref{vd2}.

The associated components of $\delta {\tilde T}_\subB^{\mu\nu} (\tilde{\bx},\lambda)$ are found to scale\footnote{This scaling is not pointwise, but is instead associated with the action of $\delta {\tilde T}_\subB^{\mu\nu}$ on $\lambda$-independent test functions; see appendix \ref{Sect:VarDerivs}.} like $\lambda^0$ if $n \neq 4$ and like $\ln \lambda$ if $n=4$. The logarithmic terms which appear at leading order in four spatial dimensions can be written as two total derivatives of a quantity with compact support, and therefore affect the quadrupole renormalization but not, e.g., the mass renormalization. Although logarithmic terms affect the quadrupole renormalization for all even $n$, their effects are subleading when $n>4$. We emphasize, however, that four spatial dimensions are not special except with regards to the quadrupole moment. Logarithmic terms can appear at leading order in the renormalizations of other multipole moments in other numbers of dimensions.

\subsection{Mass renormalization}
\label{Sect:massR}

The effective or renormalized mass $m$ of a body is given by, from (\ref{dS}), (\ref{padef}) and (\ref{mDef}),
\be
m = - \int_{\Sigma_t} H_{a'b}
 \nabla_a X^b
u^a u_{b'}
 {\hat T}_\subB^{a'b'}
\, dV_{\bperp}'.
\ee
This can now be split via $m = m_0 + \delta m$, where $m_0$ denotes the ``bare'' mass computed using $T^{ab}_\subB$ instead of $\hat{T}^{ab}_\subB$. The mass renormalization $\delta m$ then arises only from $\delta \tilde{T}^{ab}_\subB$. Using (\ref{deltaTdef}), \eqref{big}, and (\ref{vd2}) for the renormalized stress-energy tensor, and noting that $\nabla_a X^b = - \delta^b_a + O(\lambda^2)$ and $u^b H^{a'}{}_{b} = u^{a'} + O(\lambda^2)$,
         the mass shift is therefore
\begin{equation}
  \delta m = \frac{\lambda^n}{2 (n-2)} \int d^n \tilde{\bfv} \int d^n \tilde{\bfw} \left(  \frac{ {\tilde \rho}(\tilde{\bfv}) {\tilde \rho}(\tilde{\bfw}) }{ \tilde{r}^{n-2} } \right) ,
  \label{mShift}
\end{equation}
where $\tilde{\bfr} = \tilde{\bfv} - \tilde{\bfw}$, $\tilde{r} = | \mathbf{r} |$, and higher-order terms in $\lambda$ have been omitted.
It follows from \eqref{THatem} and \eqref{EMVD} that the same formula also applies in the electromagnetic case with the replacement ${\tilde \rho} \to {\tilde J}$. The familiar formula\footnote{Our mass shift (\ref{mShift}) has the ``conventional'' self-energy sign in the electrostatic case, and the opposite sign to the conventional one in the scalar case. Our definition for $\delta m$ can in the simplest settings be written as an integral of the stress-energy tensor due to $\phi_\subS$ or $\Phi_\subS$, which cannot be negative and can differ from some notions of self-energy.}
 for the electrostatic self-energy is therefore recovered when $n=3$.

\subsection{Stress-energy quadrupole renormalization}
\label{Sect:Quad}

It follows from \eqref{gExtensions}, \eqref{GenForceMultipole}, and \eqref{Imoments} that the generalized gravitational quadrupole force is in general
\be
        {\cal F}_{\rm quad} = - \frac{1}{6} N \hat{J}^{abcd} {\cal L}_\xi R_{abcd}.
\label{qforce}
\ee
The renormalized stress-energy quadrupole ${\hat J}^{abcd}$ is given by (\ref{Jmoments})
specialized to $p=2$.  To compute this, first note from \eqref{Adef} that
\begin{align}
  \stackrel{ (2) }{ A } \!\! {}_{aba'b'} = H_{a'}^{\ c} H_{b'}^{\ d} \Big[3 u_a u_b u_{c} u_{d} - 4 u_{(a} h_{b)(c} u_{d)}
+ h_{a(c}  h_{d)b} \Big]
  \label{A2s}
\end{align}
to leading order in $\lambda$. In terms of the rescaled coordinates $\tilde{\bx}$, the spatial components of the quadrupole moment are thus
\be
{\hat J}^{iklj}  = -\lambda^{n+2} \int \tilde{x}^{[k} ( {\tilde T}_{\subB}^{i][j} + \delta {\tilde T}_{\subB}^{i][j} ) \tilde{x}^{l]} d^n \tilde{\bx},
\ee
where (\ref{deltaTdef}) has been used and terms which are higher order in $\lambda$ have again been dropped. The portion $\delta J^{iklj}$ of this which is due to $\delta {\tilde T}_{\subB}^{ab}$ now follows from \eqref{big} and \eqref{vd1}, after integrating over $\tilde{\bx}$ and then over $s$.  If $n \neq 4$, we find through leading nontrivial order that
\begin{align}
        \delta J^{iklj} = \frac{\lambda^{n+2}}{4(n-2)}  \int d^n \tilde{\bfv} \int
d^n \tilde{\bfw} \left( \frac{ {\tilde \rho}(\tilde{\bfv}) {\tilde \rho}(\tilde{\bfw}) }{
  \tilde{r}^n} \right)
  \nonumber
  \\
  ~ \times \Big[ 2(n-2) \tilde{v}^{[i} \tilde{w}^{k]} \tilde{v}^{[j} \tilde{w}^{l]} -  \tilde{r}^2 \big( \tilde{v}^{[i} \delta^{k][l} \tilde{w}^{j]}
  \nonumber
  \\
  ~ + \tilde{w}^{[i} \delta^{k][l} \tilde{v}^{j]} \big) -  g_n \tilde{r}^4 \delta^{[i[j} \delta^{k]l]} \Big] ,
\label{qqq2}
\end{align}
where $g_n$ is defined by \eqref{gndef} and $\delta^{[i[j} \delta^{k]l]}$ denotes separate antisymmetrizations on the index pairs $(i,k)$ and $(j,l)$. If $n=4$, we have instead that
\begin{align}
  \delta J^{iklj} = - \frac{ 1 }{8} \lambda^6 \ln \lambda \tilde{q}^2 \delta^{[i[j} \delta^{k]l]} + O(\lambda^6),
  \label{dJsp4}
\end{align}
where
\be
        {\tilde q} = \int {\tilde \rho}(\tilde{\bx}) d^4 \tilde{\bx}
\ee
describes the leading-order scaling of the total charge $q = \lambda^3 \tilde{q} + O(\lambda^4)$; cf. \eqref{qScale}.

A similar calculation shows that the leading-order mixed components of the quadrupole moment are
\begin{align}
        {\hat J}^{tijt}  = - \frac{3}{4} \lambda^{n+2} \int  \tilde{x}^{i} \tilde{x}^{j}
( {\tilde T}_{\subB}^{tt} + \delta {\tilde T}_{\subB}^{tt} ) d^n \tilde{\bfx}.
\end{align}
Combining this with \eqref{big} and \eqref{vd2} gives
\begin{align}
        \delta J^{tijt} = -\frac{3 \lambda^{n+2}}{8(n-2)}  \int d^n \tilde{\bfv} \int
        d^n \tilde{\bfw} \left( \frac{ {\tilde \rho}(\tilde{\bfv}) {\tilde \rho}(\tilde{\bfw}) }{
  \tilde{r}^{n-2}} \right)
  \nonumber
  \\
  ~ \times \big( \tilde{w}^i \tilde{w}^j  - \frac{1}{2} g_n
    \tilde{r}^2 \delta^{ij} \big)
    \label{dJtt}
\end{align}
when $n \neq 4$ and
\begin{align}
  \delta J^{tijt}  = \frac{3}{32} \lambda^6 \ln \lambda \tilde{q}^2 \delta^{ij} + O(\lambda^6)
  \label{dJtt4}
\end{align}
otherwise.

Note that these expressions could not be obtained by naively computing a quadrupole moment associated with the stress-energy tensor of $\phi_\subS$. Such attempts would generically result in divergent integrals, and would also depend on properties of the scalar field (and the geometry) at large distances. None of these undesirable properties are shared by the quadrupole shift $\delta J^{abcd}$ which appears naturally in our formalism.

We also note that our leading-order quadrupole renormalizations enter the laws of motion at the same order as subleading corrections to the mass, which we have not computed explicitly.

\subsubsection{Vacuum regions of spacetime}

If the background spacetime satisfies the vacuum Einstein equation $R_{ab} = 0$, only the ``trace-free components'' of $\hat{J}^{abcd}$ can affect the quadrupole force ${\cal F}_{\rm quad}$. To be more precise, note that the quadrupole moment has the same algebraic symmetries as the Riemann tensor, and can therefore be decomposed into trace parts and trace-free components just as $R_{abcd}$ may be decomposed into its Weyl and Ricci components. The analogous decomposition results in
\begin{align}
        \hat{J}^{abcd} = \hat{J}_\subTF^{abcd} + \frac{2}{n-1} \left( g^{a[c} \hat{J}^{d]b} - g^{b[c} \hat{J}^{d]a} \right) \nonumber \\
        ~ - \frac{2  g^{a[c} g^{d]b} }{n(n-1)} (g^{ef} \hat{J}_{ef}) ,
\end{align}
where $\hat{J}_{ac} \equiv g^{bd} \hat{J}_{abcd}$ and $\hat{J}_\subTF^{abcd}$ is trace-free on all pairs of indices. In vacuum regions, it follows from \eqref{LieXiZ} that if $C_{abcd}$ denotes the spacetime Weyl tensor, the quadrupolar gravitational force (\ref{qforce}) reduces to
\begin{align}
        \mathcal{F}_{\rm quad} =   - \frac{N}{6} (J_\subTF^{abcd} + \delta J_\subTF^{abcd})
        \mathcal{L}_\xi C_{abcd}.
\label{qc11}
\end{align}

The piece of the quadrupole shift which dominates in the Newtonian limit is the piece which couples to the electric component of the Weyl tensor. Explicitly, this component of $\delta J^{abcd}_\subTF$ is
\begin{align}
\delta J_\subTF^{tijt} = \frac{1}{n-1} [ (n-2) \delta J^{tijt} + \delta_{kl} \delta J^{iklj} ]_\subTF ,
\end{align}
where ``TF'' denotes the trace-free component of the quantity in brackets. Using \eqref{qqq2} and \eqref{dJtt}, it explicitly evaluates to
\begin{align}
        \delta J_\subTF^{tijt} = -\frac{\lambda^{n+2}}{8 (n-1)} \int d^n \tilde{\bfv} \int d^n \tilde{\bfw}
                \left( \frac{ {\tilde \rho}( \tilde{\bfv} ) {\tilde \rho}( \tilde{\bfw} ) }{
                \tilde{r}^{n}} \right)
  \nonumber \\
        ~ \times \big( 2 |\tilde{\bfv}|^2 \tilde{r}^{(i} \tilde{w}^{j)} + 3 \tilde{r}^2 \tilde{w}^i \tilde{w}^j  \big)_\subTF
\end{align}
when $n \neq 4$. If $n = 4$, we have instead that $\delta J^{tijt}_\subTF = 0$ through leading $O(\lambda^6 \ln \lambda)$ order. Indeed, all components of $\delta J^{abcd}_\subTF$ vanish at this order in four spatial dimensions.


\subsubsection{Dependence on lengthscale $\ell$ in propagator.}

As we have mentioned, the Hadamard parametrix generically involves the arbitrary lengthscale $\ell$ for all even $n$; see \eqref{GHadamardSC}. Changing $\ell$ implicitly changes the definition of the quadrupole moment, and in particular the quadrupole shift $\delta J^{abcd}$. Although this effect occurs at subleading order, it is easily computed using our existing expressions when $n=4$: These shifts may be obtained merely by replacing the $\ln \lambda$ in the quadrupole shifts \eqref{dJsp4} and \eqref{dJtt4} by $-\ln \ell$. To leading nontrivial order, a change $ \ell \to e^\varpi \ell$ in lengthscale is therefore accompanied by the quadrupole shifts
\begin{subequations}
\begin{gather}
        \delta J^{tijt} \to \delta J^{tijt} - \frac{3}{32} \lambda^6 \delta^{ij} {\tilde q}^2 \varpi, \\
\delta J^{iklj} \to \delta J^{iklj} + \frac{1}{8} \lambda^6 \delta^{[i[j} \delta^{k]l]} {\tilde q}^2 \varpi.
\end{gather}
\end{subequations}
Now the gravitational force associated with the quadrupole moment $\delta J^{abcd}$ is given by \eqref{qforce}. Shifting $\ell$ (in a not-necessarily-vacuum $n=4$ spacetime) therefore shifts the quadrupole force via
\begin{align}
        {\cal F}_{\rm quad} &\rightarrow {\cal F}_{\rm quad}+
  \frac{N}{48} \lambda^6 \varpi \, {\tilde q}^2 ({\cal L}_\xi R - u^a u^b {\cal L}_\xi R_{ab} ) \nonumber \\
&\rightarrow {\cal F}_{\rm quad} + \frac{N}{48} \lambda^6 \varpi \, {\tilde q}^2 {\cal L}_\xi (R_{\bperp} -  3 N^{-1} D^2 N)
\label{elll}
\end{align}
to leading order, where $R$ and $R_{ab}$ are the five-dimensional (spacetime) Ricci scalar and tensor, $R_{\bperp}$ is the four-dimensional spatial Ricci scalar, and we have used (\ref{RicciProjections}) in the second line. This shift vanishes in vacuum but not in general\footnote{A gravitational coupling via the trace of the quadrupole is familiar from Newtonian physics, for example it occurs for a star moving in a cloud of non-interacting dark matter of nonuniform density.}, is canceled by similar $\ell$-dependencies in the scalar self-force and in the subleading mass renormalization.

Superficially similar dependencies of the self-force on a choice of lengthscale in a logarithm were previously encountered in the computations of Beach, Poisson, and Nickel \cite{Poisson5dSF} and of Taylor and Flanagan \cite{TaylorFlanagan} [although those compututations were specialized to vacuum spacetimes for which the force shift (\ref{elll}) vanishes].  The physical relevance of those dependencies is clarified by the results of this paper: No physically measurable quantities depend on the arbitrary choice of lengthscale $\ell$, because of the renormalization of body parameters. However, that renormalization also implies that there is a sense in which the total force acting on the body for $n=4$, at the order at which the self-force first appears, does indeed depend on the internal structure of the body, even for spherically symmetric bodies, as originally suggested by Beach, Poisson, and Nickel \cite{Poisson5dSF}. See section \ref{Sect:ST} below for further discussion of these issues.

\section{Comparison with previous work and applications to specific spacetimes}
\label{Sect:Examples}

This paper provides the first rigorous understanding of the self-force in dimensions $n \neq 3$. We are not the first, however, to comment on this subject; see \cite{TaylorFlanagan, Poisson5dSF, FrolovZelnikovSF, Shuryak, Kosyakov, Galtsov, Galakhov, Kazinski}. Previous work has approached it heuristically, restricting to point particle contexts where the self-force is asserted to be $q \phi_a^\mathrm{reg}$ for some regularization $\nabla_a \phi_\mathrm{self} \mapsto \phi_a^\mathrm{reg}$. We have shown in section \ref{Sect:pp} that appropriate point particle regularizations have the form $\phi_a^\mathrm{reg} = \nabla_a \hat{\phi}_\mathrm{self}$, where $\hat{\phi}_\mathrm{self}$ is given by \eqref{phiSelfpp}. Various other procedures have nevertheless been proposed. Unlike ours, these were not obtained from first principles. The reasoning used in much of the prior literature is stated only in passing (if at all), so it is difficult for us to provide detailed comments on all approaches.

Nevertheless, one persistent theme is the inordinate attention which has been paid to the detailed structure of the point particle self-field. It has been common to argue that certain terms in the gradients of these fields should be discarded or ``smoothed out'' based largely on the way they diverge. Even though the existence of an appropriate regularization is perhaps necessary if a well-behaved point particle limit is expected to exist, removing all singularities is far from sufficient: Given one regularization, it is trivial to build others which ``predict'' any finite answer whatsoever. Although this is widely acknowledged, it is often expected to be irrelevant in practice; appropriate selection principles might be expected to arise from physical reasoning or analogies with other, better-understood systems. What has actually occurred, however, is that different authors have drawn different conclusions from known cases, and thus suggested inequivalent regularizations.

We now provide more detailed comments on prior work by considering in detail two static problems which have been discussed in the literature: point charges in Schwarzschild-Tangherlini \cite{Poisson5dSF, TaylorFlanagan} and Rindler \cite{FrolovZelnikovSF} spacetimes. We apply the formalism developed in this paper to these problems and then contrast with earlier analyses. This also serves to illustrate how our formalism can be applied using concrete examples.

Separately, there has also been prior work on the non-static self-force problem in various numbers of dimensions \cite{Shuryak, Kosyakov, Galtsov, Galakhov, Kazinski}. We do not attempt to discuss this in any detail, although see section \ref{Sect:noStatic} for other comments on the dynamical problem.

\subsection{Rindler spacetime}
\label{Sect:Rindler}

Let $\tau^a$ be a boost-type Killing field in flat spacetime with $n \geq 2$. Introducing Minkowski coordinates $(\textsf{T}, \textsf{X}^1, \cdots, \textsf{X}^n)$ and a constant $a > 0$ with dimensions of inverse length, suppose that $\tau^a$ describes a boost in the $\textsf{X}^n$ direction so
\begin{equation}
  \tau^a = a \left( \textsf{X}^n \frac{\partial}{\partial \textsf{T} } + \textsf{T} \frac{\partial}{\partial \textsf{X}^n } \right).
\end{equation}
Restricting to the wedge $\textsf{X}^n > |\textsf{T}|$ then recovers the $n+1$ dimensional Rindler spacetime together with the everywhere-static Killing field $\tau^a$. We introduce Rindler coordinates $(t,\bx)$ on this wedge such that $\textsf{X}^i = x^i$ for $i \neq n$ and
\begin{align}
        \textsf{T} = y \sinh at, \qquad \textsf{X}^n = y \cosh at,
\end{align}
where $y \equiv x^n > 0$. Then $\tau^a = \partial/\partial t$, so the lapse and spatial metric are explicitly
\begin{align}
        N = a y, \qquad h_{ij} = \delta_{ij}.
\end{align}
It follows from \eqref{accel} that that the acceleration of a (static) worldline at fixed $\bx$ is $y^{-1} \partial/\partial y$.

Now consider the fields which must be imposed so that a charged object does not evolve with $t$. We suppose for simplicity that it is only scalar, and not electric, charge which is involved. The self-field $\phi_\mathrm{self}$ which is described in section \ref{Sect:ExtFields} is then given by \eqref{phiSelfConv} for some Green function $G_\mathrm{self}$ which satisfies $D^a (N D_a G_\mathrm{self}) = - \omega_n \delta_\Sigma$. The particular Green function which is used depends on which boundary conditions are physically appropriate. One possibility has been computed by Frolov and Zelnikov \cite{FrolovZelnikovSF}, and is adopted here:
\begin{align}
        G_\mathrm{self} = \frac{ \sqrt{\pi} \Gamma ( \frac{n-1}{2} ) }{a\, \Gamma( \frac{n}{2} ) } \frac{ P_{ \frac{1}{2} (n-3) } ( \coth\eta) }{ (2 y y' \sinh\eta )^{\frac{1}{2} (n-1)} } .
\label{Gfrolov}
\end{align}
Here $P_{\frac{1}{2} (n-3)}$ is a Legendre function of the first kind and $\eta$ is defined to satisfy
\begin{align}
\label{eq:eta}
        \cosh\eta = 1+\frac{|\bx - \bx'|^2}{2 y y'}.
\end{align}
With this choice and any compact (not necessarily pointlike) charge density $\rho(\bx)$, the self-field goes to zero as $|\bx| \rightarrow \infty$ and is finite on the $y = 0$ boundary of the Rindler wedge.

Computing a self-force now requires an appropriate propagator $G$ with which to compute $\phi_\subS$ and $\hat{\phi}_\mathrm{self} = \phi_\mathrm{self} - \phi_\subS$.  We let
\begin{equation}
  G = G_\mathrm{self},
  \label{GS}
\end{equation}
a choice which satisfies the generalized version of our propagator assumptions discussed immediately after assumption \ref{paramAssume} in section \ref{Sect:ForceScalar}. Those generalized assumptions allow the consideration of a single propagator $G$ in a single geometry---which is the case of interest here---rather than a family of propagators which are specified functionals of $N$ and $h_{ab}$. It is then required only that $G$ be a parametrix, that it be symmetric in its arguments, and that there exist some quasilocal three-point functions for which all Lie derivatives take the form \eqref{LieGen}.
That $G_\mathrm{self}$ is a Green function immediately
implies that it is also a parametrix. That it is also symmetric in its
arguments follows immediately from inspection of \eqref{Gfrolov} and
\eqref{eq:eta}.
Noting that $G_\mathrm{self}$ depends only on $N N' = a^2 y y'$ and the spatial world function $\sigma(\bx,\bx') = \frac{1}{2} |\bx-\bx'|^2$, it follows that $\Lie_\psi G_\mathrm{self}$ has the appropriate form for all spatial vector fields $\psi^a$. The choice \eqref{GS} is therefore justified\footnote{Analogous identifications are not always possible. Depending on boundary conditions and other factors, $G_\mathrm{self}$ might not satisfy all conditions required of $G$.}.

Using it, \eqref{phiSelfConv}, \eqref{phiSDef}, and \eqref{phiHatSelfDef} immediately imply that
\begin{equation}
  \hat{\phi}_\mathrm{self} = 0.
  \label{phiHatRindler}
\end{equation}
\textit{Self-forces and self-torques determined by \eqref{SFscalar} therefore vanish to all orders} and for all $n \geq 2$. This statement is true for any compact and static extended charge in Rindler spacetime; it holds even without imposing a point particle limit.

As we have emphasized in section \ref{Sect:Ginvariance}, this self-force is not particularly interesting on its own. What is much more relevant is the external force which must be imposed in order to hold a body ``fixed''---which in the Rindler context corresponds to a uniform acceleration. Rindler spacetime is flat, so the generalized Killing fields $\xi^a \in K_\subG$ are all genuine Killing fields and $\mathcal{L}_\xi g_{ab} = 0$. Gravitational forces and torques determined by \eqref{GravForce} therefore vanish identically\footnote{Alternatively, the gravitational force and torque depend on the metric extensions $g_{ab,c_1 \cdots c_p}$ for all $p \geq 2$, but these vanish in flat spacetime.}. The only remaining effects which must be considered are those associated with the holding field $\phi_\hold$. It follows from \eqref{Fhold} and \eqref{Fconsistency2} that the force exerted by this field must be
\begin{align}
\label{rindlerR1}
  F_a^\mathrm{hold} &= m a \nabla_a y
  \nonumber
  \\
  &= \sum_{p=0}^\infty \frac{ay}{p!} q^{b_1 \cdots b_p} \nabla_a \nabla_{b_1} \cdots \nabla_{b_p} \phi_\hold.
\end{align}
Similarly, \eqref{Nmultipole} and \eqref{Nconsistency2} show that a body's rotational degrees of freedom are constrained by
\begin{align}
\label{rindlerR2}
        N_{ab}^\mathrm{hold} &= 0
        \nonumber
        \\
        &= \sum_{p=1}^\infty \frac{ 2 ay }{(p-1)!} q^{c_1 \cdots c_{p-1} }{}_{[a} \nabla_{b]} \nabla_{c_1} \cdots \nabla_{c_{p-1}} \phi_\hold.
\end{align}
These relations are exact even without applying a point particle limit.
In the monopole approximation, they yield that the gradient of the holding field is $\nabla_a \phi_\hold = (m/q) \nabla_a \ln y$ if $q \neq 0$.

The simplicity of the results (\ref{rindlerR1}) and (\ref{rindlerR2}) stems largely from our ability to use the propagator freedom outlined in section \ref{Sect:Ginvariance} to impose \eqref{GS}. We have emphasized, however, that other choices for $G$ are nevertheless possible. Although such transformations can result in nontrivial self-forces, these are implicitly compensated by differing values for $m$; the propagator freedom in this case corresponds to a physically-irrelevant degeneracy between what one might label inertial versus self-interaction effects. Appendix \ref{Sect:RindlerAppendix} considers an explicit example of this degeneracy by choosing $G$ to be the Hadamard parametrix described in appendix \ref{Sect:Hadamard} (instead of $G_\mathrm{self}$). The resulting calculation is considerably more complicated in that case, underlining how the flexibility in our choice of propagator may be leveraged to simplify calculations.

We now compare our results in Rindler spacetime with those of Frolov and Zelnikov \cite{FrolovZelnikovSF}, who discussed the self-force acting on static scalar and electric charges in Rindler spacetimes with spatial dimensions ranging from $n = 3$ to $8$. The specific procedure which they advocated was motivated by Lagrangian considerations\footnote{A point particle action was postulated, but as usual for such methods, the corresponding Euler-Lagrange equations have no solutions. Such arguments are therefore formal.} and analogies to quantum field theory. A force was eventually obtained by computing detailed point particle self-fields, dropping some singular terms, absorbing others into the mass, and also introducing an infrared cutoff. Their final result was that the self-force depends on the logarithm of the cutoff parameter. Their suggested explanation for this divergence was that the cutoff might describe the scale over which the eternal acceleration of the Rindler model breaks down.

It is difficult to compare the methods of Frolov and Zelnikov directly to our own. Their calculation nevertheless results in a very different answer; in our approach, one natural definition for the self-force vanishes and no auxiliary parameters appear. Other definitions are consistent, however.

\subsection{Schwarzschild-Tangherlini spacetime}
\label{Sect:ST}

In a recent article, Beach, Poisson, and Nickel \cite{Poisson5dSF} discussed pointlike scalar and electric charges held fixed outside a 5-dimensional Schwarzschild-Tangherlini black hole. They found a logarithmic dependence of the self-force on a cutoff parameter, which they interpreted as a dependence on the charge's internal structure.

The primary assumption underlying their calculation was that the total force can be computed using a two-step regularization: First, focusing on the scalar case for concreteness, the ill-defined gradient of the point particle field is replaced by its surface average $\langle \nabla_\mu \phi \rangle_r$ over a sphere of radius $r$ in Riemann normal coordinates $x^\mu$ centered on the particle. The result is not finite as $r \rightarrow 0$, but some diverging terms\footnote{It does not appear to have been noticed in \cite{Poisson5dSF} that \textit{all} divergent terms in $\langle \nabla_\mu \phi \rangle_r$ were proportional to the acceleration. Indeed, no other direction is possible given the symmetries of the problem. It is likely, however, that some divergent terms would not be proportional to the acceleration in more complicated spacetimes.} were shown to be proportional to the acceleration and were absorbed into the mass. This was said to result in a ``regularized average'' $\langle \nabla_a \phi \rangle^\mathrm{reg}_r$ from which the force was claimed to follow. The result still diverged, however, like $\ln r$ as $r \rightarrow 0$.

The use of an average by Beach, Poisson, and Nickel \cite{Poisson5dSF} was motivated by appealing to the Quinn-Wald axioms \cite{QuinnWald, Quinn}. These axioms provide a somewhat different prescription, however. Although it was not mentioned explicitly in \cite{Poisson5dSF}, the use of an average is sometimes also motivated by the claim that it is (``mostly'') equivalent to computing the force on a small spherical shell \cite{PoissonLR}. We make two main comments: The first is that while surface averaging can be used to compute forces---see the end of section \ref{Sect:limitoutside}---the version described in \cite{Poisson5dSF} has not been justified. In particular, simple averaging of point particle fields does not generically correspond to the force on a shell. Second, it was not realized that the self-force should renormalize not just the mass, but also the stress-energy quadrupole moment for point particles in four spatial dimensions.

Taylor and Flanagan \cite{TaylorFlanagan} again considered the Schwarzschild-Tangherlini spacetime, but using different methods. They did not employ a cutoff, but instead considered a one-parameter family of regularizations. Each of these resulted in a different (but finite) force. The regularizations used were special cases of the ones derived in this paper: Green functions were obtained and used to define $\phi_\subS$ and $\Phi_\subS$, and these were subtracted away from $\phi$ and $\Phi$, respectively, to compute a force.  The Green functions used satisfied the assumptions of this paper, and so the self-forces thus obtained fit within the framework of this paper.

However, the interpretation of the results given by Taylor and Flanagan was incomplete. Different choices for the $G$ and $\mathcal{G}$ which were considered there were noted to result in different self-forces, and the reason for this was not understood\footnote{It was incorrectly assumed that a more detailed analysis would reveal the existence of a preferred, correct choice of propagator.}. The discussion of this paper shows that non-uniqueness of the self-force is related to an incomplete accounting of the forces involved. The self-force is only one component, and the inertial and gravitational quadrupole forces must be taken into account as well. Different choices for $G$ and $\mathcal{G}$ result in different effective quadrupole moments---an example of which is illustrated explicitly in section \ref{Sect:Quad} above---and also different masses.

Finally, we comment on the suggestion of Beach, Poisson, and Nickel that the point particle self-force depends on internal structure when $n=4$ \cite{Poisson5dSF}.  This is not the case for what we are calling the scalar self-force.  However, it is the case for the sum of the scalar self-force and the $S$-field renormalization of the gravitational multipole couplings (which one might call a total self-force). In this sense Beach, Poisson, and Nickel were correct.
On the other hand, an analogous result is true even in flat space and even when $n=3$.  In that context, the self-interaction contribution to the mass depends on the details of a body's internal structure.  This dependence is usually not considered to be physically significant since the final equation of motion depends only on the renormalized mass and the bare mass is typically impractical to measure.  Similarly, when $n=4$, the final equation of motion depends only on the renormalized mass and the renormalized quadrupole, and these are not easily separated from the bare equivalents.

\section{Generalization to dynamical bodies and spacetimes}
\label{Sect:noStatic}

In this section, we discuss how our results on static systems can be generalized to dynamical bodies and dynamical spacetimes. The general strategy used above, where self-forces are obtained from identities analogous to \eqref{GenForceFin}, can also be adopted in the dynamical case; see Harte \cite{HarteReview} for such an analysis when $n=3$. We first note that those $3+1$ dimensional results generalize immediately for all odd $n$, resulting in $S$-fields generated by generalizations of the Detweiler-Whiting Green function. For even $n$, Detweiler-Whiting Green functions do not appear to exist. Modifications are therefore needed, and we conjecture what those might be. Our discussion is restricted for simplicity to scalar self-interaction.

\subsection{Odd number of spatial dimensions: The Detweiler-Whiting
  prescription}

In the usual case with four spacetime dimensions, it is known that the self-force can generically---even in dynamical cases---be found by following what has come to be known as the Detweiler-Whiting prescription. Generalizing early ideas due to Dirac \cite{Dirac}, Detweiler and Whiting proposed \cite{DetweilerWhiting2003} that the physical field $\phi$ around a point particle could be regularized via
\begin{equation}
  \hat{\phi}(x) = \phi(x) - \int G_\subDW (x,x') \rho(x') dV',
  \label{phiHatDyn}
\end{equation}
in which case comparison with previously-obtained expressions \cite{Mino:1997nk,QuinnWald, Quinn, DeWittBrehme} showed that the force on a point charge reduces to $q \nabla_a \hat{\phi}$ (plus perhaps test-body type dipole terms \cite{HarteEM, GrallaHarteWald}). The spacetime bidistribution $G_\subDW$ which appears in this prescription is known as the ($S$-type) Detweiler-Whiting Green function, and is uniquely characterized \cite{PoissonLR} by the three properties
\begin{enumerate}
\item \label{DW1} $\nabla^a \nabla_a G_\subDW (x,x')=-4 \pi \delta(x,x')$,
\item \label{DW2} $G_\subDW (x,x')=G_\subDW (x',x)$,
\item \label{DW3} $G_\subDW (x,x')=0$ if $x,x'$ are timelike-separated.
\end{enumerate}

That self-forces can be computed in this way was later derived directly from first principles, and also generalized to hold non-perturbatively for arbitrary extended bodies \cite{HarteScalar, HarteEM, HarteGrav, HarteReview, HarteTrenorm}. Moreover, it was shown to hold for torques as well as forces, and to remain valid to all multipole orders. The methods used to establish these results are the same as those used in this paper, so it is straightforward to compare results even at the non-perturbative level. Without going into details, it was shown that a fully dynamical extended body with scalar charge in $n=3$ spatial dimensions admits a renormalized momentum $\hat{P}_t$ and a renormalized stress-energy $\hat{T}_{\subB}^{ab}$ such that
\begin{equation}
  \frac{d \hat{P}_t}{dt} = \int_{\Sigma_t} \left( \frac{1}{2} \hat{T}_{\subB}^{ab} \Lie_\xi g_{ab} + \rho \Lie_\xi \hat{\phi} \right) t^a dS_a,
  \label{ForceDyn}
\end{equation}
where $t^a$ is a time evolution vector field for a family $\Sigma_t$ of hypersurfaces which have been chosen to foliate the body's worldtube. The field $\hat{\phi}$ which appears here is given by \eqref{phiHatDyn}.

As briefly hinted at in \cite{HarteReview}, the derivation of the force law \eqref{ForceDyn} generalizes trivially to spacetimes with arbitrary $n$, at least if a $G_\subDW$ satisfying the above axioms is assumed to exist (with the obvious rescaling $4 \pi \rightarrow \omega_n$ on the right-hand side of the field equation in property \ref{DW1}). It follows that the Detweiler-Whiting prescription is valid for all dimensions in which there exists a Detweiler-Whiting Green function.

Such Green functions do indeed exist for all odd $n$, and so \textit{the Detweiler-Whiting prescription remains valid in all such cases}. Explicitly, $G_\subDW$ has the form
\begin{align}
\label{eq:GDW}
        G_\subDW = U \delta^{(\frac{1}{2}(n-3))}(X) + V \Theta(X)
\end{align}
for odd $n \geq 3$, where $U$ and $V$ are well-defined smooth bitensors and $X$ is Synge's world function on spacetime. If $n=1$, it is instead $G_\subDW = U \Theta(X)$.




The renormalized force law \eqref{ForceDyn} is very similar to our static result \eqref{GenForceFin}, the only differences being that $\hat{P}_t$, $\hat{T}_{\subB}^{ab}$, and $\hat{\phi}$ are defined somewhat differently. Comparing the last of these quantities, for example, \eqref{phiSDef}, \eqref{phiHatDef}, and \eqref{phiHatDyn} imply that all differences lie in the underlying propagators as well as a time integral in the dynamical setting. This suggests that in a static spacetime, a time integral of the Detweiler-Whiting Green function should result in a spatial propagator which is of the type considered in section \ref{Sect:mf}. It is shown in appendix \ref{Sect:DW} that this is indeed the case; the time integral of $G_\subDW$ is a symmetric, geometrically-constructed Green function for the static problem. The Detweiler-Whiting construction is therefore consistent with the general framework we have derived to understand the static self-force.

Indeed, the time integral of $G_\subDW$ has more specifically been shown by Casals, Poisson, and Vega \cite{CasalsPoissonVega} to coincide with the Hadamard parametrix $G_\subH$ discussed in appendix \ref{Sect:Hadamard}, at least for ultrastatic $3+1$ dimensional spacetimes. They also give evidence that it holds more generally in this number of dimensions, and we suspect that it is true in general static spacetimes with odd $n$.

Before moving to cases with even $n$, recall that we have emphasized in this paper that different propagators can reasonably be chosen in the static regime. This remains true in the dynamical setting---with a somewhat reduced space of possibilities---so the Detweiler-Whiting prescription is but one of many possibilities when $n$ is odd. It is nevertheless useful.



\subsection{Even number of spatial dimensions}

Detweiler-Whiting Green functions do not appear to exist when $n$ is even. Progress may nevertheless be made by noting that the renormalized force law \eqref{ForceDyn} remains valid for a wide variety of propagators other than $G_\subDW$. Excluding integral convergence issues which can sometimes arise, it holds for any bidistribution which is symmetric in its arguments and is quasilocally constructed only from $g_{ab}$.

The simplest such example is the sum of the retarded and advanced Green functions: Considering scalar charges in Minkowski spacetime for simplicity,
\begin{equation}
        \frac{1}{2} ( G_\mathrm{ret} + G_\mathrm{adv} ) \propto \frac{ \Theta(-X) }{ (-X)^{ \frac{1}{2} (n-1) } }.
        \label{Gadvret}
\end{equation}
This is indeed symmetric and geometrically-constructed. It is also a Green function. It does not, however, vanish when its arguments are timelike-separated. Applying the renormalized force law with this propagator would result in ``effective momenta'' which depend on an object's entire past and future---a clearly unphysical situation\footnote{This dependence is not unphysical in static systems, which explains in part why we have had no difficulty finding useful static propagators for even $n$.}.

One might initially suspect that the problem could be resolved by substituting $X \rightarrow -X$, thus producing the symmetric propagator $\Theta(X)/X^{ \frac{1}{2} (n-1) }$
which does vanish when its arguments are timelike-separated. Unfortunately, the result is no longer a Green function; worse, it is homogeneous, $\nabla^a \nabla_a [ \Theta(X)/X^{ \frac{1}{2} (n-1) }] = 0$ \cite{GelfandShilov}. Although \eqref{ForceDyn} is again valid in this case, it is again unhelpful; computing an associated $S$-field results in $\phi_\subS = (\mbox{constant})$ for static point charges, so $\hat{\phi} = \phi - \phi_\subS$ fails to admit a regular point particle limit.

Having rejected these two possibilities, we demand that an acceptable propagator be symmetric and geometrically-constructed, that it vanish when its arguments are timelike-separated,  and also that the associated $\hat{\phi}$ remain smooth even when $\rho$ is not. It is only the last of these constraints which is nontrivial to verify, and we conjecture that
\begin{equation}
        G_\mathrm{dyn} \equiv d_n \left[ \frac{ \ln (X/\ell^2) }{ X^{ \frac{1}{2} (n-1) } } \right]\Theta( X )
        \label{Gdyn}
\end{equation}
is an appropriate choice in flat spacetime, where $d_n$ is a normalization constant and $\ell$ is arbitrary. This is not a Green function for $\nabla^a \nabla_a$, nor even a parametrix. Nevertheless, it is compatible with the renormalized force law and vanishes when its arguments are timelike-separated. Integrating $G_\mathrm{dyn}$ against a Minkowski time coordinate may be shown (for appropriate $d_n$) to recover ordinary static Green functions. Indeed, this statement generalizes also to the Rindler context. Well-behaved point particle limits therefore result at least for uniformly-accelerated charges in flat spacetime.

It is unclear whether or not point particle limits associated with $G_\mathrm{dyn}$ remain regular more generally. While this question could be decided by directly computing the relevant point particle fields, it would be far less tedious to instead find a general principle which directly guaranteed the desired result. Recall that the appropriate principle in the static regime was elliptic regularity. This immediately implied that for propagators which were parametrices, the relevant effective fields must be well-behaved for all $\rho$. For odd $n$ where $G_\subDW$ exists, dynamical effective fields instead satisfy the hyperbolic equation $\nabla^a \nabla_a \hat{\phi} = 0$. Elliptic regularity does not apply in this case, and indeed, singular solutions do exist---impulsive waves, for example. General theorems on the propagation of singularities \cite{Hormander} may nevertheless be used to show that any singularities which might be present can propagate only in null directions. They may therefore be viewed as ignorable peculiarities which quickly pass through a body's timelike worldtube. For even $n$ where dynamical effective fields are generated by $G_\mathrm{dyn}$, we do not know of an analogous statement. Finding one would likely be critical to constructing a curved spacetime generalization, and we leave both of these issues for future work.

Incidentally, $G_\mathrm{dyn}$ can be generated by performing the $X \rightarrow -X$ substitution in $G_\mathrm{ret} + G_\mathrm{adv}$ and then adding a multiple of this to its derivative with respect to $n$. The variation with respect to dimension suggests that using $G_\mathrm{dyn}$ in a point particle limit might be equivalent to a kind of dimensional regularization. We do not attempt, however, to make this precise.

\section{Discussion}
\label{Sect:Conclusions}

An overview of some results of this paper is given in table \ref{tab:propagators}, which summarizes some of the propagators considered in this paper and elsewhere, their properties and interrelationships, and how they are used here.

\renewcommand{\arraystretch}{1.5}

\begin{table*}[t]
\centering
\footnotesize
\begin{tabular}{|C{0.8cm}||C{1cm}|C{1cm}|C{5.2cm}||C{1cm}|C{1cm}|C{5.2cm}|}
\hline
&\multicolumn{3}{c||}{\textbf{Odd number of spatial dimensions}}  & \multicolumn{3}{c|}{\textbf{Even number of spatial dimensions}} \\
\hline
&Symbol& Section & Description &Symbol& Section & Description \\
\hline
  \raisebox{0pt}{\parbox[t]{2mm}{\multirow{25}{*}{\rotatebox[origin=c]{90}{\textbf{Static}}}} }& $G_\mathrm{self}$ &
\ref{Sect:ExtFields}
 & Green function used to compute a body's self-field. Can be, e.g., the time integral of the retarded Green function.& $G_\mathrm{self}$ &
\ref{Sect:ExtFields}
 & Same as for odd $n$.
\\
\cline{2-7}
& $G$ &
\ref{Sect:ForceScalar}
 & Generic propagator used to compute scalar $S$-fields in static spacetimes. Affects the self-force, effective momenta, and effective stress-energy moments. Appropriate choices must satisfy the five assumptions listed at the end of section \ref{Sect:ForceScalar}: They are geometrically and quasilocally constructed, symmetric, and parametrices.
& $G$ &
\ref{Sect:ForceScalar}
 & Same as for odd $n$.
\\
\cline{2-7}
& ${\cal G}$ &
\ref{Sect:ForceEM}
 & Same as $G$ but for electrostatic fields
& ${\cal G}$ &
\ref{Sect:ForceEM}
 & Same as for odd $n$.
\\
\cline{2-7}
 & $G_\subH$ &
\ref{Sect:bodyR} App.\ \ref{Sect:Hadamard} App.\ \ref{Sect:VarDerivs}
 & Scalar Hadamard parametrix. A specific bidistribution obtained
from Hadamard's procedure which satisfies all properties required of $G$. Detailed form differs for odd and even $n$.

~

$\displaystyle{G_\subH = c_n \sqrt{\frac{\Delta}{N N'}} \frac{\mathsf{U}_{\textrm{sc}}}{\sigma^{\frac{n}{2}-1}}}$

& $G_\subH$ &
\ref{Sect:bodyR} App.\ \ref{Sect:Hadamard} App.\ \ref{Sect:VarDerivs}
 & Scalar Hadamard parametrix.

~

$\displaystyle{G_\subH = c_n \sqrt{\frac{\Delta}{N N'}} \left[\frac{\mathsf{U}_{\mathrm{sc}}}{\sigma^{\frac{n}{2}-1}} + \mathsf{V}_{\mathrm{sc}} \ln \left(\frac{\sigma}{\ell^2}\right) \right]}$

\\
\cline{2-7}
& ${\cal G}_\subH$ &
App.\ \ref{Sect:Hadamard}
 & Same as $G_\subH$ but for electrostatic fields
& ${\cal G}_\subH$ &
App.\ \ref{Sect:Hadamard}
 & Same as $G_\subH$ but for electrostatic fields
\\
\cline{2-7}
& $\displaystyle{\int G_{\subDW}}$ &
\ref{Sect:noStatic}, App.\ \ref{Sect:DW}
 & Time integral of Detweiler-Whiting Green function (see below).  Satisfies our five assumptions for $G$, coincides with $G_\subH$ at least for ultrastatic spacetimes when $n=3$ \cite{CasalsPoissonVega}, and conjectured to coincide with $G_\subH$ for general static spacetimes with odd $n$.
&  $\displaystyle{\int G_{\mathrm{dyn}}}$ & \ref{Sect:noStatic}
 & Time integral of our conjectured spacetime propagator $G_{\mathrm{dyn}}$ in flat spacetime (see below). Coincides with $G_{\subH}$ at least for static charges in Minkowski and Rindler spacetimes.
\\
\cline{2-7}
\hline
\hline
 \raisebox{30pt}{\parbox[t]{2mm}{\multirow{2}{*}{\rotatebox[origin=c]{90}
 {\textbf{Dynamical}}}} }
  &$G_\subDW$ &
\ref{Sect:noStatic}
&
Detweiler-Whiting Green function. For all odd $n \geq 3$,

~

$\displaystyle{G_\subDW = U \delta^{( \frac{1}{2} (n-3) )}(X) + V \Theta(X)}$
 &
 $G_\mathrm{dyn}$ &
\ref{Sect:noStatic}
& Conjectured replacement for $G_\subDW$ when $n$ is even. Useful at least for uniform acceleration in flat spacetime; the general case requires further analysis.

~

$\displaystyle{G_{ \mathrm{dyn} } = d_n \left[  \frac{  \ln(X/\ell^2) }{ X^{ \frac{1}{2} (n-1) } } \right] \Theta(X) }$

\\
\hline\hline
\end{tabular}

\caption{
A key role in our analysis is played by {\it propagators}, by which we
mean bidistributions on spacetime or on spatial slices that can be integrated
against charge densities to produce
various kinds of self-fields.
This table lists some of the propagators we discuss, where in the paper they are located, their properties, and interrelationships. There is some dependence on the parity of the number $n$ of spatial dimensions.}
%
%
\label{tab:propagators}
\end{table*}

\normalsize

Our main objective has been to understand static extended charges in static spacetimes. While forces and torques can be directly computed using the spacetime metric $g_{ab}$, the scalar and electromagnetic potentials $\phi$ and $\Phi$, a body's charge densities $\rho$ and $J^a$, and its stress-energy tensor $T^{ab}_\subB$, complete knowledge of these quantities is often unavailable. One might instead have access only to a body's mass, net charge, and perhaps a handful of additional multipole moments. It is well-known that these parameters can accurately describe the forces and torques which act on sufficiently small test bodies, and a similar result might be expected to hold more generally. We show that this is indeed the case: Multipole expansions for the force and torque are derived for strongly self-interacting charges, and these are shown to be formally identical to expansions which had previously been known for extended test bodies. Our expressions differ, however, in that the definitions for the various multipole moments and fields are renormalized with respect to their test body counterparts.

These results follow from the identities \eqref{GenForceFin} and \eqref{GenForceFinEM}, which show that generalized forces can be computed not only from the physical fields $\phi$ and $\Phi$, but also from appropriately-defined ``effective fields'' $\hat{\phi}$ and $\hat{\Phi}$. In many cases of physical interest---point particle limits, for example---the effective fields are simpler; forces due to them can admit simple multipole expansions even when those involving the physical fields do not. We use this to show that the force and torque necessary to hold a body fixed must satisfy \eqref{FNgrav}, \eqref{FNhold}, \eqref{Fconsistency2}, \eqref{Nconsistency2}, and \eqref{FNconsistency}, expressions which are valid through all multipole orders and in all dimensions.

More precisely, our expressions involve certain two-point propagators which determine, via \eqref{phiSDef} and \eqref{phiSDefEM}, the differences between the physical and effective fields. Although the forces due to these fields are not necessarily identical, all disagreements are of a special form which can be absorbed into an effective shift $T^{ab}_\subB \rightarrow \hat{T}^{ab}_\subB$ in a body's stress-energy tensor---shown explicitly by \eqref{THat} and \eqref{THatem} to depend on functional derivatives of the propagators with respect to the lapse $N$ and the spatial metric $h_{ab}$. A body's linear and angular momenta are thus renormalized, as well as its quadrupole and higher couplings to the spacetime curvature. This mixes effects which might be  labeled as ``gravitational,'' ``inertial,'' ``scalar,'' or  ``electromagnetic.''

Such mixings are particularly relevant in light of our result that the propagators are not unique. Our formalism applies for all $G$ and $\mathcal{G}$ which satisfy the five properties summarized at the end of section \ref{Sect:ForceScalar} (or their electromagnetic analogs), and we have emphasized that many possibilities exist. Different choices generically results in different effective fields, different self-forces, different gravitational forces, and so on. While these ambiguities could be ``removed'' by convention, perhaps by restricting only to Hadamard parametrices---table \ref{tab:propagators} summarizes these and other propagators---we stress the importance of observables which remain invariant under all allowable transformations. In the static systems considered here, the natural observables are the forces and torques which must be supplied to maintain staticity.

Specializing our results to ``point particles'' corresponds to considering appropriate one-parameter families of extended charges. The properties of such a family depend on the number of spatial dimensions $n$, and on a scaling parameter $\lambda$. The point particle limit then corresponds to taking this  parameter to zero, in which case sizes scale like $\lambda$, charges like $\lambda^{n-1}$, and masses like $\lambda^n$. Expanding our expressions in powers of $\lambda$ shows that the leading-order self-force scales like $\lambda^{2(n-1)}$, which is comparable for all $n \geq 2$ to test body forces which involve a body's $2^{(n-2)}$-pole moments. In this sense, the relative magnitude of the self-interaction becomes progressively smaller as the number of dimensions increases; if one is interested in self-force effects in higher dimensions, extended body effects must also be taken into account. In lower dimensions, the self-force can instead be comparable to the  monopole test body interaction, and it would be interesting to understand if this type of enhancement could have experimental consequences in lower-dimensional condensed matter or fluid systems.

Another possible direction for future work is to move beyond the static regime, considering dynamical self-force problems in different numbers of dimensions. We remark in section \ref{Sect:noStatic} that prior results in the literature immediately generalize to describe fully-generic self-interaction effects in all even-dimensional spacetimes. In particular, the Detweiler-Whiting scheme remains valid in those cases. The situation is less clear in odd-dimensional spacetimes, in which case there does not appear to exist any ``Detweiler-Whiting'' Green function (or parametrix) which is symmetric, geometrically-constructed, and quasi-local. Progress may nevertheless be made by noting that our results allowing forces to be computed using effective instead of physical fields apply even if some of these constraints are weakened. The parametrix constraint in particular may be dropped, and we conjecture that the propagator \eqref{Gdyn} provides a useful self-force prescription for dynamical bodies in all flat, odd-dimensional spacetimes. We note that the force law \eqref{ForceDyn} remains valid in this context, and also that the expected results---including well-behaved point particle limits---are recovered in the static limit. What remains to be shown is whether this choice guarantees well-behaved point particle limits for arbitrarily-accelerated bodies, and also how it can be generalized to curved spacetimes. We hope to address these issues in a future article.

\acknowledgments
We wish to thank Jordan Moxon for his suggestions on the point particle limits used in this paper. PT is supported by the Irish Research Council under the ELEVATE scheme which is co-funded by the European Commission under the Marie Curie Actions program.  EF acknowledges support from NSF grant PHY-1404105.

\appendix

\section{NOTATION AND CONVENTIONS}
\label{app:notation}

Throughout this paper, units are chosen in which $G=c=1$, the metric signature is positive, abstract indices are denoted by $a,b,\ldots$, spacetime coordinate indices by $\mu, \nu, \ldots$, and spatial coordinate indices by $i,j,\ldots$. Covariant derivatives on the spacetime $(\Sigma \times I, g_{ab})$ are denoted by $\nabla_a$, and on its spatial sections $(\Sigma, h_{ab})$ by $D_a$. Riemann tensors on spacetime and on space are $R_{abc}{}^{d}$ and $R^{\bperp}_{abc}{}^{d}$ respectively, with similar conventions for Ricci tensors and Ricci scalars. Signs are such that $R_{ab} = R_{acb}{}^{c}$, and $R_{abc}{}^{d} \omega_d = 2 \nabla_{[a} \nabla_{b]} \omega_c$ for arbitrary 1-forms $\omega_a$.

We assume wherever necessary that for every pair of points in appropriate regions, there exists exactly one geodesic which passes through that pair. Although this is generically false on large scales, we require it only in finite regions, typically the interior of the body of interest. Throughout, hatted symbols denote renormalized versions of unhatted quantities (e.g., $\hat{\phi}$ is the renormalized scalar field). Certain renormalized quantities such as the mass are nevertheless written without hats for brevity.

Various propagators are used in this paper and summarized in table \ref{tab:propagators}. For any propagator, a tilde above the symbol, as in ${\tilde G}$, indicates a version obtained by multiplying by powers of the lapse function, as in \eqref{GtildeScalar} and \eqref{GtildeEM}. To aid the reader, we list other symbols that occur throughout the paper in table \ref{tab:symbols2}.

\renewcommand{\arraystretch}{1.3}


\begin{table*}[h!]
\centering
\footnotesize
\begin{tabular}{| L{3.6cm} || L{9cm} | L{3cm} | }
\hline\hline
\textbf{Symbol}\qquad & \textbf{Meaning} & \textbf{Relevant equations} \\
\hline\hline
$N(x),\,h_{ab}(x)$  & The lapse function and the spatial metric.    &       \eqref{NDef}, \eqref{hDef}      \\ \hline
$\displaystyle{dV, \, dV_{\bperp}}$    & Volume elements with respect to the spacetime and spatial metrics.        &       \eqref{dS}      \\ \hline
$\displaystyle{T^{ab}_{\subB}(x),\,\hat{T}_{\subB}^{ab}(x)}$    & A body's bare and renormalized stress-energy tensors.   &       \eqref{THat}, \eqref{THatem}
\\ \hline
$\displaystyle{\phi(x),\,\phi_\mathrm{self}(x), \, \phi_{\textrm{hold}}(x) }$      & Total (physical) scalar field, scalar self-field, and the scalar holding field required to maintain staticity; related by $\phi = \phi_\mathrm{self} + \phi_\mathrm{hold}$. Electrostatic equivalents are denoted by $\Phi_{\cdots}$.   &       \eqref{phiEqn}, \eqref{phiSelf}, \eqref{phiExtDef}
\\ \hline
$\displaystyle{ \phi_{\subS} (x),\, \hat{\phi}(x), \, \hat{\phi}_\textrm{self} (x) }$      & The scalar $S$-field and the effective (or renormalized) physical and self-fields, related by $\hat{\phi}= \phi - \phi_\subS$ and $\hat{\phi}_{\textrm{self}}=\phi_{\textrm{self}}-\phi_{\subS}$. Electrostatic equivalents are again denoted by $\Phi_{\cdots}$.  & \eqref{phiSDef}, \eqref{phiHatDef}, \eqref{phiHatSelfDef},
\\  \hline
$\displaystyle{\rho(x),\,J(x)}$         & Scalar and electrostatic charge densities, respectively.  &       \eqref{phiEqn}, \eqref{EMsimps} \\  \hline
$\displaystyle{\xi^{a}(x),\,\psi^{a}(x)}$       & A generalized Killing vector field on spacetime and an arbitrary spatial vector field, respectively.   (Sometimes $\psi^{a}=h^{a}{}_{b}\xi^{b}$ is a projection of a generalized Killing field).        & \eqref{LieXiZ}, \eqref{KillingTransport}, \eqref{LieX}        \\  \hline
$\displaystyle{\mathcal{Z}, \, z_t}$         & Timelike worldline used to construct the generalized Killing fields and a point on that worldline at time $t$. Sometimes chosen to be a body's center of mass.  &     \eqref{LieXiZ},  \eqref{Centroid} \\  \hline
$\displaystyle{P_{t}(\xi),\,\hat{P}_{t}(\xi)}$  & Bare and renormalized generalized momenta. &       \eqref{PDef}, \eqref{PHat}    \\  \hline
$\displaystyle{p^{a}(t),\,S^{ab}(t)}$   & Linear and angular momenta, defined to be components of the renormalized generalized momentum.   &       \eqref{PtopS}   \\  \hline
$\displaystyle{F^{a}(t),\,N^{ab}(t)}$   & Net force and torque, defined to be components of the generalized force.     &       \eqref{PDotToFN}        \\  \hline
$\displaystyle{X(x,x'),\,\sigma(x,x')}$    & Synge's world function with respect to $g_{ab}$ and $h_{ab}$, respectively. &      \\  \hline
$\displaystyle{X^{a}(x,x')}$    & Spacetime separation vector defined via the exponential map. Also related to $X$ via $X^a = -\nabla^{a}X$ (note the unconventional minus sign).
&       \eqref{XDef}    \\  \hline
$\displaystyle{g_{ab,c_{1}\cdot\cdot\cdot c_{p}}(x), \, \hat{\phi}_{,a_{1}\cdot\cdot\cdot a_{p}}(x)}$         & Tensor extensions for the metric and the effective scalar field (usually evaluated on the central worldline $\mathcal{Z}$).       &       \eqref{phiExtendDef}, \eqref{phiExtensions}, \eqref{phiSyms}, \eqref{MetricExtensions}, \eqref{gSyms}, \eqref{gExtensions}    \\  \hline
$\displaystyle{\hat{J}^{a_{1}\cdot\cdot\cdot a_{p}bc}(z_{t}),\,\hat{I}^{a_{1}\cdot\cdot\cdot a_{p}bc}(z_{t})}$ & Multipole moments of a body's effective stress-energy tensor. These hold equivalent information but have different index symmetries.   &       \eqref{Jmoments}, \eqref{Imoments}      \\  \hline
$\displaystyle{Q^{a_{1}\cdot\cdot\cdot a_{p}}(z_{t}),\,q^{a_{1}\cdot\cdot\cdot a_{p}}(z_{t})}$  & Electrostatic and scalar multipole moments, respectively. &       \eqref{chargeMoments}   \\  \hline
$\displaystyle{\lambda}$        & Scaling parameter for one-parameter families used to define point particle limits.    &       \eqref{Scalings}        \\  \hline
$\displaystyle{\beta,\,\gamma}$ & Scaling exponents for the one-parameter families of source densities and stress-energy tensors used to define point particle limits.   &       \eqref{Scalings}, \eqref{scaleExps}     \\  \hline
$\displaystyle{\mathsf{L},\,L_{\textrm{sc}},\,L_{\textrm{em}}}$  & Differential operator for the static field equations corresponding to, respectively, an arbitrary potential, the potential for the rescaled static scalar field, and the potential for the rescaled electrostatic field. &       \eqref{GtildeFieldEqn}, \eqref{LDef}, \eqref{LDefEM}    \\  \hline
$\displaystyle{\mathsf{U}(x,x'),\,\mathsf{V}(x,x'), \,\mathsf{W}(x,x')}$   & Spatial biscalars appearing in the Hadamard Green function associated with $\mathsf{L}-\Lambda$. When the potential $\Lambda$ is the scalar or electrostatic one, we use $U_{\textrm{sc}}$, $U_{\textrm{em}}$, etc.      &       \eqref{Useries}, \eqref{UWints}, \eqref{V0int}, \eqref{xPortV}, \eqref{xPortWeven}    \\  \hline
$\Delta(x,x')$  & The spatial van Vleck-Morette determinant.   &       \eqref{vanVleckDef}     \\  \hline
\hline
\end{tabular}
\caption{In this table, for the aid of the reader, we list some commonly occurring symbols that appear in the paper.
We do not list symbols whose meaning is very conventional, or symbols
which are used only in the immediate vicinity of where they are
introduced.  For each item listed, we give a brief description, and
also a reference to the equation in the text where the symbol is defined, or after which the symbol is first introduced.
We do not include propagators (which are listed separately in table \ref{tab:propagators}).
}
\label{tab:symbols2}
\end{table*}

\section{EXISTENCE OF AN APPROPRIATE PROPAGATOR}
\label{Sect:Hadamard}

It is convenient to have a two-point distribution $G[N,h_{ab}](x,x')$ on $\Sigma \times \Sigma$ which satisfies properties \ref{functionalAssume}-\ref{paramAssume} summarized at the end of section \ref{Sect:ForceScalar}. This appendix shows that one possibility is to use the Hadamard parametrix \cite{Hadamard, FriedlanderWave}. We first review what this is, and then show that it possesses all required properties. Constructing Hadamard parametrices is algorithmic, and therefore conceptually (but not necessarily calculationally) straightforward.

As a matter of notation, all quantities in this appendix are purely spatial. Events $x, x', \ldots$ are to be interpreted as elements of $\Sigma$, and all indices may be viewed as referring to $n$-dimensional tangent or cotangent spaces in this manifold. For example, much of this appendix uses a spatial version of Synge's world function $\sigma(x,x') = \sigma(x',x)$ \cite{Synge, FriedlanderWave, PoissonLR}, which returns one half of the squared geodesic distance between its arguments as computed in $(\Sigma,h_{ab})$.

\subsection{The Hadamard construction in general}
\label{Sect:Hadamardc}

Before describing the Hadamard parametrix, it is convenient to first apply a transformation which places the scalar and electromagnetic problems on the same footing. This is accomplished by defining the rescaled propagator
\begin{align}
        \tilde{G} (x,x') \equiv [N(x) \, N(x')]^{1/2} G(x,x')
    \label{GtildeScalar}
\end{align}
in the scalar case, and
\begin{align}
        \tilde{ \mathcal{G} } (x,x') \equiv [N(x)\,N(x')]^{-1/2} \mathcal{G}(x,x')
    \label{GtildeEM}
\end{align}
in the electromagnetic case. Substituting the first of these definitions into \eqref{GDefparam} shows that if $G$ is a parametrix, $\tilde{G}$ must satisfy
\begin{align}
        \label{GtildeFieldEqn}
        L_\mathrm{sc} \tilde{G} (x,x') = -\omega_{n}\,\delta_\Sigma(x,x') + \mathcal{S}_\mathrm{sc}(x,x'),
\end{align}
where
\begin{equation}
  L_\mathrm{sc} \equiv D^2-\Lambda_\mathrm{sc} , \qquad \Lambda_\mathrm{sc} \equiv N^{-1/2} D^2 N^{1/2}
  \label{LDef}
\end{equation}
and $\mathcal{S}_\mathrm{sc} \equiv (N'/N)^{1/2} \mathcal{S}$. The rescaled electromagnetic propagator $\tilde{\mathcal{G}}$ satisfies analogous equations with $L_\mathrm{em} = D^2 - \Lambda_\mathrm{em}$, although the potential is then
\begin{align}
        \Lambda_\mathrm{em} \equiv N^{1/2} D^2 N^{-1/2}.
        \label{LDefEM}
\end{align}
For the purposes of this section, parametrices with the form \eqref{GtildeFieldEqn} are considered with general potentials, so there is no need to distinguish between the scalar and electromagnetic cases. We use sans serif font for the general case, so, e.g., $\tilde{\textsf{G}}(x,x')$ is a parametrix of $\textsf{L} \equiv D^{2}- \mathsf{ \Lambda} $.

The first step in building a Hadamard parametrix is to isolate the most singular components of $\tilde{\textsf{G}}(x,x')$ in terms of the distance between its arguments, as represented by the world function $\sigma(x,x')$. Introducing convenient constants $c_n$ and $\ell$ together with certain non-singular biscalars\footnote{We denote the scalar and electromagnetic versions of these scalars by $U_{\rm sc}$, $V_{\rm sc}$, $W_{\rm sc}$ and $U_{\rm em}$, $V_{\rm em}$ and $W_{\rm em}$.} $\Delta(x,x')$, $\mathsf{U} (x,x')$, $\mathsf{V} (x,x')$, and $\mathsf{W} (x,x')$, it may be shown that\footnote{The inverse powers of $\sigma$ appearing here are not necessarily classically integrable, and must therefore be defined properly as distributions. The prescription we adopt may be described by starting with the (unique) distribution which corresponds to an integrable power of $\sigma$, and then reducing this power by iteratively applying the coordinate Laplacian $\delta^{ij} \partial_i \partial_j$ \cite{GelfandShilov} as a distributional operator in Riemann normal coordinates.}
\begin{align}
  \tilde{\textsf{G}}(x,x') = c_n \Delta^{1/2} \bigg[ \frac{ \mathsf{U} }{ \sigma^{\frac{1}{2} n-1} }
  + \mathsf{V} \ln \left( \sigma / \ell^2  \right)  + \mathsf{W} \bigg].
  \label{GtildeSing}
\end{align}
The first of the biscalars appearing here is known as the van Vleck-Morette determinant, and is defined by
\begin{equation}
  \Delta(x,x') \equiv \det [ - h^{a'}{}_{a}(x,x') D^b D_{a'} \sigma(x,x') ] ,
  \label{vanVleckDef}
\end{equation}
where $h^{a'}{}_{a}$ denotes the parallel propagator on $\Sigma$. It is included here to simplify the remaining terms, and has a simple geometric interpretation in terms of the focusing of geodesic congruences \cite{PoissonLR}. This determinant can be shown to be symmetric in its arguments: $\Delta(x,x') = \Delta(x',x)$.

The remaining biscalars which appear in \eqref{GtildeSing} cannot typically be written in terms of any simple, closed-form expressions involving $\sigma$ and $h^{a'}{}_{a}$. The Hadamard procedure instead supposes that
\begin{align}
        \mathsf{U} (x,x')=\sum_{p=0}^{\infty} \mathsf{U}_{p} (x, x') \sigma^{p}(x, x'),
\label{Useries}
\end{align}
along with similar ans\"{a}tze for $\mathsf{V}$ and $\mathsf{W}$. Note that the ``Hadamard coefficients'' $\mathsf{U}_p$ appearing here are not constants, but can themselves be nontrivial biscalars. Hadamard's construction demands that they be determined by substituting \eqref{GtildeSing} into \eqref{GtildeFieldEqn}, equating explicit powers of $\sigma$, and  demanding regularity.

The result of this procedure is that each Hadamard coefficient must satisfy an ordinary differential equation (or  ``transport equation'') between its arguments. These equations have the general form
\begin{equation}
  (\sigma_a D^a  + \kappa) f = F,
  \label{genXport}
\end{equation}
where $f(x,x')$ denotes some Hadamard coefficient, $F(x,x')$ is a regular biscalar, $\kappa \geq 1/2$ is a constant, and $\sigma_a \equiv D_a \sigma(x,x')$. That this is a transport equation may be seen by considering the affinely-parameterized (spatial) geodesic $\gamma(s)$ with endpoints $x' = \gamma(0)$ and $x = \gamma(1)$. In terms of this, the differential operator appearing in \eqref{genXport} reduces to
\begin{equation}
        \sigma_a ( \gamma(s), x' ) D^a f( \gamma(s), x') = s \frac{d}{ds} f ( \gamma(s), x').
        \label{xPortGen}
\end{equation}
It follows that the only solution to \eqref{genXport} which is well-behaved as $x \rightarrow x'$ is
\begin{equation}
  f(x,x') =  \int_0^1 s^{\kappa-1} F( \gamma(s), x') ds,
  \label{xPortGenSoln}
\end{equation}
showing explicitly that $f(x,x')$ depends on $F(x,x')$ only along the geodesic connecting $x$ to $x'$.

\subsubsection{Odd spatial dimensions}

If a Green function is desired [so $\mathcal{S} = 0$ in \eqref{GtildeFieldEqn}] and $n \geq 1$ is odd, the right-hand side of \eqref{GtildeSing} is determined as follows:
\begin{align}
        c_n = \frac{1}{ 2^{\frac{1}{2} n-1} (n-2) },
        \label{c}
\end{align}
all $\mathsf{V}_p$ vanish, $\mathsf{U}_0 = 1$, and
\begin{subequations}
\label{xPortOdd}
        \begin{align}
                \big( \sigma_a D^a + p+1 \big) \mathsf{U}_{p+1}  = - \frac{ \textsf{L} (\Delta^{1/2} \mathsf{U}_{p}) }{ (2p+4-n) \Delta^{1/2} }
                \label{xPortUodd}
        \\
        \big( \sigma_a D^a + p + \frac{1}{2}n \big) \mathsf{W}_{p+1}  = - \frac{ \textsf{L} (\Delta^{1/2} \mathsf{W}_p) }{ 2 (p+1) \Delta^{1/2} }
        \label{xPortWodd}
        \end{align}
\end{subequations}
for all $p \geq 0$. These are transport equations with the form \eqref{xPortGen}. Applying \eqref{xPortGenSoln}, the appropriate solutions are explicitly
\begin{subequations}
\label{UWints}
        \begin{align}
                \mathsf{U}_{p+1} = - \int_0^1 \frac{ s^p }{ 2p+4-n } \left[ \frac{ \textsf{L} (\Delta^{1/2} \mathsf{U}_{p}) }{ \Delta^{1/2} }  \right] ds ,
                \label{UintOdd}
                \\
                \mathsf{W}_{p+1} = - \int_0^1 \frac{ s^{\frac{1}{2} n + p -1 } }{ 2 (p+1) }\left[ \frac{ \textsf{L} (\Delta^{1/2} \mathsf{W}_p) }{ \Delta^{1/2} } \right] ds .
                \label{WintOdd}
        \end{align}
\end{subequations}
$\mathsf{U}_0$ is given, so \eqref{UintOdd} can be iterated order by order to obtain all $\mathsf{U}_p$. The same cannot be said for the $\mathsf{W}_p$. These coefficients depend on $\mathsf{W}_0$, which is not constrained by Hadamard's procedure. If a choice is made, however, all higher $\mathsf{W}_p$ can be computed by iteratively applying \eqref{WintOdd}. The freedom to choose $\mathsf{W}_0$ corresponds to the many distinct solutions which exist to $\textsf{L} \tilde{\textsf{G}} = - \omega_n \delta_\Sigma$ (in the absence of any boundary conditions or other constraints).

It is a particular characteristic of the odd-dimensional case that the $\mathsf{W}$ term in \eqref{GtildeSing} is a linear functional of $\mathsf{W}_0$. Moreover, $\mathsf{W}$ describes a homogeneous solution in the sense that $\textsf{L} (\Delta^{1/2} \mathsf{W}) = 0$. Neither of these properties holds when $n$ is even.

\subsubsection{Even spatial dimensions greater than three}

Now suppose that $\tilde{\textsf{G}}$ is a Green function and that $n > 3$ is even. The constant $c_n$ in these cases is still given by \eqref{c}, $\mathsf{U}_0 = 1$, and the $\mathsf{U}_{p}$ satisfy \eqref{UintOdd} for $p=0, \ldots, \frac{1}{2} n - 2$. Unlike when $n$ is odd, however, $\mathsf{U}_p = 0$ for all $p > \frac{1}{2} n-2$. Additionally, $\mathsf{V} \neq 0$ in general. Its first Hadamard coefficient is
\begin{align}
  \mathsf{V}_0 = - \frac{1}{2} \int_0^1 s^{\frac{1}{2} n - 2} \left[ \frac{ \textsf{L} (\Delta^{1/2} \mathsf{U}_{\frac{1}{2} n-2} ) }{ \Delta^{1/2} } \right] ds ,
  \label{V0int}
\end{align}
while the remaining coefficients follow by iteratively applying
\begin{equation}
  \mathsf{V}_{p+1} = - \int_0^1 \frac{ s^{\frac{1}{2} n + p - 1} }{ 2(p+1) } \left[ \frac{ \textsf{L} (\Delta^{1/2} \mathsf{V}_p) }{ \Delta^{1/2} } \right] ds
  \label{xPortV}
\end{equation}
for all $p \geq 0$. $\mathsf{W}_0$ is again arbitrary, while the higher-order $\mathsf{W}_p$ satisfy
\begin{align}
  \mathsf{W}_{p+1} = \int_0^1 s^{\frac{1}{2} n + p - 1} \Bigg\{ \frac{ \textsf{L} [\Delta^{1/2} ( \mathsf{V}_p - (p+1) \mathsf{W}_p ) ] }{ 2 (p+1)^2 \Delta^{1/2} }
  \nonumber
  \\
  ~ - \mathsf{V}_{p+1} \Bigg\} ds
  \label{xPortWeven}
\end{align}
for all $p \geq 0$. Note that $\mathsf{W}$ is generically nonzero even if $\mathsf{W}_0 = 0$.

\subsubsection{Two spatial dimensions}

The case $n = 2$ is slightly different from the other even-dimensional possibilities: The power law portion of \eqref{GtildeSing} vanishes and
\begin{equation}
  \tilde{\textsf{G}} = -\frac{1}{2} \Delta^{1/2} \left[ \mathsf{V} \ln ( \sigma / \ell^2 ) + \mathsf{W}\right].
\end{equation}
Here, $\mathsf{V}_0 = 1$, the higher-order $\mathsf{V}_p$ are determined by \eqref{xPortV}, $\mathsf{W}_0$ remains arbitrary, and the higher-order $\mathsf{W}_p$ satisfy \eqref{xPortWeven}.

\subsection{The Hadamard parametrix}
\label{Sect:HadamardDef}

The above discussion provides an algorithmic method to construct Green functions for the differential operators $L_\mathrm{sc}$ and $L_\mathrm{em}$ defined by \eqref{LDef} and \eqref{LDefEM}. The construction is not unique, however. Different choices for $\mathsf{W}_0$ lead to different Green functions, and the majority of these \textit{do not} satisfy the properties demanded in section \ref{Sect:mf}. In particular, it is difficult to choose $\mathsf{W}_0$ so that the symmetry condition \eqref{symassumption} is satisfied when $n$ is even; simple choices such as $\mathsf{W}_0 = 0$ generically fail. While conditions may be imposed which perturbatively guarantee symmetry up to some given order---see section \ref{Sect:symGreen}---it is not clear how to accomplish this more generally.

The simplest way to make progress\footnote{Similar issues arise in quantum field theory in curved spacetime in the point-splitting method of computing the renormalized expected stress-energy tensor \cite{WaldQFT}. There, as here, one needs to find a locally-constructed, bidistributional solution to the field equation. In that context, one chooses $\mathsf{W}_0=0$ since it is possible to accommodate a nonsymmetric Green function.} is to ignore $\mathsf{W}$ altogether. Removing it from \eqref{GtildeSing} defines the\footnote{If $\mathsf{V} \neq 0$, the lengthscale $\ell$ may be varied arbitrarily to produce a one-parameter family of Hadamard parametrices. We nevertheless refer to ``the'' Hadamard parametrix for simplicity. Although employing different values of $\ell$ to describe the same system might lead to, e.g., different ``self-forces,'' observables remain invariant as emphasized in section \ref{Sect:Ginvariance}.}  ``Hadamard parametrix'' $\tilde{\textsf{G}}_\subH$: It is explicitly
\begin{align}
  \tilde{\textsf{G}}_\subH \equiv c_n \Delta^{1/2} \bigg[ \frac{ \mathsf{U} }{  \sigma^{\frac{1}{2} n-1} }
  + \mathsf{V}  \ln ( \sigma/ \ell^2)  \bigg]
  \label{GParam}
\end{align}
if $n \neq 2$, and
\begin{equation}
  \tilde{\textsf{G}}_\subH \equiv -\frac{1}{2} \Delta^{1/2} \mathsf{V} \ln ( \sigma / \ell^2 )
  \label{GParam2D}
\end{equation}
otherwise. All biscalars here are the same as in the Green function case. That $\tilde{\textsf{G}}_\subH$ is indeed a parametrix follows from noting that $\Delta^{1/2} \mathsf{W}$ is smooth and that it remains smooth when acted on by $\textsf{L}$. If $n$ is odd, $\textsf{L} (\Delta^{1/2} \mathsf{W}) = 0$ so $\tilde{\textsf{G}}_\subH$ is actually a Green function in those cases.

\subsection{Suitability of the Hadamard parametrix for computing forces and torques}
\label{Sect:HadamardWorks}

We now explain why the Hadamard parametrix is an explicit example of a propagator which satisfies all constraints imposed in section \ref{Sect:mf}. More precisely, we consider the scalar and electromagnetic bidistributions
\begin{equation}
        G_\subH = (N N')^{-1/2} \tilde{G}_\subH, \qquad \mathcal{G}_\subH = (N N')^{1/2} \tilde{\mathcal{G}}_\subH,
        \label{HadamardProp}
\end{equation}
where $\tilde{G}_\subH$ and $\tilde{\mathcal{G}}_\subH$ are the Hadamard parametrices for $L_\mathrm{sc} = D^2 - \Lambda_\mathrm{sc}$ and  $L_\mathrm{em} = D^2 - \Lambda_\mathrm{em}$, respectively.

We first remark that these propagators depend only on $N$ and $h_{ab}$. That this is so is intuitively clear given that no non-geometric choices have been made\footnote{The lengthscale $\ell$ is not determined by the geometry, but is a constant and therefore does not affect our statement.}. It is also clear that our propagators transform appropriately under spatial diffeomorphisms. To see that they transform correctly under time rescalings with the form \eqref{tScale}, first note that the differential operator $\textsf{L}$ is independent of these scalings, and that $\tilde{\mathsf{G}}_\subH$ is as well. The required transformation laws \eqref{GScaleScalar} and \eqref{GscaleEM} are instead recovered by the factors of $N$ in \eqref{HadamardProp}.

Next, we verify that our propagators are quasilocal in $N$ and $h_{ab}$. This follows from  noting that the bitensors $\sigma$, $\Delta$, $\mathsf{U}$, and $\mathsf{V}$ from which $\tilde{\mathsf{G}}_\subH (x,x')$ is constructed depend only on quantities along the geodesic which connects $x$ and $x'$: The definition of $\sigma$ in terms of geodesic distance implies this immediately for the world function. The above integrals for the Hadamard coefficients show that it is also true for $\mathsf{U}$ and $\mathsf{V}$. Similarly, the van Vleck-Morette determinant satisfies the transport equation \cite{PoissonLR}
\begin{equation}
        \sigma_a D^a \ln \Delta  = ( n - D^2 \sigma ) ,
        \label{xPortVanVleck}
\end{equation}
together with $\Delta(x,x) = 1$, and can therefore be written as a similar integral with similar dependencies.

Lastly, it follows immediately from \eqref{GtildeScalar}, \eqref{GtildeEM}, and \eqref{HadamardProp} that $G_\subH$ and $\mathcal{G}_\subH$ are parametrices for $D^a (N D_a)$ and $D^a (N^{-1} D_a)$, respectively. Properties \ref{functionalAssume}-\ref{paramAssume} found at the end of section \ref{Sect:ForceScalar} are therefore satisfied for propagators defined in terms of Hadamard parametrices. This is true for both odd and even $n$.

As discussed in the body of the paper, those five properties are satisfied by many different choices of propagator; the Hadamard parametrix is just one example.  Other examples are straightforward to obtain. For example, in the scalar case for even $n$, one may consider
\be
G_\subH + \zeta \sigma^{n/2} (D_a D^a + D_{a'} D^{a'})^n \sigma,
\ee
where $\zeta$ is a dimensionless constant.  This example does not work for odd $n$ since the additional term is not smooth.  For odd $n$, an alternative propagator is
\be
G_\subH + \zeta \sigma^{(n-1)/2} (D_a D^a + D_{a'} D^{a'})^n \sigma,
\ee
where $\zeta$ is now a constant with dimensions of length.  There do not seem to be any natural examples for odd $n$ that do not involve the specification of a dimensionful parameter.

\subsection{Constructing a symmetric Green function}
\label{Sect:symGreen}

While useful propagators can always be constructed from Hadamard parametrices using \eqref{HadamardProp}, other choices are possible. In particular, it can sometimes be convenient to consider Green functions instead of more general parametrices. This would, e.g., allow effective fields to be computed exactly---and not only to leading order in the point particle limit---using surface integrals exterior to the body of interest [cf. \eqref{phiHatSelfsurf}]. The Hadamard parametrix is already a Green function for odd $n$, but not in general for even $n$, so we now describe how to systematically construct appropriate Green functions in even spatial dimensions.

As alluded to at the beginning of section \ref{Sect:HadamardDef}, the difficulty when starting from the general Hadamard procedure is to pick a $\mathsf{W}_0$ such that the resulting $\tilde{\mathsf{G}}$ is symmetric in its arguments\footnote{It is straightforward to find symmetric Green functions as solutions to boundary-value problems, although it is difficult in those contexts to enforce a quasilocal dependence only on $N$ and $h_{ab}$.}. We do not know how to do so non-perturbatively, but can derive appropriate constraints order by order in a Taylor expansion. These constraints become increasingly complicated at higher orders, so we illustrate the procedure only in the simplest cases. Our method expands on a calculation by Brown \cite{SymHadamard} which was in the context of quantum field theory.

First, we note that $\mathsf{W}(x,x')$ is a regular biscalar and suppose that it has a covariant Taylor expansion with the form
\begin{align}
\label{eq:TaylorExp}
        \mathsf{W} (x,x')= \mathsf{w} (x)+\sum_{p=1}^{\infty}\frac{1}{p!} \mathsf{w}_{a_{1} \cdots a_{p}}(x) \sigma^{a_{1}}\cdots\sigma^{a_{p}},
\end{align}
where $\mathsf{w}(x) = \mathsf{W} (x,x) = \mathsf{W}_{0} (x,x)$ and the higher-order coefficients are ordinary tensors at $x$. In the language of section \ref{Sect:Taylor}, these coefficients are, up to sign, tensor extensions evaluated using $h_{ab}$: $\mathsf{w}_{a_1 \cdots a_p} = (-1)^p \mathsf{W}_{,a_1 \cdots a_p}$. They may be found by, e.g., differentiating both sides of \eqref{eq:TaylorExp} and applying the coincidence limit $x' \rightarrow x$. If we require $\mathsf{W}(x,x')$ to be symmetric in its arguments, equating its Taylor series to that of $\mathsf{W}(x',x)$ yields the constraints
\begin{align}
        [(-1 & )^p  -1] \mathsf{w}_{a_{1} \cdots a_{p}} = D_{(a_{1}} \cdots D_{a_{p})} \mathsf{w}
        \nonumber\\
        & ~ + \sum_{m=1}^{p-1}\binom{p}{m} D_{(a_{1}} \cdots D_{a_{p-m}} \mathsf{w}_{ a_{p-m+1} \cdots a_{p}) },
\end{align}
which determine all odd-order coefficients in terms of the lower-order coefficients.

Since all of the $\mathsf{W}_{p}$ are fixed once $\mathsf{W}_{0}$ is specified, the symmetry constraints on $\mathsf{W}$ can be translated into constraints on the Taylor coefficients of $\mathsf{W}_{0}$. If we require symmetry only through second order in $\sigma^a$, choices with the form
\begin{align}
        \mathsf{W}_{0} (x,x') = \mathsf{w}(x)-\frac{1}{2} D_{a} \mathsf{w}(x) \sigma^{a} + \frac{1}{2} \mathsf{w}_{ab}^0 (x) \sigma^a \sigma^b + \ldots
\end{align}
guarantee symmetry for any $\mathsf{w}(x)$ and any $\mathsf{w}_{ab}^0 (x)$. The remaining propagator requirements are then satisfied if these functions depend only on $N$ and $h_{ab}$, and only quasilocally. Although it is consistent here to simply let both functions vanish, doing so can lead to inconsistencies at third order in $\sigma^a$. Expanding through that order requires the solution of a nontrivial constraint involving $D_a \mathsf{w}^0_{bc}$ together with $\mathsf{w}$ and its first three derivatives \cite{SymHadamard}. Although conceptually straightforward, some dedication is required to find analogous constraints at higher orders. In most cases, it is far more efficient to use the Hadamard parametrices described above.


%

\section{TIME INTEGRAL OF DETWEILER-WHITING GREEN FUNCTION IS AN APPROPRIATE STATIC PROPAGATOR}
\label{Sect:DW}

As discussed in section \ref{Sect:noStatic}, in even spacetime dimensions, there exists a Detweiler-Whiting Green function. In a static system, the Detweiler-Whiting prescription would require that scalar forces and torques be computed by removing the $S$-field
\begin{equation}
  \phi_\subS(x) = \int_I dt' \int_{\Sigma_t} \! dV'_{\bperp}  \rho(x') N(x') G_\subDW(x,x')
\end{equation}
from the physical one. This equation is equivalent to \eqref{phiSDef} if the static propagator $G$ is identified with a time integral of $G_\subDW$. More precisely, we note that the Detweiler-Whiting Green function depends only on $g_{ab}$ and set
\begin{align}
        \label{eq:GstaticDW}
        G[N,h_{ab}](x,x') = \int_I G_{\subDW}[ h_{ab} - N^2 \nabla_a t \nabla_b t ](x,x') dt'
\end{align}
for all static metrics $g_{ab} = h_{ab} - N^2 \nabla_a t \nabla_b t$ on the manifold $\Sigma \times I$. The time coordinate $t$ is assumed to be fixed. We now show that this propagator satisfies the five properties summarized at the end of section \ref{Sect:ForceScalar}.

Our first task is to show that $G$ is spatial. To see this, note that in all static spacetimes, the Detweiler-Whiting Green function must satisfy $\mathcal{L}_\tau G_\subDW(x,x') = 0$. It can therefore depend on $t$ and $t'$ only in the combination $t-t'$. For fixed $\bx$ and $\bx'$, it also vanishes for sufficiently large $|t-t'|$. As long as $t$ is not too close to a boundary of $I$, the time integral in \eqref{eq:GstaticDW} is independent of $t$ and so the left-hand side can be interpreted as a bidistribution on $\Sigma \times \Sigma$.

The propagator $G$ is manifestly well-behaved under spatial diffeomorphisms, and also transforms appropriately under the time rescalings \eqref{tScale}. That the Detweiler-Whiting Green function is symmetric in its arguments additionally implies the symmetry of $G$. Furthermore, applying $N \nabla^a \nabla_a$ to the left-hand side of \eqref{eq:GstaticDW} shows that $G$ satisfies \eqref{GDef}, and is therefore a parametrix---really a Green function---for $D^a (N D_a)$.

Lastly, it is clear by construction that $G$ is functionally dependent only on $N$ and $h_{ab}$. That this dependence is quasilocal follows from the fact that the $G_\subDW(x,x')$ can be expanded in a Hadamard series in a fashion analogous to what was described in appendix \ref{Sect:Hadamard}, and arguments similar to those used there show that it can depend on the geometry only along the (spacetime) geodesic which connects $x$ to $x'$. It follows that $G(\bx,\bx')$ can depend on $N$ and $h_{ab}$ only along the spatial projections of all spacetime geodesics connecting points $(t,\bx)$ to $(t',\bx)$ which are not timelike-separated. For fixed $\bx$ and $\bx'$, the set of all such paths has finite size, so the dependence on the geometry is indeed quasilocal.

\section{SELF-FORCE IN RINDLER USING HADAMARD PARAMETRICES}
\label{Sect:RindlerAppendix}

We showed in section \ref{Sect:Rindler} that in Rindler spacetime, the Frolov-Zelnikov Green function (\ref{Gfrolov}) satisfies our criteria to be a valid  propagator with which to construct effective self-fields. If the boundary conditions are such that this Green function generates the physical self-field, it immediately followed that the associated self-forces and self-torques must vanish. In this appendix, we revisit the problem of static scalar charges in Rindler spacetime using a different $G$. For all $n > 2$, we identify this with the Hadamard parametrix discussed in appendix \ref{Sect:Hadamard}. The associated self-force no longer vanishes in this case, although it is compensated by an appropriate shift in the effective mass.

We first change the argument of the Legendre function in \eqref{Gfrolov} from $\coth \eta$ to $\cosh \eta$, which, recalling that $\eta$ is defined by \eqref{eq:eta}, provides a much simpler representation in terms of $\sigma = \frac{1}{2} |\bx-\bx'|^2$ and $y y'$: Employing the Whipple transformation for Legendre functions found in, e.g., 3.3(14) of \cite{Erdelyi} shows that
\begin{align}
G_{\textrm{self}}=\frac{e^{-i \pi m}}{\Gamma(m+1)\sqrt{NN'}}\frac{Q_{-\frac{1}{2}}^{ m }(\cosh\eta )}{(2 y y' \sinh\eta)^m },
\label{GRindlerQ}
\end{align}
where $m \equiv n/2-1$ (which is not to be confused with a mass).

\subsection{Odd spatial dimensions}

If $n$ is odd, $G_\mathrm{self}$ can be written in the form of a convergent series by noting that for $\nu-\mu$ a negative integer, $Q^{\mu}_{\nu}(\zeta)$ has the hypergeometric representation
\begin{align}
e^{-i\,\mu\pi}Q_{\nu}^{\mu}(\zeta) = \frac{ \Gamma(\mu) }{ 2 }\Big(\frac{\zeta+1}{\zeta-1}\Big)^{\frac{\mu}{2}}F(-\nu,1+\nu,1-\mu,\tfrac{1-\zeta}{2}),
\end{align}
so
\begin{align}
\label{GRindlerEven}
G_{\textrm{self}} = \frac{c_n  }{\sqrt{N N'}}\sum_{p=0}^{\infty}\frac{ \Gamma( p+\frac{1}{2} )^{2} \Gamma( m - p) }{ \pi p! \Gamma(m) (2 y y')^{p}}\sigma^{p-m},
\end{align}
where $c_n$ is given by \eqref{c}.

Let us turn now to $G$, which we identify in this appendix with the Hadamard parametrix \eqref{HadamardProp}. More precisely, we let $G = (N N')^{-1/2} \tilde{G}_\subH$, where $\tilde{G}_{\subH}$ is the Hadamard parametrix for the differential operator $L_{\textrm{sc}}=D^{2}+1/4y^{2}$. This has the explicit form \eqref{GParam}, where $V_\mathrm{sc}$ vanishes for all odd $n$ and $U_\mathrm{sc}$ is determined by the series \eqref{Useries} in terms of the Hadamard coefficients $U_{p}^\mathrm{sc}$. These coefficients in turn satisfy \eqref{UintOdd}, which can be evaluated in closed form to yield
\begin{align}
\label{eq:UkClosed}
        U_{p}^\mathrm{sc} =\frac{\Gamma(p+\tfrac{1}{2})^{2}\Gamma(m-p)}{\pi p! \Gamma( m) } \frac{1}{(2 y y')^{p}}.
\end{align}
This results in a series for $G$ which is identical to the series \eqref{GRindlerEven} for $G_\mathrm{self}$. The Hadamard parametrix is therefore identical to the Frolov-Zelnikov Green function we use to generate the self-field, and the prescription adopted here is identical to the one discussed in section \ref{Sect:Rindler}.

Another potential approach to this problem could be to identify $G$ with the time integral of the Detweiler-Whiting Green function $G_\subDW$. A straightforward calculation shows that this too recovers $G_\mathrm{self}$; the Detweiler-Whiting field for a static charge in Rindler spacetime charge coincides with the field obtained from our Hadamard Green function. As remarked in section \ref{Sect:noStatic}, we believe this agreement holds also in more general spacetimes.

\subsection{Even spatial dimensions}

If $n$ is even, $m=n/2-1$ reduces to an integer. For $p\ge m$, the $U_{p}^\mathrm{sc}$ coefficients vanish identically, while for $0\le p\le m-1$, they are given by (\ref{eq:UkClosed}). The Hadamard function $V_\mathrm{sc}$ is nonzero in this context, and the associated Hadamard coefficients $V_p^\mathrm{sc}$ are determined by the integrals \eqref{V0int} and \eqref{xPortV}. Evaluating these integrals yields
\begin{align}
        V_{p}^\mathrm{sc}  = \frac{(-1)^{p+1}\Gamma(m+p+\tfrac{1}{2})^{2}}{\pi p! (m-1)! (m+p)!}\frac{1}{(2 y y')^{m+p}},
\end{align}
which results in a Hadamard series for $V_\mathrm{sc}$ which can summed in closed form by comparing with the hypergeometric series representation for $P^{m}_{-1/2}(\cosh\eta)/(2 y y' \sinh\eta)^{m}$. This representation is identical, up to an overall constant, to our Hadamard series for $V_\mathrm{sc}$, and results in
\begin{align}
\label{eq:VRindler}
        V_\mathrm{sc} = \frac{(-1)^{m-1}2^{m}P^{m}_{-\frac{1}{2}}(\cosh\eta)}{(m-1)! (2 y y' \sinh\eta)^{m}}.
\end{align}
Combining these results finally shows that the Hadamard parametrix in Rindler spacetime is explicitly
\begin{align}
G = \frac{c_n}{\sqrt{N N'}} \Bigg[\sum_{p=0}^{m-1} \frac{ \Gamma( p+\tfrac{1}{2} )^{2} (m-p-1)! }{ \pi p! (m-1)!} \frac{ \sigma^{p-m} }{ (2y y')^p }\nonumber\\
~ -\frac{(-1)^{m} 2^{m} P^{m}_{-\frac{1}{2}}(\cosh\eta)}{(m-1)!(2 y y' \sinh\eta)^{m} }\ln \left( \sigma/\ell^2 \right) \Bigg].
\end{align}

We now use the fact that for $m$ a non-negative integer, the associated Legendre function of the second kind can be expressed as (cf. 3.6(11) in \cite{Erdelyi})
\begin{widetext}
\begin{align}
Q_{-\frac{1}{2}}^{m}( \zeta ) = P_{ -\frac{1}{2} }^{m} (\zeta) \left[ \frac{1}{2} \ln \left(\frac{\zeta+1}{\zeta-1} \right) - \gamma - \psi( m + \tfrac{1}{2} ) \right] + \frac{\Gamma(\tfrac{1}{2}+m)}{\Gamma(\tfrac{1}{2}-m)}\left( \frac{\zeta-1}{\zeta+1} \right)^{\frac{m}{2}}\sum_{p=0}^{\infty}\frac{(-1)^{p}\Gamma(p+\tfrac{1}{2})^{2}H_{m+p}}{\pi 2^{p+1} p! (m+p)!}(\zeta-1)^{p}\nonumber\\
~ +\left(\frac{\zeta+1}{\zeta-1}\right)^{\frac{m}{2}} \left[ \sum_{p=0}^{m-1}\frac{(-1)^{m}\Gamma(p+\tfrac{1}{2})^{2}(m-p-1)!}{\pi 2^{p+1}p!}(\zeta-1)^{p}+\sum_{p=1}^{\infty}\frac{(-1)^{m+p}\Gamma(m+p+\tfrac{1}{2})^{2}H_{p}}{\pi 2^{m+p+1} p! (m+p)!} (\zeta-1)^{m+p} \right] ,
\end{align}
which permits the representation
\begin{align}
\label{eq:GsRindler}
        G = G_\mathrm{self} + \frac{c_n}{\sqrt{N N'}}\Bigg\{ \left[ \ln \left( \frac{ y y'}{ \ell^2 } \left( 2+ \sigma/yy' \right) \right) - 2 \gamma - 2 \psi( m + \tfrac{1}{2} )\right] V_\mathrm{sc} - \sum_{p=1}^{\infty}\frac{ \Gamma(m+p+\tfrac{1}{2})^{2}H_{p} }{\pi p!(m-1)!(m+p)! } \frac{ (-\sigma)^{p} }{ (2 y y')^{p+m} }
        \nonumber\\
        ~ -\sum_{p=0}^{\infty}\frac{2^m \Gamma(m+\tfrac{1}{2})^{2} \Gamma(p+\frac{1}{2})^{2}H_{p+m} }{ \pi^{2} p! (m-1)! (m+p)! (1+ \sigma/y y')^m} \frac{ (-\sigma)^{p} }{ (2 y y')^{p+m} } \Bigg\}.
\end{align}
\end{widetext}
In these expressions, $V_\mathrm{sc}$ is given explicitly by (\ref{eq:VRindler}), $\gamma$ is Euler's constant, $\psi(\zeta)\equiv\Gamma'(\zeta)/\Gamma(\zeta)$ is the Digamma function, and $H_{p}\equiv\sum_{j=1}^{p}j^{-1}$ is the $p^{\textrm{th}}$ Harmonic number. We have also used \eqref{GRindlerQ} for $G_\mathrm{self}$. The effective self-field $\hat{\phi}_\mathrm{self}$ is now generated by the propagator $G_\mathrm{self} - G$, which is nonsingular throughout the interior of the Rindler wedge.

Recall that in the point particle limit, the self-force is given by $q \nabla_a \hat{\phi}_\mathrm{self}$ through leading nontrivial order. This depends explicitly on $G_\mathrm{self} - G$ via \eqref{phiSelfpp}, and is easily computed using \eqref{eq:VRindler} and \eqref{eq:GsRindler}. The result is not particularly enlightening, although we do note that it is nonzero in general. Moreover, it depends logarithmically on the arbitrary parameter $\ell$. The various parametrices defined by different values of $\ell$ (and the parametrix $G_\mathrm{self}$ used for $G$ in section \ref{Sect:Rindler}) are each associated with different definitions for the mass, and the holding force remains invariant under these transformations even while the self-force does not.

We also note that the results of section \ref{Sect:Rindler} can be nonperturbatively recovered from the perspective of the Hadamard construction. This is simplest to see by allowing for a nontrivial $W_\mathrm{sc}$ in \eqref{GtildeSing}, which can be chosen in this context so that the resulting Green function coincides with $G_\mathrm{self}$. That $W_\mathrm{sc}$ is symmetric in its arguments and quasilocally constructed from $N$ and $h_{ab}$. It is not clear, however, which $W_0^\mathrm{sc}$ would be associated with it.

\section{VARIATIONAL DERIVATIVES OF THE HADAMARD PARAMETRIX}
\label{Sect:VarDerivs}

The renormalization of a body's stress-energy tensor as derived in section \ref{Sect:mf} depends on the variational derivatives of the propagator $G[N,h_{ab}]$ with respect to the spatial metric and the lapse function. In this appendix, we compute those variational derivatives within a certain approximation which is sufficient to describe shifts in the mass and the stress-energy quadrupole moment to leading order in a point particle limit---shifts which are computed explicitly in section \ref{Sect:bodyR}.

We specialize to the scalar case and to $n>2$, and also set the propagator to be the Hadamard parametrix described in appendix \ref{Sect:Hadamard}; hence, $G = G_\subH$. Following (\ref{GtildeScalar}) and (\ref{GParam}), the scalar Hadamard parametrix is explicitly
\begin{equation}
G_\subH[N,h_{ab}] = c_n \sqrt{\frac{\Delta}{N N'}} \left[ \frac{ U_\mathrm{sc} }{ \sigma^{\frac{n}{2} - 1} } + V_\mathrm{sc} \ln (\sigma/\ell^2) \right],
  \label{GHadamardSC}
\end{equation}
where $c_n$ is the constant \eqref{c}, $\Delta[h_{ab}]$ is the spatial van Vleck-Morette determinant \eqref{vanVleckDef}, $\sigma[h_{ab}]$ the spatial world function, $U_\mathrm{sc}[N,h_{ab}]$ and $V_\mathrm{sc}[N,h_{ab}]$ are appropriate biscalars, and $\ell$ is an arbitrary constant with dimensions of length. Each choice of $\ell$ technically defines a different parametrix, and therefore a different renormalization.

We now compute the variational derivatives of the propagator (\ref{GHadamardSC}) using two simplifications.  First, we
specialize to linear perturbations about a flat spatial geometry
with a trivial lapse function, and we specialize the coordinates
so that the background quantities are $h_{ij} = \delta_{ij}$ and $N=1$.
At the end of the appendix we will
discuss variational derivatives about more general backgrounds,
and explain why the flat space, unaccelerated variational derivatives
are sufficient for our renormalization computations.

The second simplification involves a truncation of the Hadamard series:
The biscalars $U_\mathrm{sc} [N,h_{ab}]$ and $V_\mathrm{sc} [N,h_{ab}]$ in \eqref{GHadamardSC} have been defined only via the Hadamard series \eqref{Useries}, so the only clear way to compute their variational derivatives is to vary the Hadamard coefficients $U^\mathrm{sc}_p [N,h_{ab}]$ and $V_p^\mathrm{sc} [N,h_{ab}]$, and then to sum---or to approximate the sum of---the resulting series. Attempting to formally carry this out results in a series which involves arbitrarily-many derivatives of Dirac $\delta$-distributions.  We truncate this series by omitting all terms which involve more than two derivatives of Dirac distributions.  The justification for this is discussed at the end of the appendix.

Using static Minkowski coordinates $(t,\bx)$, we denote by $\bfv$ and $\bfw$ the spatial coordinates of the two arguments of the propagator (\ref{GHadamardSC}). Varying with respect to the spatial metric, $h_{ij} \to \delta_{ij} + \delta h_{ij}$, shows that
\begin{align}
\delta G_\subH(\bfv,\bfw) =  c_n \bigg[ \frac{ \delta (\Delta^{1/2})}{\sigma^{\frac{n}{2}-1}} + \frac{\delta {U}_{\rm sc}}{\sigma^{\frac{n}{2}-1}}
 - \frac{(n-2) \delta \sigma }{2 \sigma^{\frac{n}{2} } }
\nonumber \\
 ~ +  \delta V_{\rm sc} \ln (\sigma/\ell^2)  \bigg],
\label{zzz}
\end{align}
where we have used the fact that the unvaried biscalars are $\Delta[\delta_{ij}] = 1$, $U_{\rm sc} [1,\delta_{ij}] =1$, and $V_{\rm sc} [1,\delta_{ij}] =0$ in the flat, unaccelerated background which has been assumed. We now evaluate these terms one by one.

The second-to-last term in (\ref{zzz}) is the simplest to compute.
The variation of Synge's world function is
\be
        \delta \sigma = \frac{1}{2} \int_0^1 ds \, r^i r^j \delta h_{ij}( \bx_s ),
        \label{dsigma}
\ee
where $\bfr \equiv \bfv - \bfw$ and
\be
\bx_s \equiv \bfw + s \bfr
\label{flatgeodesic}
\ee
is the affinely-parametrized geodesic joining the points $\bfw = \bx_0$ and $\bfv = \bx_1$ in the background space. Also note that the unvaried world function is explicitly $\sigma(\bfv,\bfw) = r^2/2$, where $r \equiv | {\bf r}|$.  The formula (\ref{dsigma}) can be obtained by directly varying the definition of $\sigma$ given by Eq.\ (3.1) of Ref.\ \cite{PoissonLR},
or by varying the identity $D^a \sigma[h_{cd}] D_a \sigma[h_{cd}] = 2\sigma[h_{cd}]$ \cite{PoissonLR} to obtain a transport equation for $\delta\sigma$.

Similarly, from the definition (\ref{vanVleckDef}) of the van Vleck-Morette determinant we obtain
\begin{align}
  & \delta \Delta =  \int_0^1 ds \bigg[ \delta h(\bx_s) +  (2s-1) r^j D^i \delta h_{ij} (\bx_s)
  \nonumber
  \\
  & ~ - \frac{1}{2} s (1-s) r^i r^j D^2 \delta h_{ij} (\bx_s) \bigg] - \frac{1}{2} [ \delta h(\bfv) + \delta h(\bfw) ],
  \label{dDelta}
\end{align}
where $D^2 = \delta^{ij} \partial_i \partial_j$ and $\delta h \equiv
\delta^{ij} \delta h_{ij}$.

Next we vary $U_\mathrm{sc}$ with respect to the spatial metric. Recalling \eqref{UintOdd} and our aforementioned criterion regarding the retained derivatives of Dirac distributions, it follows that only the zeroth and first-order terms in the Hadamard series must be varied. The zeroth-order term is $U^\mathrm{sc}_0 [N,h_{ab}] = 1$ for all $N$ and all $h_{ab}$, so its variation trivially vanishes. Using \eqref{xPortUodd}, the variation of the first-order term is instead
\begin{align}
  \delta U_1^\mathrm{sc} &= \frac{1}{2 (n-4)} \int_0^1 ds  D^2 \delta
  \Delta (\bx_s, \bfw)
  \label{dU1}
\end{align}
if $n \neq 4$, where the Laplacian is understood to act on the first argument of $\delta \Delta$. If $n=4$ however, $U^\mathrm{sc}_1 [N,h_{ab}]=0$ for all metrics and so $\delta U^\mathrm{sc}_1 = 0$. Defining
\be
g_n =
\left\{
        \begin{array}{lll}
                1/(4-n)  && \mbox{if } n \ne 4, \\
                0 && \mbox{if } n = 4,
        \end{array}
\right.
\label{gndef}
\ee
it follows from \eqref{dDelta} and \eqref{dU1} that
\begin{widetext}
\begin{align}
  \delta U_\mathrm{sc} = \frac{ g_n r^2 }{ 16 } \int_0^1 ds \{ s^2 (1-s)^2 r^i r^j D^4 \delta h_{ij}(\bx_s) + 4 s (1-s) (1-2s) r^i D^j D^2 \delta h_{ij} (\bx_s) +  [2-4s(1-s)] D^2 \delta h (\bx_s)
  \nonumber
  \\
  ~ - 8 s (1-s) D^i D^j \delta h_{ij} (\bx_s) \}.
  \label{second}
\end{align}
Terms involving second and higher-order terms in the Hadamard series have been omitted here.

If $n \neq 4$, these expressions are all that are needed to evaluate $\delta G_\subH/\delta h_{ij}$ in the approximation used in section \ref{Sect:bodyR}. The variation of $V_\mathrm{sc}$ is important only if $n=4$, and approximating it in that case by $\delta V_0^\mathrm{sc}$, it follows from \eqref{V0int} that $\delta V_\mathrm{sc}$ is proportional to a line integral of $D^2 \delta \Delta$. More precisely, it is given by the right-hand side of \eqref{second} with the $g_n r^2$ prefactor removed.  Now inserting \eqref{c}, \eqref{dsigma}, \eqref{dDelta}, and \eqref{second} into \eqref{zzz} shows that for all $n > 2$,
\begin{align}
        \frac{\delta G_\subH(\bfv,\bfw)}{\delta h_{ij}(\bfx)} = -\frac{1}{4 (n-2) r^n} \bigg\{  r^2 \delta^{ij} [ \delta_\Sigma (\bfx, \bfv) + \delta_\Sigma (\bfx, \bfw) ] -
                2 \int_0^1 ds \Big( r^2 \delta^{ij} - r^i r^j \big[ (n-2) + \frac{1}{2} s (1-s) r^2 D_x^2 \big]
\nonumber
\\
~  - (2s-1) r^2 r^{(i} D^{j)}_x  \Big) \delta_\Sigma (\bx, \bfw + s \bfr) + \frac{r^n}{4} \left[ \frac{ 2g_n }{ r^{n-4} } + \delta_{n,4} \ln \left( \frac{r^2}{2 \ell^2} \right) \right] \int_0^1 ds \Big( 4 s (1-s) D^i_x D^j_x
\nonumber
\\
~  + \big[ 2s (1-s) -1 \big] \delta^{ij} D^2_x \Big) \delta_\Sigma (\bx, \bfw + s \bfr) \bigg\} ,
\label{vd1}
\end{align}
where again, $\bfr = \bfv - \bfw$ and third and higher derivatives of Dirac distributions have been omitted.

Next we turn to the variational derivative of $G_\subH$ with respect to the lapse function $N$, varied so that $N \rightarrow 1 + \delta N$.  The Hadamard parametrix depends on the lapse through the explicit prefactors in \eqref{GHadamardSC}, and also through the potential $\Lambda_\mathrm{sc}$ in (\ref{LDef}) that enters into the differential operator $L_\mathrm{sc}$ which affects $U_\mathrm{sc} [N,h_{ab}]$ and $V_\mathrm{sc} [N,h_{ab}]$ via (\ref{UintOdd}) and (\ref{V0int}). Taking this into account, a calculation similar to the one for $\delta G_\subH/\delta h_{ij}$ gives
\begin{align}
        \frac{\delta G_\subH(\bfv,\bfw)}{\delta N(\bfx)} = - \frac{1}{ 2 (n-2) r^{n-2} }  \bigg\{ \delta_\Sigma (\bx, \bfv) + \delta_\Sigma (\bx,\bfw) - \frac{ r^{n-2} }{ 4 } \left[ \frac{ 2 g_n }{ r^{n-4} } + \delta_{n,4} \ln \left( \frac{r^2}{2 \ell^2} \right) \right] \int_0^1 ds D^2_x \delta_\Sigma (\bx, \bfw + s \bfr) \bigg\},
\label{vd2}
\end{align}
\end{widetext}
where higher-derivatives terms have again been omitted.

Although these calculations have all been performed for scalar fields, they are easily adapted to the electromagnetic case:  We note that the electromagnetic Hadamard parametrix $\mathcal{G}_\subH [N,h_{ab}]$, which is given by \eqref{GParam} and \eqref{HadamardProp}, can be obtained from the scalar parametrix $G_\subH [N,h_{ab}]$ using the substitution $N \to 1/N$. It follows that the electromagnetic variational derivatives in the flat, unaccelerated limit are just
\be
\label{EMVD}
\frac{\delta \mathcal{G}_\subH(\bfv,\bfw)}{\delta h_{ij}(\bfx)}=
\frac{\delta G_\subH(\bfv,\bfw)}{\delta h_{ij}(\bfx)},
\quad
\frac{\delta \mathcal{G}_\subH(\bfv,\bfw)}{\delta N(\bfx)}=
-\frac{\delta G_\subH(\bfv,\bfw)}{\delta N(\bfx)}.
\ee

We now explain why the simplifications used in the above computations---specialization to flat, unaccelerated backgrounds and truncation at two derivatives---are sufficient for the renormalization computations in the body of the paper.  For this explanation it is helpful to consider variational derivatives about general backgrounds $(h_{ij},N)$.
For each integer $p \ge 0$,
we define the set ${\mathcal F}_p$ of functionals $F = F[h_{ij},N]$ that
are symmetric bidistributions on $\Sigma$ by the requirement that the variation
of $F$ under $h_{ij} \to h_{ij} + \delta h_{ij}$, $N \to N + \delta N$ is given by
\begin{eqnarray}
&&\delta F(x,x') = \sum_{q=0}^p \int_0^1 ds \bigg[
H^{K''}(x,x',x_s'',s) D_{K''} \delta N(x_s'')
\nonumber \\
&&
+{\tilde H}^{i''j'' K''}(x,x',x_s'',s) D_{K''} \delta h_{i''j''}(x_s'') \bigg].
\label{formal}
\end{eqnarray}
Here $K''$ means the sequence of indices $k_1'' \ldots k_q''$,
$D_{K''} = D_{k_1''} \ldots D_{k_q''}$, $x_s''$ for $0 \le s \le 1$ is the affinely parameterized geodesic joining $x$ to $x'$, and $H^{K''}(x,x',x'',s)$ and ${\tilde H}^{i''j'' K''}(x,x',x'',s)$ are some smooth tritensors on $\Sigma$.
In other words, functionals in ${\mathcal F}_p$ have variations which consist of integrals along the geodesic joining $x$ and $x'$ of derivatives of the variations $\delta h_{ij}$ and $\delta N$ up to $p$th order.
One can show that ${\mathcal F}_p$ is closed under simple algebraic operations, and taking covariant derivatives with respect to $x$ or $x'$ maps ${\mathcal F}_p$ to ${\mathcal F}_{p+1}$.  Finally one can show that
the type of operation on functionals given in (\ref{UWints})
(\ref{V0int}) and (\ref{xPortV}) maps ${\mathcal F}_p$ to ${\mathcal F}_{p+2}$.

The variation of Synge's world function is still given by (\ref{dsigma}) (with $r^i$ replaced by $dx^i/ds$), and so $\sigma$ is an element of ${\mathcal F}_0$.  It follows using the definition (\ref{vanVleckDef}) that $\Delta$ is an element of ${\mathcal F}_2$, and we obtain from the Hadamard construction that $U^{\mathrm{sc}}_p$ lies in ${\mathcal F}_{2p+2}$ and $V^{\mathrm{sc}}_p$ lies in ${\mathcal F}_{2p+4}$.

Consider now the evaluation of stress-energy moments using the expression (\ref{big}) for the renormalization of the stress-energy tensor.
We wish to consider the limit $\lambda \to 0$ of such moments.
Note that this involves a weak limit of the distributional quantities which appear in the third line of (\ref{big}), not a pointwise limit.
The explicit expression for a renormalized stress-energy moment
will be given by inserting a variational derivative obtained from
an expression of the form (\ref{formal})
into (\ref{big}) and then into (\ref{Jmoments}).
The arguments of the tritensors $H^{K''}$ and ${\tilde H}^{i''j''K''}$ in (\ref{formal}) will then contain explicit factors of $\lambda$. Therefore, by local flatness and by smoothness of the background lapse function, to leading order in $\lambda$ (in a weak limit sense) these tritensors can be replaced by their flat space, unaccelerated limits.  Similarly the geodesic $x_s''$ can be replaced by its flat space version (\ref{flatgeodesic}).  In other words, one can use the flat space, unaccelerated variational derivatives (\ref{vd1}) and (\ref{vd2}) computed above, to leading order in $\lambda$, interpreting the coordinates $(t,\bx)$ in these expressions to be the Fermi normal coordinates defined after (\ref{ffamily0}).

Finally, we can omit all terms in $\delta G_\subH/\delta h_{ij}$ or $\delta G_\subH/\delta N$ which involve more than two derivatives of Dirac distributions.  This is because integrations by parts with respect to $\tilde {\bx}$ in multipole expressions obtained from (\ref{big})
show that such terms cannot contribute to renormalizations of the quadrupole and lower-order moments.

As a final remark, we note that one might have naively attempted to avoid Hadamard parametrices by instead building a family $G_0[N,h_{ab}]$ of propagators using the methods of perturbation theory. Suppose that $G_0[1,\delta_{ij}] = c_n \sigma[\delta_{ij}]^{1- \frac{n}{2}}$, so the usual propagator is recovered in a flat, unaccelerated background. Also suppose that this family of propagators is more generally a symmetric Green function in the sense that it satisfies \eqref{GDef} for all $N$ and $h_{ab}$. Varying this equation with respect to $N$ off of a background in which $N=1$ and $h_{ij} = \delta_{ij}$ shows that $D^2 \delta G_0 = \omega_n \delta_\Sigma(x,x') \delta N - D^a \delta N D_a G_0 $. From the viewpoint of perturbation theory, perhaps the most natural solution this this equation treats the entire right-hand side as a source and integrates it against the background $G_0[1,\delta_{ij}]$. Using such a procedure to define $G_0[1 + \delta N, \delta_{ij}]$, it follows that
\begin{align}
  \frac{ \delta G_0 (\bfv, \bfw) }{ \delta N (\bx) } = - \frac{1}{\omega_n}  \delta_{ij} D^i_x G_0 (\bfv,\bx) D_x^j G_0 (\bx,\bfw)
\end{align}
A similar expression may also be obtained for variations with respect to $h_{ij}$. In either case, fixing $\bfv$ and $\bfw$ results in variations which do not have compact support in $\bx$. The family of propagators which is obtained in this way therefore fails to satisfy the constraints summarized at the end of section \ref{Sect:ForceScalar}, implying that the boundary conditions implicit in such a construction are inappropriate for our purposes. The Hadamard family of propagators is different and does not share this problem.

\bibliographystyle{apsrev4-1}
\bibliography{selfforce}

\end{document}